\documentclass[twocolumn]{aastex63}
\usepackage{amsmath}
\usepackage{here}

\newcommand{\gas}{\mathrm{g}}
\newcommand{\dst}{\mathrm{d}}

\newcommand{\sigmag}{\Sigma_{\gas}}
\newcommand{\sigmagup}{\Sigma_{\gas,0}}
\newcommand{\sigmad}{\Sigma_{\dst}}
\newcommand{\sigmadup}{\Sigma_{\dst,0}}
\newcommand{\tstop}{t_{\mathrm{stop}}}
\newcommand{\edd}{\mathrm{eddy}}
\newcommand{\st}{\mathrm{St}}
\newcommand{\taus}{\tau_{\mathrm{s}}}
\newcommand{\tausup}{\tau_{\mathrm{s},0}}
\newcommand{\cs}{c_{\mathrm{s}}}

\newcommand{\hd}{H_{\mathrm{d}}}

\newcommand{\up}{\mathrm{unp}}
\newcommand{\app}{\mathrm{ap}}
\newcommand{\gdl}{\mathrm{gdl}}
\newcommand{\vpp}{v_{\mathrm{pp}}}
\newcommand{\mpar}{m_{\mathrm{p}}}
\newcommand{\rhoint}{\rho_{\mathrm{int}}}

\newcommand{\ddx}[1]{\frac{\partial #1}{\partial x}}
\newcommand{\ddt}[1]{\frac{\partial #1}{\partial t}}
\newcommand{\Lddt}[1]{\frac{d #1}{dt}}
\newcommand{\wmod}{W_{\mathrm{mod}}}
\newcommand{\epmid}{\varepsilon_{\mathrm{mid}}}
\newcommand{\epmidup}{\varepsilon_{\mathrm{mid},0}}
\newcommand{\sigsub}{\Sigma_{\gas,\mathrm{s}}}
\newcommand{\sigsubup}{\Sigma_{\gas,\mathrm{s},0}}

\submitjournal{ApJ}

\shorttitle{coagulation instability}
\shortauthors{Tominaga et al.}
\begin{document}

\title{Coagulation Instability in Protoplanetary Disks: \\A Novel Mechanism Connecting Collisional Growth and Hydrodynamical Clumping of Dust Particles}

\correspondingauthor{Ryosuke T. Tominaga}

%\author[0000-0002-8596-3505]{Ryosuke T. Tominaga}
\author{Ryosuke T. Tominaga}
\affil{Department of Physics, Nagoya University, Nagoya, Aichi 464-8692, Japan}
\email{rttominaga@nagoya-u.jp}

\affil{RIKEN Cluster for Pioneering Research, 2-1 Hirosawa, Wako, Saitama 351-0198, Japan}
\email{ryosuke.tominaga@riken.jp}

%\author[0000-0003-4366-6518]{Shu-ichiro Inutsuka}
\author{Shu-ichiro Inutsuka}
\affiliation{Department of Physics, Nagoya University, Nagoya, Aichi 464-8692, Japan}

%\author[0000-0001-8808-2132]{Hiroshi Kobayashi}
\author{Hiroshi Kobayashi}
\affiliation{Department of Physics, Nagoya University, Nagoya, Aichi 464-8692, Japan}



\begin{abstract}
We present a new instability driven by a combination of coagulation and radial drift of dust particles. We refer to this instability as ``coagulation instability" and regard it as a promising mechanism to concentrate dust particles and assist planetesimal formation in the very early stages of disk evolution. Because of dust-density dependence of collisional coagulation efficiency, dust particles efficiently (inefficiently) grow in a region of positive (negative) dust density perturbations, which lead to a small radial variation of dust sizes and as a result radial velocity perturbations. The resultant velocity perturbations lead to dust concentration and amplify dust density perturbations. This positive feedback makes a disk unstable. The growth timescale of coagulation instability is a few tens of orbital periods even when dust-to-gas mass ratio is of the order of $10^{-3}$. In a protoplanetary disk, radial drift and coagulation of dust particles tend to result in dust depletion. The present instability locally concentrates dust particles even in such a dust-depleted region. The resulting concentration provides preferable sites for dust-gas instabilities to develop, which leads to further concentration. Dust diffusion and aerodynamical feedback tend to stabilize short-wavelength modes, but do not completely suppress the growth of coagulation instability. Therefore, coagulation instability is expected to play an important role in setting up the next stage for other instabilities to further develop toward planetesimal formation, such as streaming instability or secular gravitational instability.
\end{abstract}



\section{Introduction}\label{sec:intro}
Formation of kilometer-sized planetesimals is an important step during growth from dust grains to planets in protoplanetary disks. Formation mechanisms are still under debate because of some barriers against dust growth. One of the barriers is the ``radial drift barrier". Dust particles are subject to significant radial drift toward a central star as a result of aerodynamical interaction with gas once they grow into intermediate-sized grains \citep[][]{Weidenschilling1977,Nakagawa1986}. The sizes of those grains are of the order of centimeters to meters in minimum mass solar nebula \citep[][]{Hayashi1981}. Fast radial drift generally prevents further dust growth although it depends on disk models including an initial dust-to-gas mass ratio \citep[][]{Brauer2008}. 

Several mechanisms have been proposed to explain the origin of planetesimals.  \citet[][]{Okuzumi2012} showed that collisional growth of fluffy dust aggregates can overcome the drift barrier and forms planetesimals in an inner disk region. Resultant fluffy aggregates are compressed by both surrounding gas and self-gravity of aggregates toward kilometer-sized planetesimals with its density of $0.1\;\mathrm{g/cm}^3$ \citep[][]{Kataoka2013a,Kataoka2013b}.

Hydrodynamical instabilities are also possible mechanisms of planetesimal formation. Hydrodynamical processes may be more important to form outer planetesimals and larger bodies than direct coagulation because coagulation takes longer time at outer radii. Streaming instability is one possible instability to assist planetesimal formation \citep[][]{Youdin2005,Youdin2007,Jacquet2011}. The streaming instability can lead to dust clumping, and resultant clumps eventually collapse self-gravitationally once those mass densities exceed Roche density \citep[e.g.,][]{Johansen2007nature,Simon2016}. The validity of streaming instability depends on dust-to-gas mass ratio and dimensionless stopping time $\taus\equiv\tstop\Omega$, where $\Omega$ is orbital frequency of a disk \citep[e.g.,][]{Johansen2009b,Carrera2015,Yang2017}. \citet{Carrera2015} and \citet{Yang2017} numerically investigated conditions for streaming instability to result in strong dust clumping required for subsequent gravitational collapse. They showed that dust-to-gas mass ratio should be larger than 0.02 for dust of $\taus=0.1$, and higher dust abundance is necessary for smaller dust ($\taus=10^{-2}-10^{-3}$). In the presence of turbulent diffusion of dust particles, strong clumping via streaming instability requires much higher dust abundance \citep[][]{Chen2020,Umurhan2020,Gole2020}. Global gas pressure profile also affects the critical dust-to-gas mass ratio, and small radial pressure gradient is preferable for strong clumping \citep[e.g.,][]{Bai2010c,Abod2019,Gerbig2020}.

Secular gravitational instability (GI) is also a possible mechanism leading to planetesimal formation \citep[][]{Ward2000,Youdin2011,Takahashi2014}. The instability can develop even in self-gravitationally stable disks and create multiple dust rings that will azimuthally fragment \citep[][]{Tominaga2018,Tominaga2020}. Linear analyses based on two-fluid equations show that secular GI requires high dust-to-gas mass ratio as in the case of streaming instability. The required dust-to-gas mass ratio is larger than 0.02 for $\taus=0.1$ in weakly turbulent disks \citep[see Figure 11 in][]{Tominaga2019}. \citet{Tominaga2019} presented another secular instability called two-component viscous GI (TVGI). Although TVGI is operational in relatively dust-poor disks where secular GI can not operate, high dust-to-gas ratio ($>0.02$) seems to be required for TVGI to develop within hundreds orbits.

As mentioned above, the hydrodynamical instabilities require large dust particles ($\taus\sim0.1$) with its abundance higher than 0.01 that is the value usually assumed. However, according to numerical studies of dust coagulation, dust surface density significantly decreases as dust grows into sizes of $\taus\sim0.1$ \citep[e.g.,][]{Brauer2008,Okuzumi2012}. Although fragmentation replenishes dust that avoid significant radial drift if its critical velocity is very low \citep[a few m/s;][]{Birnstiel2009}, it may be inefficient to trigger the above instabilities because those fragments have not only low $\taus$ but also depleted to some extent. Therefore, for the above dust-gas instabilities to operate, dust particles have to avoid the depletion and accumulate via other mechanisms in advance unless solid material are supplied from a surrounding envelope. 

Pressure bumps and zonal flows in gas disks are possible sites of dust concentration \citep[e.g.,][]{Whipple1972,Kretke2007,Dzyurkevich2010,Johansen2009a,Bai2014,Flock2015}. Streaming instability in pressure bumps has been investigated both analytically \citep[][]{Auffinger2018} and numerically \citep[][]{Taki2016,Carrera2020}. \citet{Auffinger2018} and \cite{Carrera2020} showed that streaming instability develops for dust-to-gas ratio of 0.01 if a relative amplitude of a pressure bump is larger than 10-20\%. We should note that deformation of a bump via frictional backreaction potentially inhibits subsequent gravitational collapse \citep{Taki2016}. Vortices also trap dust particles \citep[e.g.,][]{Barge1995,Chavanis2000,Lyra2013,Raettig2015}, which observations of lopsided structures suggest \citep[e.g.,][]{Fukagawa2013,van-der-Marel2013,Casassus2015}.

Solid concentration can also occur near water snow line \cite[e.g.,][]{Stevenson1988,Drazkowska2014,Drazkowska2017,Schoonenberg2017,Schoonenberg2018}. Water ice evaporates when it passes the snow line. Release of fragile silicate can result in traffic jam and solid enhancement inside the snow line. For the traffic jam to operate, fragmentation of silicate whose critical velocity is $\sim$ a few m/s \citep{Drazkowska2017} or disintegration of aggregates and release small silicate cores \citep{Saito2011} is required.  At the same time, the water vapor diffuses outward and recondenses onto icy solids, which increases solid abundance outside the snow line. \cite{Schoonenberg2017} showed that the solid enhancement outside the snow line is more efficient. Besides, recent experiments suggest that dry silicate grains are less fragile than previously considered \citep{Kimura2015,Steinpilz2019}, which might make the solid enhancement inside the snow line inefficient.

In this work, we present a new instability that can be another promising mechanism to accumulate dust particles that are subject to depletion via radial drift. The new instability that we refer to as ``coagulation instability" is triggered by dust coagulation, which is completely distinct from the other dust-gas instabilities previously studied. Coagulation instability is distinct from self-induced dust traps proposed by \citet{Gonzalez2017}. The present instability does not require backreaction to gas as shown in Section \ref{sec:simpleana} while the latter mechanism requires the backreaction. We show that coagulation instability operates even when dust abundance decreases down to the order of $10^{-3}$. In the new mechanism, dust particles actively play a role in concentrating themselves in contrast to mechanisms involving pressure bumps, zonal flow, and vortices.

This paper is organized as follows. Section \ref{sec:simpleana} gives physical understanding of coagulation instability based on one-fluid linear analyses. In Section \ref{sec:linana_woExF}, we perform more comprehensive linear analyses based on both dust and gas equations. Although we only consider turbulence-driven coagulation in diffusion-free disks in Section \ref{sec:simpleana} and \ref{sec:linana_woExF}, we discuss effects of dust diffusion and other components of relative velocities in Sections \ref{subsec:diffusion} and \ref{subsec:disp_drift}. In Section \ref{subsec:BR_midplane}, we discuss how the drag feedback (backreaction) at the midplane affects coagulation instability. We also compare a growth timescale and a traveling time of perturbations in Section \ref{subsec:vs_drift}. Section \ref{subsec:fragmentation} discusses effects of fragmentation. Section \ref{subsec:SI_SGI} gives speculation on coevolution with other dust-gas instabilities. We conclude in Section \ref{sec:conclusion}.

\section{Simplified analyses for physical understanding}\label{sec:simpleana}
We first give the physical interpretation of the coagulation instability based on simplified linear analyses. Analyses presented here only consider the motion of dust in a steady gas disk. This one-fluid analysis provides clear understanding of the coagulation instability.

\subsection{Basic equations}
We describe dust motion using the vertically integrated continuity equation. In the Cartesian local shearing sheet \citep{Goldreich1965} whose radial distance $r$ from a star is $R$ and Keplerian orbital frequency is $\Omega$, the dust continuity equations is  
\begin{equation}
\ddt{\sigmad}+\ddx{\sigmad v_x}=0,\label{eq:dsiddt}
\end{equation}
where $x$ and $v_x$ are the radial coordinate and velocity, respectively. The radial velocity of the dust is given by the terminal drift velocity \citep{Adachi1976,Nakagawa1986}:
\begin{equation}
v_x=-\frac{2\taus}{1+\taus^2}\eta R\Omega,\label{eq:vxNSH}
\end{equation}
where $\eta$ is given as a function of radial distance $r$ and gas density $\rho_{\gas}$:
\begin{equation}
\eta\equiv-\frac{1}{2\rho_{\gas}r\Omega^2}\frac{\partial\cs^2\rho_{\gas}}{\partial r},\label{eq:eta}
\end{equation}
where $\cs$ is the sound speed. Another important physical value in the dust dynamics is Stokes number $\st\equiv\tstop/t_{\edd}$, where $t_{\edd}$ is turnover time of the largest eddy. Hereafter, considering cases of $t_{\edd}=\Omega^{-1}$, we assume $\taus=\st$.

Dust coagulation is described by the Smoluchowski equation for column number density of dust per unit particle mass $m$ \citep[e.g.,][]{Smoluchowski1916,Schumann1940,Safronov1972,Brauer2008,Okuzumi2012,Birnstiel2012}. Since analytical approaches of such a coagulation equation are difficult, we use a moment equation describing size evolution of dust particles that dominate the total dust mass \citep[see][]{Estrada2008,Ormel2008,Sato2016}. The mass of such dust particles is called the peak mass \citep{Ormel2008}, which is denoted by $\mpar$ in the following. The time evolution of $\mpar$ is governed by
\begin{equation}
\Lddt{\mpar}=\frac{2\sqrt{\pi}a^2\Delta\vpp}{\hd}\sigmad,\label{eq:dmpdt}
\end{equation}
where $a$ is the particle radius, $\Delta\vpp$ is the particle-particle collision velocity, and $\hd$ is the dust scale height. In this work, we only consider spherical dust particles, and $\mpar$ is thus given by $\mpar=4\pi\rhoint a^3/3$, where $\rhoint$ is internal mass density of a dust particle. In addition, our analyses focus on small particles that are in the Epstein regime
\begin{equation}
\taus=\sqrt{\frac{\pi}{8}}\frac{\rhoint a}{\rho_{\gas}\cs}\Omega.
\end{equation}
If one assumes the vertically Gaussian profile of the gas density
\begin{equation}
\rho_{\gas}(z)=\frac{\sigmag}{\sqrt{2\pi}H}\exp\left(-\frac{z^2}{2H^2}\right),
\end{equation}
where $H\equiv\cs/\Omega$ is the gas scale height, the normalized stopping time at the midplane can be written in terms of the gas surface density $\sigmag$
\begin{equation}
\taus=\frac{\pi}{2}\frac{\rhoint a}{\sigmag}.\label{eq:stEp}
\end{equation}
Although $\taus$ varies in the vertical direction, we use the midplane value of $\taus$ to refer the representative dust size in our vertically integrated formalism because most of the mass resides around the midplane. The simplified analyses given in this section assume the gas surface density to be constant: $\sigmag=\sigmagup$. In such a case, Equations (\ref{eq:dmpdt}) and (\ref{eq:stEp}) give the evolutionary equation of $\taus$:
\begin{equation}
\ddt{\taus}+v_x\ddx{\taus}=\frac{\sqrt{\pi}}{4}\left(\frac{\sigmad}{\sigmagup}\right)\left(\frac{\Delta\vpp}{\taus\hd}\right)\taus.\label{eq:dstdt1f_formal}
\end{equation}
Giving a specific expression of $\Delta\vpp$ closes the set of the equations. In Section \ref{sec:simpleana} and \ref{sec:linana_woExF}, we focus on the turbulence-induced collisions. Although the turbulence-induced velocity generally depends on Reynolds number and eddy classes \citep{Volk1980}, we use simplified expression $\Delta\vpp=\sqrt{C\taus\alpha}\cs$ valid for intermediate-sized dust particles \citep{Ormel2007}, where $C$ is a numerical factor, and $\alpha$ is the strength of turbulence \citep{Shakura1973}. The simplified expression allows us to easily perform further analyses. \citet{Sato2016} effectively considered a size dispersion by introducing a size ratio to calculate $\Delta\vpp$. They showed that Equation (\ref{eq:dmpdt}) with the size ratio of 0.5 well reproduces the peak-mass evolution obtained from simulations with a size distribution. We thus follow their formalism. In such a case, the numerical factor $C$ is about $2.3$ \citep[see Equation (28) in][]{Ormel2007}. We also use the simplified expression for the dust scale height to proceed analytical discussions: $\hd=\sqrt{\alpha/\taus}H$ \citep[][]{Dubrulle1995,Carballido2006,YL2007}. These simplifications give the following equation for $\taus$:
\begin{equation}
\ddt{\taus}+v_x\ddx{\taus}=\frac{\sigmad}{\sigmagup}\frac{\taus}{3t_0},\label{eq:dstdt1f}
\end{equation}
where $t_0\equiv(4/3\sqrt{C\pi})\Omega^{-1}\simeq0.49\Omega^{-1}$ is mass growth time of a dust particle. Equations (\ref{eq:dsiddt}) and (\ref{eq:dstdt1f}) are the governing equations in the simplified analyses.

Note that the coagulation rate $\mpar^{-1}\partial\mpar/\partial t\propto\taus^{-1}\partial\taus/\partial t$ is independent from the value of $\alpha$ because the following two processes are canceled out: (1) stronger turbulence stirs dust grains up more and increases $\hd\propto\alpha^{1/2}$, and thus the midplane dust density decreases with $\alpha$, (2) in contrast, the dust-to-dust collision velocity $\Delta\vpp$ increases with $\alpha$ ($\Delta \vpp\propto\alpha^{1/2}$). As a consequence, the coagulation rate is determined by the dust-to-gas surface density ratio as already mentioned in previous studies \citep[e.g.,][]{Brauer2008,Okuzumi2012}. More specifically, the rate is related to the ``total" gas surface density $\sigmag$ ($\sigmagup$), not a gas density averaged within a dust sublayer. This is also the case if turbulence is too weak and the settling velocity determines the collision rate \citep[see][]{Nakagawa1981}.

% ==
\begin{figure}[htp]%[htp] or [H]
	\begin{center}
		%\hspace{-20pt}\raisebox{0pt}{
		%\hspace{100pt}\raisebox{20pt}{
		\includegraphics[width=\columnwidth]{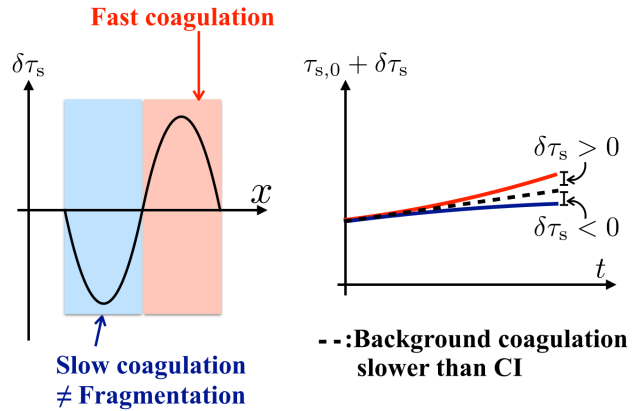}
		%}
	\end{center}
	%\vspace{-30pt}
\caption{Schematic picture to show the interpretation of wave-like perturbations on $\taus$ and time evolution of $\tausup+\delta\taus$ along with the background coagulation and coagulation instability (CI). The sign of $\delta\taus$ means whether dust sizes are larger or smaller than those expected from the background coagulation. In other words, positive $\delta\taus$ means fast coagulation while negative $\delta\taus$ means slow coagulation. We note that the negative $\delta\taus$ does not mean fragmentation because we only consider coagulation in Equation (\ref{eq:dstdt1f}). We also note that our linear analyses focus on shorter-term evolution via CI than $3t_0/\varepsilon$ so that the background evolution is insignificant although the right figure depicts the background evolution.}
\label{fig:delta_taus}
\end{figure}

\subsection{Linear analyses}\label{subsec:1flinana}
We first construct the unperturbed state. For simplicity, we ignore the evolution of the background state via coagulation, and assume uniform surface density $\sigmadup$ with a uniform dimensionless stopping time $\tausup$ and the velocity $v_{x,0}=-2\tausup\eta R\Omega/(1+\tausup^2)$. This assumption is valid if the instability is faster than coagulation. As shown below, coagulation instability satisfies this condition at wavelengths shorter than $3t_0|v_{x,0}|/\varepsilon$, where $\varepsilon\equiv\sigmadup/\sigmagup$ is unperturbed dust-to-gas surface density ratio.

%\begin{figure*}
%\gridline{
%\fig{Compare_r20_1f_in_Omega_H.eps}{0.8\textwidth}{ }
%}
%%\plotone{Compare_r20_1f_in_Omega_H.eps}
%\vspace{-20pt}
\begin{figure*}[htp]%[htp] or [H]
	\begin{center}
		%\hspace{-20pt}\raisebox{0pt}{
		%\hspace{100pt}\raisebox{20pt}{
		\includegraphics[width=2.0\columnwidth]{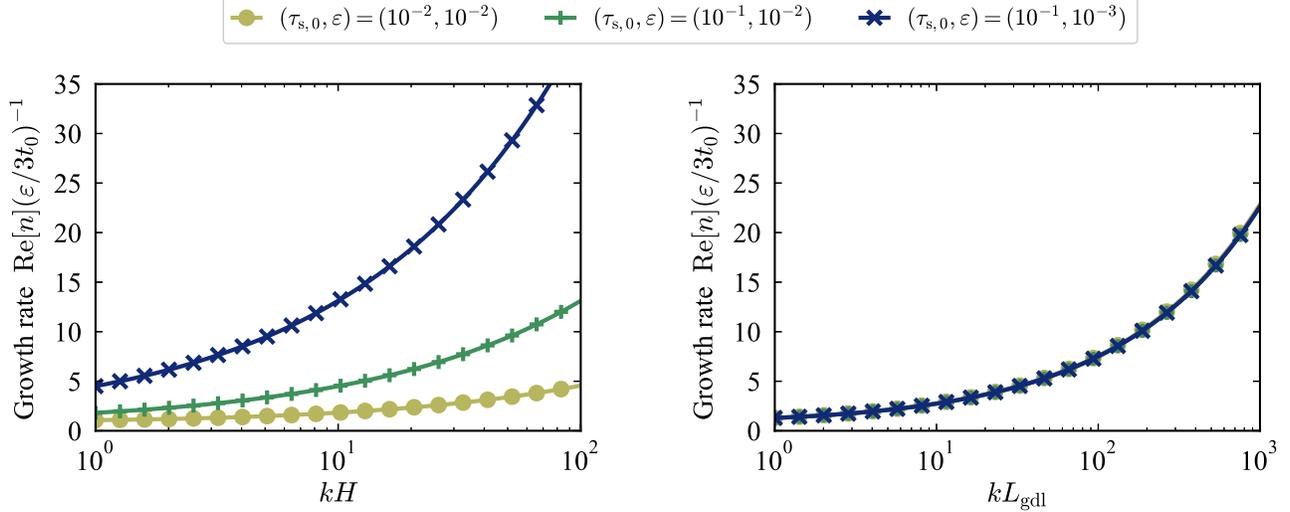}
		%}
	\end{center}
	%\vspace{-30pt}
\caption{Growth rate of coagulation instability obtained from the simplified linear analyses. The left panel shows the growth rate normalized by $\varepsilon/3t_0$ as a function of $kH$. As expected from Equation (\ref{eq:nap_re_ap}), the coagulation instability grows faster than background coagulation, which is more remarkable for $\varepsilon=10^{-3}$. The right panel shows the growth rates as a function of $kL_{\gdl}$. The growth rate is almost self-similar as shown by Equations (\ref{eq:normal_n_ap_pre}) and (\ref{eq:normal_n_ap}). In both panels, yellow, green, and blue lines show growth rates for $(\tausup,\varepsilon)=(10^{-2},10^{-2}),\;(10^{-1},10^{-2}),$ and $(10^{-1},10^{-3})$, respectively.}
\label{fig:1fluid_growthrate}
\end{figure*}

\subsubsection{Dispersion relation}

Based on the above unperturbed state, we obtain the following linearized equations from Equations (\ref{eq:dsiddt}), (\ref{eq:vxNSH}) and (\ref{eq:dstdt1f})
\begin{equation}
(n+ikv_{x,0})\delta\sigmad+ik\sigmadup\delta v_x=0,\label{eq:simpl_eoc}
\end{equation}
\begin{equation}
\delta v_x=\frac{1-\tausup^2}{1+\tausup^2}\frac{\delta\taus}{\tausup}v_{x,0},\label{eq:simpl_vx}
\end{equation}
\begin{equation}
(n+ikv_{x,0})\delta\taus=\frac{\delta\sigmad}{\sigmagup}\frac{\tausup}{3t_0}+\frac{\sigmadup}{\sigmagup}\frac{\delta\taus}{3t_0},\label{eq:simpl_coag}
\end{equation}
where $n$ and $k$ are the complex growth rate and the wavenumber of perturbations, and we assume the exponential form of perturbations, i.e., $\delta\sigmad\propto\delta v_x\propto\delta\taus\propto\exp(ikx+nt)$. The wave-like perturbation on $\taus$ represents fast or slow coagulation (see Figure \ref{fig:delta_taus}). The negative $\delta\taus$ means not fragmentation but ``delayed" coagulation compared to the background state. This interpretation is supported by the fact that the coagulation rate is proportional to dust surface density, $\sigmadup+\delta\sigmad$. Lower surface density delays coagulation and results in smaller dust sizes than the background sizes $\tausup$ although dust still grows in size (see Figure \ref{fig:delta_taus}).

Solving the eigenvalue problem with Equations (\ref{eq:simpl_eoc}) - (\ref{eq:simpl_coag}), we obtain the following dispersion relation
\begin{equation}
n=n_{\app,\pm}\equiv-ikv_{x,0}+\frac{\varepsilon}{6t_0}\left(1\pm\sqrt{1-\frac{12t_0}{\varepsilon}\frac{1-\tausup^2}{1+\tausup^2}ikv_{x,0}}\right).\label{eq:1fdispersion}
\end{equation}
There is one growing mode, $n=n_{\app,+}$. For short-wavelength perturbations, the complex growth rate of the growing mode can be approximated as
\begin{align}
n_{\app,+}&\simeq-ikv_{x,0}+\frac{\varepsilon}{6t_0}\left(1+\sqrt{\frac{12t_0}{\varepsilon}\frac{1-\tausup^2}{1+\tausup^2}ik|v_{x,0}|}\right)\notag\\
&=-ikv_{x,0}+\frac{\varepsilon}{6t_0}\left(1+(1+i)\sqrt{\frac{6t_0}{\varepsilon}\frac{1-\tausup^2}{1+\tausup^2}k|v_{x,0}|}\right).\label{eq:shortwave_n}
\end{align}
Thus, the growth rate, i.e. the real part of $n$, at large $k>0$ is
\begin{align}
\mathrm{Re}[n_{\app,+}] &\simeq \frac{\varepsilon}{6t_0}\left(1+\sqrt{\frac{6t_0}{\varepsilon}\frac{1-\tausup^2}{1+\tausup^2}k|v_{x,0}|}\right)\notag\\
&\simeq \frac{\varepsilon}{3t_0}\sqrt{\frac{3t_0}{2\varepsilon}\frac{1-\tausup^2}{1+\tausup^2}k|v_{x,0}|}.\label{eq:nap_re_ap}
\end{align}
Equation (\ref{eq:nap_re_ap}) shows that a disk is unconditionally unstable. The growth rate is much larger than $\varepsilon/3t_0$ at wavenumbers that satisfy $t_0k|v_{x,0}|/\varepsilon\gg1$. Therefore, this instability develops faster than the unperturbed state that evolves at the timescale of $3t_0/\varepsilon$. The left panel of Figure \ref{fig:1fluid_growthrate} shows the growth rate as a function of $kH$ for $(\tausup,\varepsilon)=(10^{-2},10^{-2}),\;(10^{-1},10^{-2}),$ and $(10^{-1},10^{-3})$. The vertical axis is normalized by coagulation timescale and thus shows how fast coagulation instability grows compared to background coagulation. Equation (\ref{eq:nap_re_ap}) and Figure \ref{fig:1fluid_growthrate} show that coagulation instability proceeds faster than background coagulation especially for low dust-to-gas ratio.

The relative amplitude of $\delta \taus$ to $\delta \sigmad$ is 
\begin{equation}
\frac{\delta\taus/\tausup}{\delta\sigmad/\sigmadup}=\frac{n+ikv_{x,0}}{ik|v_{x,0}|}\frac{1+\tausup^2}{1-\tausup^2}
\end{equation}
At short wavelengths, the relative amplitude is approximately given as
\begin{align}
\frac{\delta\taus/\tausup}{\delta\sigmad/\sigmadup}&\simeq(1-i)\frac{\varepsilon}{6t_0}\sqrt{\frac{6t_0}{\varepsilon k|v_{x,0}|}\frac{1+\tausup^2}{1-\tausup^2}},\notag\\
&=\exp\left(-i\frac{\pi}{4}\right)\sqrt{\frac{\varepsilon}{3t_0 k|v_{x,0}|}\frac{1+\tausup^2}{1-\tausup^2}}\label{eq:4pi_ts_sigmad}.
\end{align}
This shows that $\delta\taus/\tausup$ is smaller than $\delta\sigmad/\sigmadup$ at short wavelengths ($k^{-1}<3t_0|v_{x,0}|/\varepsilon$), and its phase is shifted outward by $\pi/4$.

Characteristic length of the coagulation instability is $L_{\gdl}\equiv3t_0|v_{x,0}|/\varepsilon$, which we call a ``growth-drift length". The growth-drift length represents a distance that dust grains move with the unperturbed velocity during the $e-$folding timescale of dust-size growth. The ratios $L_{\gdl}/H$ and $L_{\gdl}/R$ are given by
\begin{equation}
\frac{L_{\gdl}}{H}=3\left(\frac{t_0\Omega}{0.5}\right)\left(\frac{\varepsilon}{0.01}\right)^{-1}\left(\frac{|v_{x,0}|/\cs}{0.02}\right),\label{eq:Lgdl_H}
\end{equation}
\begin{align}
\frac{L_{\gdl}}{R}&=0.15\left(\frac{t_0\Omega}{0.5}\right)\left(\frac{\varepsilon}{0.01}\right)^{-1}\left(\frac{|v_{x,0}|/R\Omega}{10^{-3}}\right),\notag\\
&\simeq0.15\left(\frac{t_0\Omega}{0.5}\right)\left(\frac{\varepsilon}{0.01}\right)^{-1}\left(\frac{\tausup}{0.1}\right)\left(\frac{\eta}{5\times10^{-3}}\right).
\end{align}
In the last equality, we approximate $|v_{x,0}|$ as $2\tausup\eta R\Omega$. One can obtain a universal form of the dispersion relation by scaling wavelengths by the growth-drift length. The right panel of Figure \ref{fig:1fluid_growthrate} shows the growth rate as a function of $kL_{\gdl}$, showing the universal relation between $\mathrm{Re}[n](\varepsilon/3t_0)^{-1}$ and $kL_{\gdl}$. Using $L_{\gdl}$ to normalize a wavenumber $\tilde{k}\equiv kL_{\gdl}$ and assuming $\tilde{k}\gg1$, we obtain
\begin{equation}
n_{\app,+}\simeq i\frac{\varepsilon}{3t_0}\tilde{k}+\frac{\varepsilon}{6t_0}\left(1+(1+i)\sqrt{\frac{1-\tausup^2}{1+\tausup^2}2\tilde{k}}\right).
\end{equation}
The growth rate normalized by $\varepsilon/3t_0$ is then given by
\begin{equation}
\mathrm{Re}[n_{\app,+}]\left(\frac{3t_0}{\varepsilon}\right) = \frac{1}{2}+\frac{1}{2}\sqrt{\frac{1-\tausup^2}{1+\tausup^2}2\tilde{k}} \label{eq:normal_n_ap_pre}
\end{equation}
When dust particles are small and $\tausup\lesssim0.1$, we can approximate $1\pm\tausup^2\simeq1$ with only a few percent errors. This approximation gives the growth rate apparently independent of $\tausup$:
\begin{equation}
\mathrm{Re}[n_{\app,+}]\left(\frac{3t_0}{\varepsilon}\right) = \frac{1}{2}+\sqrt{\frac{\tilde{k}}{2}}.\label{eq:normal_n_ap}
\end{equation}  
We note that $k$ is normalized by $L_{\gdl}$ that depends on $|v_{x,0}|$, and hence on $\tausup$. Equation (\ref{eq:normal_n_ap}) shows that the coagulation instability of different size particles can grow at the same rate relative to background coagulation but at different wavelengths $\lambda\propto L_{\gdl}$. 

\begin{figure*}
\gridline{
\fig{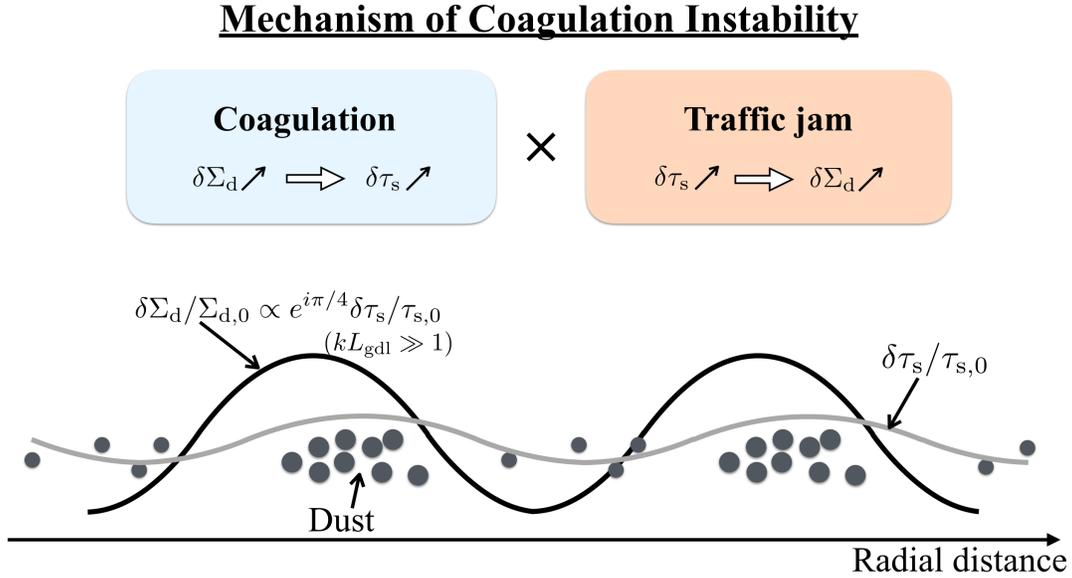}{0.8\textwidth}{ }
}
\vspace{-20pt}
\caption{Schematic picture of eigenfunctions of coagulation instability for $kL_{\gdl}\gg1$. Black and gray lines depict sinusoidal perturbations of $\delta\sigmad$ and $\delta\taus$, respectively. The sizes of filled circles represent dust sizes.  Coagulation instability is driven by a combination of coagulation and traffic jam. Because coagulation efficiency is proportional to dust surface density $\sigmadup+\delta\sigmad$ (Equation (\ref{eq:dstdt1f})), dust surface density perturbations lead to radial variation of dust sizes and $\delta\taus$. The radial variation of $\delta\taus$ leads to traffic jam and amplifies $\delta\sigmad$. This positive feedback process triggers coagulation instability.}
\label{fig:eigenfunc}
\end{figure*}
% ==

\subsubsection{Physical mechanism of coagulation instability}
Coagulation instability is triggered by a combination of coagulation and traffic jam. In fact, the growth rate at short wavelengths is determined by the geometric mean of coagulation rate, $\varepsilon/3t_0$, and a rate of traffic jam, $\left[(1-\tausup^2)/(1+\tausup^2)\right]k|v_{x,0}|$ (see Equation (\ref{eq:nap_re_ap})). Figure \ref{fig:eigenfunc} shows a schematic picture of eigenfunctions of $\delta\sigmad$ and $\delta\taus$. When there is a perturbation in dust surface density, dust particles grow more efficiently in regions of $\delta\sigmad>0$. On the other hand, dust particles grow less efficiently at regions of $\delta\sigmad<0$. Such a spatial variation of the coagulation efficiency leads to perturbations in dust sizes and thus $\delta\taus$. Perturbations in $\delta\taus$ lead to traffic jam in the radial direction and further amplify $\delta\sigmad$. In this way, coagulation and traffic jam act as a positive-feedback process and make a disk unstable.

As noted above, the unstable mode at short wavelengths has the phase shift of $\pi/4$ between $\delta\sigmad$ and $\delta \taus$. Phase shift is sometimes important in the growth mechanism of dust-gas instabilities \citep[e.g., streaming insabilitiy, see][]{Lin2017}. The origin of the phase shift in coagulation instability can be explained based on the obtained dispersion relation and a toy-model equation described below. For simplicity, we focus on short wavelength perturbations so that we can use Equation (\ref{eq:shortwave_n}) and further simplify as follows
\begin{equation}
n_{\app,+}\simeq -ikv_{x,0}+\frac{\varepsilon (1+i)}{6t_0}\sqrt{\frac{6t_0}{\varepsilon}\frac{1-\tausup^2}{1+\tausup^2}k|v_{x,0}|}.
\end{equation}
One can see that waves of the phase velocity $v_{\mathrm{ph}}=-\mathrm{Im}[n_{\app,+}]/k$ move inward faster than dust grains drifting at $v_{x,0}$. The relative velocity $\Delta v_{\mathrm{d,w}}$ is
\begin{equation}
\Delta v_{\mathrm{d,w}}\equiv v_{x,0}-v_{\mathrm{ph}}=\frac{\varepsilon}{6t_0}\sqrt{\frac{6t_0}{\varepsilon}\frac{1-\tausup^2}{1+\tausup^2}\frac{|v_{x,0}|}{k}}.
\end{equation}
In other words, dust grains move outward at the wave-rest frame. One key property is $\mathrm{Re}[n_{\app,+}]=\Delta v_{\mathrm{d,w}}k$, meaning that $e$-folding time of $\delta\taus$ and $\delta\sigmad$ is equal to a dust-crossing time over the length of $k^{-1}$ at the short-wavelength limit. To discuss coagulation efficiency in the presence of waves of $\delta\sigmad$, we consider the following toy-model equation of $\delta \taus$ evolution at the wave-rest frame $(t,X)$:
\begin{align}
\frac{d(\delta\taus/\tausup)}{dt} &= \frac{\varepsilon}{3t_0}\frac{\delta\sigmad}{\sigmadup}\notag\\
&=\frac{\varepsilon}{3t_0}\frac{|\delta\sigmad (t=0)|}{\sigmadup}\notag\\
&\;\;\;\;\;\;\;\;\;\;\;\times\exp(\mathrm{Re}[n_{\app,+}]t)\cos(k(X_0+\Delta X))\notag\\
&=\frac{\varepsilon}{3t_0}\frac{|\delta\sigmad (t=0)|}{\sigmadup}\notag\\
&\;\;\;\;\;\;\;\;\;\;\;\times\exp(k\Delta v_{\mathrm{d,w}}t)\cos(kX_0+k\Delta v_{\mathrm{d,w}}t),\label{eq:toy_model_deltataus}
\end{align}
where $X_0$ and $\Delta X$ are the initial position of dust grains and a distance over which dust grains move. In the last equality, we utilize $\Delta X=\Delta v_{\mathrm{d,w}}t$ and $\mathrm{Re}[n_{\app,+}]=\Delta v_{\mathrm{d,w}}k$, which are noted above. We note that we neglect the growth of background state $\tausup$ because $\mathrm{Re}[n_{\app,+}]$ is larger than the coagulation rate at short wavelengths considered. The solution of Equation (\ref{eq:toy_model_deltataus}) is
\begin{equation}
\frac{\delta\taus(t)-\delta\taus(t=0)}{\tausup}=\frac{\varepsilon}{3\sqrt{2}t_0k\Delta v_{\mathrm{d,w}}}\frac{|\delta\sigmad |}{\sigmadup}F(k\Delta v_{\mathrm{d,w}}t,kX_0),
\end{equation}
\begin{align}
F(k\Delta v_{\mathrm{d,w}}t,kX_0)\equiv&\exp(k\Delta v_{\mathrm{d,w}}t)\cos\left(k\Delta v_{\mathrm{d,w}}t+kX_0-\frac{\pi}{4}\right)\notag\\
&-\cos\left(kX_0-\frac{\pi}{4}\right).\label{eq:A}
\end{align}
One can ask what is the value of $X_0$ that maximizes the size of dust grains moving across one wavelength, i.e. $k\Delta v_{\mathrm{d,w}}t=2\pi$.\footnote{The periodicity of sinusoidal perturbations justifies the choice of $k\Delta v_{\mathrm{d,w}}t=2\pi$.} Considering $0\leq kX_0<2\pi$ without loss of generality, we find that $F(2\pi,kX_0)$ is the largest for $kX_0=\pi/4$ (see Equation (\ref{eq:A})). Therefore, dust grains that start moving from $X_0=\pi/4k$ grow the most efficiently, which explains the peak position of $\delta\taus$ that is shifted by $\pi/4$ in phase relative to $\delta\sigmad$.

We note that substituting $k=0$ in Equation (\ref{eq:1fdispersion}) gives $n_{\app,+}=\varepsilon/3t_0$, i.e. the coagulation rate. The associated eigenfunction is $\delta\sigmad=0$, $\delta\taus\neq0$, and thus $\delta v_x\neq0$. This mode is just coagulation and not the pile up via instabilities. Because of its infinitely long wavelength ($k=0$), one can regard this mode as a shift in stopping time, i.e. dust sizes, from $\tausup$ to $\tausup+\delta\taus$. The size-shifted dust grains coagulate at the rate $\varepsilon/3t_0$, which is the physical explanation of the $k=0$ mode. Note, however, that our linear analyses are valid for $n>\varepsilon/3t_0$ (or $kL_{\gdl}\gg1$) as mentioned above.

\section{Two-fluid linear analyses}\label{sec:linana_woExF}

In this section, we perform two-fluid linear analyses and discuss effects of the gas density profile and nonsteady gas motion on the coagulation instability. As noted in the previous section, the coagulation rate is determined not by dust-to-gas ratio in a dust layer but by the total dust-to-gas surface density ratio $\sigmad/\sigmag$. This motivates us to first use two-fluid equations averaged over a full vertical extent of a disk. 
When the gas layer turbulence is weak and the vertical momentum exchange/transfer is very limited in a gas disk, the dust-gas momentum exchange should be averaged within a ``sublayer" around the midplane rather than for a whole disk. We will also analyze such limiting cases of weak turbulence in Appendix \ref{app:sublayer}. We find however that such a sublayer model shows insignificant differences when a sublayer height $z_{\mathrm{s}}$ is larger than $3\hd$. Thus, we describe full-disk averaging model in this section (see Section \ref{subsec:BR_midplane} for discussions on a combined effect of size-dependent settling and reduced drift velocity).

The hydrodynamic equations are 
\begin{equation}
\ddt{\sigmag}+\ddx{\sigmag u_x}=0,\label{eq:eocgas}
\end{equation}
\begin{equation}
\ddt{u_x}+u_x\ddx{u_x}=3\Omega^2x+2\Omega u_y-\frac{1}{\sigmag}\ddx{\cs^2\sigmag}+\frac{\sigmad}{\sigmag}\frac{v_x-u_x}{\taus}\Omega,\label{eq:eomxgas}
\end{equation}
\begin{equation}
\ddt{u_y}+u_x\ddx{u_y}=-2\Omega u_x+\frac{\sigmad}{\sigmag}\frac{v_y-u_y}{\taus}\Omega,\label{eq:eomygas}
\end{equation}
\begin{equation}
\ddt{\sigmad}+\ddx{\sigmad v_x}=0,\label{eq:eocdust}
\end{equation}
\begin{equation}
\ddt{v_x}+v_x\ddx{v_x}=3\Omega^2x+2\Omega v_y-\frac{v_x-u_x}{\taus}\Omega,\label{eq:eomxdust}
\end{equation}
\begin{equation}
\ddt{v_y}+v_x\ddx{v_y}=-2\Omega v_x-\frac{v_y-u_y}{\taus}\Omega,\label{eq:eomydust}
\end{equation}
where $u_x$ and $u_y$ are the radial and azimuthal gas velocities, and $v_y$ is the azimuthal dust velocity. 

Equations (\ref{eq:dmpdt}), (\ref{eq:stEp}) and (\ref{eq:eocgas}) give the evolutionary equation of $\taus$:
\begin{equation}
\ddt{\taus}+v_x\ddx{\taus}=\frac{\sigmad}{\sigmag}\frac{\taus}{3t_0}+\frac{\taus}{\sigmag}\ddx{\sigmag u_x}-\frac{\taus}{\sigmag}v_x\ddx{\sigmag},\label{eq:dstdt_full}
\end{equation}
The second term on the right hand side is the effect of the gas compression that decreases $\taus$. The third term represents a process where $\taus$ decreases as dust particles drift with $v_x$ toward inner high gas density region. Note that these terms with the gas surface density gradients originate from $\sigmag$ in the drag law adopted for the midplane dust (Equation (\ref{eq:stEp})). Equation (\ref{eq:stEp}) is valid as long as the gas density has the vertical Gaussian profile. Therefore, one can use Equation (\ref{eq:dstdt_full}) with the total gas surface density whatever vertical extent is adopted in the vertical averaging of the equations.

\subsection{Unperturbed state}\label{subsec:2f_unp}
As in the simplified one-fluid analyses, we construct unperturbed state without using Equation (\ref{eq:dstdt_full}) (see also Appendix \ref{app:unpert_state}). Neglecting Equation (\ref{eq:dstdt_full}) in unperturbed state is valid because coagulation instability grows faster than the background coagulation proceeds. The steady state is then governed by the following equations:
\begin{equation}
\ddx{\sigmag u_{x}}=0,\label{eq:eocgas_steady}
\end{equation}
\begin{equation}
u_{x}\ddx{u_{x}}=3\Omega^2x+2\Omega u_{y}-\frac{1}{\sigmag}\ddx{\cs^2\sigmag}+\frac{\sigmad}{\sigmag}\frac{v_{x}-u_{x}}{\taus}\Omega,\label{eq:steadyeomx_gas}
\end{equation}
\begin{equation}
u_{x}\ddx{u_{y}}=-2\Omega u_{x}+\frac{\sigmad}{\sigmag}\frac{v_{y}-u_{y}}{\taus}\Omega,\label{eq:steadyeomy_gas}
\end{equation}
\begin{equation}
\ddx{\sigmad v_{x}}=0,\label{eq:eocdust_steady}
\end{equation}
\begin{equation}
v_{x}\ddx{v_{x}}=3\Omega^2x+2\Omega v_{y}-\frac{v_{x}-u_{x}}{\taus}\Omega,\label{eq:steadyeomx_dust}
\end{equation}
\begin{equation}
v_{x}\ddx{v_{y}}=-2\Omega v_{x}-\frac{v_{y}-u_{y}}{\taus}\Omega.\label{eq:steadyeomy_dust}
\end{equation}

Coagulation instability requires the dust drift motion, which occurs in the presence of a global gas pressure gradient, i.e., the drift velocities are proportional to the gradient. We thus seek a dust drift solution considering a globally smooth gas disk with small radial gradients of surface density $\sigmagup'\equiv\partial\sigmagup/\partial x$ and gas pressure $(\cs^2\sigmagup)'\equiv\partial(\cs^2\sigmagup)/\partial x$. We do not consider a characteristic structure (e.g., a bump) in a disk, and assume the spatial scale of surface density and pressure variations to be of the order of radius $R$: $\sigmagup'/\sigmagup\sim R^{-1}$ and $(\cs^2\sigmagup)'/\cs^2\sigmagup\sim R^{-1}$. Linear analyses with a bump structure can be found in \citet{Auffinger2018}, where they perform linear analyses of streaming instability with dust drifting in an enforced pressure bump. In this work, we approximately derive a drift solution with the following assumptions. First, we assume the small radial size of the local frame, $\Delta R$, and $|x|/R<\Delta R/R\ll 1$. Besides the smallness of the local frame, we also assume that $(\cs^2\sigmagup)'$, $\sigmagup'$ and gradients of the other unperturbed quantities are so small that we can neglect the second- and higher-order terms of the gradients. Based on these assumptions, we find the following solution that satisfies Equations (\ref{eq:eocgas_steady})-(\ref{eq:steadyeomy_dust}) at the first order in $x/R$: 
\begin{equation}
u_x=u_{x,0}\equiv\frac{2\varepsilon\tausup}{(1+\varepsilon)^2+\tausup^2}\left(-\frac{(\cs^2\sigmagup)'}{2\Omega\sigmagup}\right),
\end{equation}
\begin{equation}
u_y=-\frac{3}{2}\Omega x+u_{y,0},
\end{equation}
\begin{equation}
u_{y,0}\equiv-\left[1+\frac{\varepsilon\tausup^2}{(1+\varepsilon)^2+\tausup^2}\right]\left(-\frac{(\cs^2\sigmagup)'}{2\Omega\sigmagup(1+\varepsilon)}\right),
\end{equation}
\begin{equation}
v_x=v_{x,0}\equiv-\frac{2\tausup}{(1+\varepsilon)^2+\tausup^2}\left(-\frac{(\cs^2\sigmagup)'}{2\Omega\sigmagup}\right),
\end{equation}
\begin{equation}
v_y=-\frac{3}{2}\Omega x+v_{y,0},
\end{equation}
\begin{equation}
v_{y,0}\equiv-\left[1-\frac{\tausup^2}{(1+\varepsilon)^2+\tausup^2}\right]\left(-\frac{(\cs^2\sigmagup)'}{2\Omega\sigmagup(1+\varepsilon)}\right),
\end{equation}
where the subscript ``0" denotes unperturbed quantities, and $\varepsilon=\sigmadup/\sigmagup$, $(\cs^2\sigmagup)'$ and $\tausup$ are input parameters in this formalism. Note that the drift velocities are of the first order in the small gradient (i.e., $(\cs^2\sigmagup)'$), and thus in $x/R$, because of the nature of the dust drift. We describe the details of the derivation in Appendix \ref{app:unpert_state}. 

The drift solution in the two-fluid height-integrated systems is characterized by
\begin{equation}
\eta_{\mathrm{2D}}\equiv-\frac{(\cs^2\sigmagup)'}{2R\Omega^2\sigmagup}.\label{eq:eta_prime}
\end{equation}
The drift velocity with $\eta_{\mathrm{2D}}$ is the reasonable solution in the two-fluid height-integrated system with the full-disk averaging. However, $\eta_{\mathrm{2D}}$ is different from $\eta$ because only the latter includes the radial profile of $H$, which prevents direct comparison with the growth rates and the phase velocities obtained in the previous section because the drift velocities are different. We thus compromise by using the value of $\eta$ with an assumed disk model and substituting it in $\eta_{\mathrm{2D}}$ of Equation (\ref{eq:eta_prime}) to calculate $(\cs^2\sigmagup)'$ and the drift velocities, i.e.,
\begin{equation}
-\frac{(\cs^2\sigmagup)'}{\sigmagup}=2\eta_{\mathrm{2D}}R\Omega^2\to 2\eta R\Omega^2.
\end{equation}
Assuming the temperature profile, we obtain $\sigmagup'$ from the following equation:
\begin{equation}
\frac{\sigmagup'}{\sigmagup}=-\frac{1}{R}\left(2\eta_{\mathrm{2D}}\frac{R^2\Omega^2}{\cs^2}+\frac{\partial \ln\cs^2}{\partial \ln R}\right).\label{eq:sigmagup_eta_relation}
\end{equation}
The definition of $\eta$ (Equation (\ref{eq:eta})) gives
\begin{equation}
\frac{\sigmagup'}{\sigmagup}=-\frac{1}{R}\left(2\eta\frac{R^2\Omega^2}{\cs^2}+\frac{\partial \ln\cs^2}{\partial \ln R}-\frac{\partial\ln H}{\partial \ln R}\right).
\end{equation}
Thus, substituting the value of $\eta$ in $\eta_{\mathrm{2D}}$ in Equation (\ref{eq:sigmagup_eta_relation}) gives smaller $\sigmagup'/\sigmagup$.\footnote{Note that we only consider a case that $\sigmagup'/\sigmag$ is negative. } 

In the following of this section, we calculate $\sigmagup'$ by using Equation (\ref{eq:sigmagup_eta_relation}) with an assumed value of $\eta$. We again note that the above calculation of $\sigmagup'$ and $\eta_{\mathrm{2D}}$ with an assumed value of $\eta$ is only for enabling the direct comparison of results obtained in one-fluid and two-fluid analyses (cf., Section \ref{subsec:BR_midplane}).

\subsection{Linearized equations}\label{subsec:linana2f}
Based on the above unperturbed state, we derive the linearized equations assuming perturbations, $\delta\sigmag,$ $\delta u_x,$ $\delta u_y,$ $\delta\sigmad,$ $\delta v_x,$ $\delta v_y,$ $\delta\taus$ proportional to $\exp(ikx+nt)$. Because of the gas density and pressure profile with $\sigmagup'x/\sigmagup\sim(\cs^2\sigmagup)'x/(\cs^2\sigmagup)\sim x/R\ll1$ (see Appendix \ref{app:unpert_state}), the amplitude of perturbations depend on $x$ in general. In this study, focusing on short-wavelength perturbations with $kR\gg1$ $(|x|<\Delta R,\; \Delta R/R\ll1)$, we assume that the spatial variation of $\exp(ikx)$ is larger than that of the perturbation amplitudes so that we can approximate $\partial \delta A/\partial x\simeq ik\delta A$, where $\delta A$ is a perturbation considered (e.g., $\delta\sigmad$). This approximation corresponds to the WKB approximation \citep[e.g., see][]{Shu1992}. We also assume that the temperature profile remains steady. The linearized equations are
\begin{equation}
(n+iku_{x,0})\delta\sigmag+\left(ik\Sigma_{\gas,\up}+\ddx{\Sigma_{\gas,\up}}\right)\delta u_x=0,
\end{equation}
\begin{align}
(n+iku_{x,0})\delta u_x=&2\Omega\delta u_y+\frac{\cs^2\delta\sigmag}{\Sigma_{\gas,\up}^2}\ddx{\Sigma_{\gas,\up}}-\frac{\cs^2}{\Sigma_{\gas,\up}}ik\delta\sigmag\notag\\
&+\frac{\sigmadup}{\Sigma_{\gas,\up}}\frac{v_{x,0}-u_{x,0}}{\tausup}\Omega\notag\\
&\times\left(\frac{\delta\sigmad}{\sigmadup}-\frac{\delta\sigmag}{\Sigma_{\gas,\up}}+\frac{\delta v_x-\delta u_x}{v_{x,0}-u_{x,0}}-\frac{\delta\taus}{\tausup}\right),
\end{align}
\begin{align}
(n+iku_{x,0})\delta u_y=&-\frac{\Omega}{2}\delta u_x+\frac{\sigmadup}{\Sigma_{\gas,\up}}\frac{v_{y,0}-u_{y,0}}{\tausup}\Omega\notag\\
&\times\left(\frac{\delta \sigmad}{\sigmadup}-\frac{\delta\sigmag}{\Sigma_{\gas,\up}}+\frac{\delta v_y-\delta u_y}{v_{y,0}-u_{y,0}}-\frac{\delta\taus}{\tausup}\right),
\end{align}
\begin{equation}
(n+ikv_{x,0})\delta\sigmad+ik\sigmadup\delta v_x=0,
\end{equation}
\begin{equation}
(n+ikv_{x,0})\delta v_x=2\Omega\delta v_y-\frac{v_{x,0}-u_{x,0}}{\tausup}\Omega\left(\frac{\delta v_x-\delta u_x}{v_{x,0}-u_{x,0}}-\frac{\delta\taus}{\tausup}\right),
\end{equation}
\begin{equation}
(n+ikv_{x,0})\delta v_y=-\frac{\Omega}{2}\delta v_x-\frac{v_{y,0}-u_{y,0}}{\tausup}\Omega\left(\frac{\delta v_y-\delta u_y}{v_{y,0}-u_{y,0}}-\frac{\delta\taus}{\tausup}\right),
\end{equation}
\begin{align}
(n+ikv_{x,0})\delta\taus&=\frac{\sigmadup}{\Sigma_{\gas,\up}}\frac{\tausup}{3t_0}\left(\frac{\delta\sigmad}{\sigmadup}-\frac{\delta\sigmag}{\Sigma_{\gas,\up}}+\frac{\delta\taus}{\tausup}\right)\notag\\
&+\frac{\tausup}{\Sigma_{\gas,\up}}\biggl[\left(ik\Sigma_{\gas,\up}+\ddx{\Sigma_{\gas,\up}}\right)\delta u_x\notag\\
&+iku_{x,0}\delta\sigmag\biggr]-\frac{\tausup v_{x,0}}{\Sigma_{\gas,\up}}ik\delta\sigmag\notag\\
&-\frac{\tausup v_{x,0}}{\Sigma_{\gas,\up}}\ddx{\Sigma_{\gas,\up}}\left(\frac{\delta\taus}{\tausup}+\frac{\delta v_x}{v_{x,0}}-\frac{\delta\sigmag}{\Sigma_{\gas,\up}}\right),
\end{align}
where $\Sigma_{\gas,\up}\equiv\sigmagup+\sigmagup'x$. The complex growth rate $n$ is derived from the above equations with $x=0$, i.e., $\Sigma_{\gas,\up}=\sigmagup$. 

We note that one finds $|\partial\Sigma_{\gas,\up}/\partial x|\ll k\Sigma_{\gas,\up}$ based on the assumption of the unperturbed gas profiles with the small gradients (e.g., $\sigmagup' x/\sigmagup\sim x/R\ll 1$) and short-wavelength perturbations ($kR\gg1$). Neglecting the gradients of the unperturbed profiles in the linearized equations, one finds that the resulting equations of motion and continuity equations are equivalent to linearized equations that can be derived with a uniform background state and an external force on gas to drive the drift motion except for $\delta\taus$ \citep[e.g., see][]{Tominaga2020}\footnote{Equations used in \citet{Tominaga2020} include self-gravity, viscosity, dust diffusion and dust velocity dispersion, which are not considered in this section. Except for the difference and $\delta \taus$ in the present equations, one finds the equivalent equations for $|\partial\Sigma_{\gas,\up}/\partial x|\ll k\Sigma_{\gas,\up}$.} Therefore, the present analyses give almost the same growth rate at $kR\gg1$ as one derived from linear analyses with the often-used external force. In other words, our analyses justify the use of the external force driving the drift to study mode properties at short wavelengths.

\subsection{Growth rate and phase velocity}\label{subsec:result2f}
%
%\begin{figure*}
%\gridline{
%\fig{grw_rate_phase_v_r20_st01_dgr1e-3_woExF.eps}{0.8\textwidth}{ }
%}
%\vspace{-20pt}
\begin{figure*}[htp]%[htp] or [H]
	\begin{center}
		%\hspace{-20pt}\raisebox{0pt}{
		%\hspace{100pt}\raisebox{20pt}{
		\includegraphics[width=2.0\columnwidth]{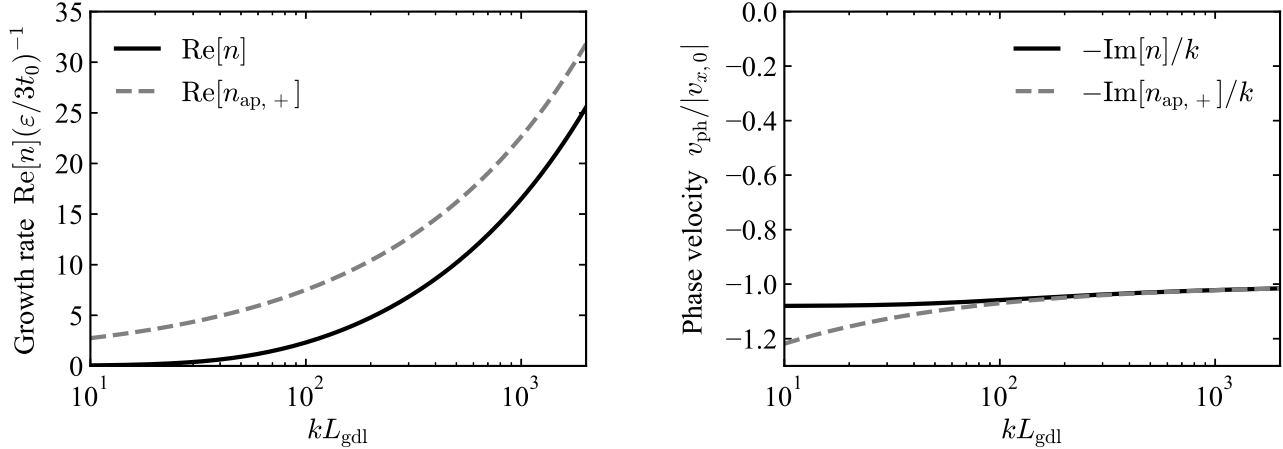}
		%}
	\end{center}
	%\vspace{-30pt}
\caption{Growth rate and phase velocity obtained from the two-fluid linear analyses for $\tausup=0.1$, $\varepsilon=10^{-3}$, and $R=20\;\mathrm{au}$ (black lines). In this case, one has $L_{\gdl}\simeq33.8H$. The gray dashed lines correspond to the growth rate and the phase velocity of the growing mode in one-fluid analyses (Equation (\ref{eq:1fdispersion})). The $k$-dependence of growth rate in two-fluid analyses is similar to that in one-fluid analyses although the two-fluid growth rate is smaller than the one-fluid growth rate. In both analyses, the phase velocities converge to the drift velocity especially at shorter wavelengths.}
\label{fig:grw_r20_st01_dgr1e-3_woExF}
\end{figure*}

%\begin{figure*}
%\gridline{
%\fig{Compare_r20_woExF_modi.eps}{0.8\textwidth}{ }
%}
%\vspace{-20pt}
\begin{figure*}[htp]%[htp] or [H]
	\begin{center}
		%\hspace{-20pt}\raisebox{0pt}{
		%\hspace{100pt}\raisebox{20pt}{
		\includegraphics[width=2.0\columnwidth]{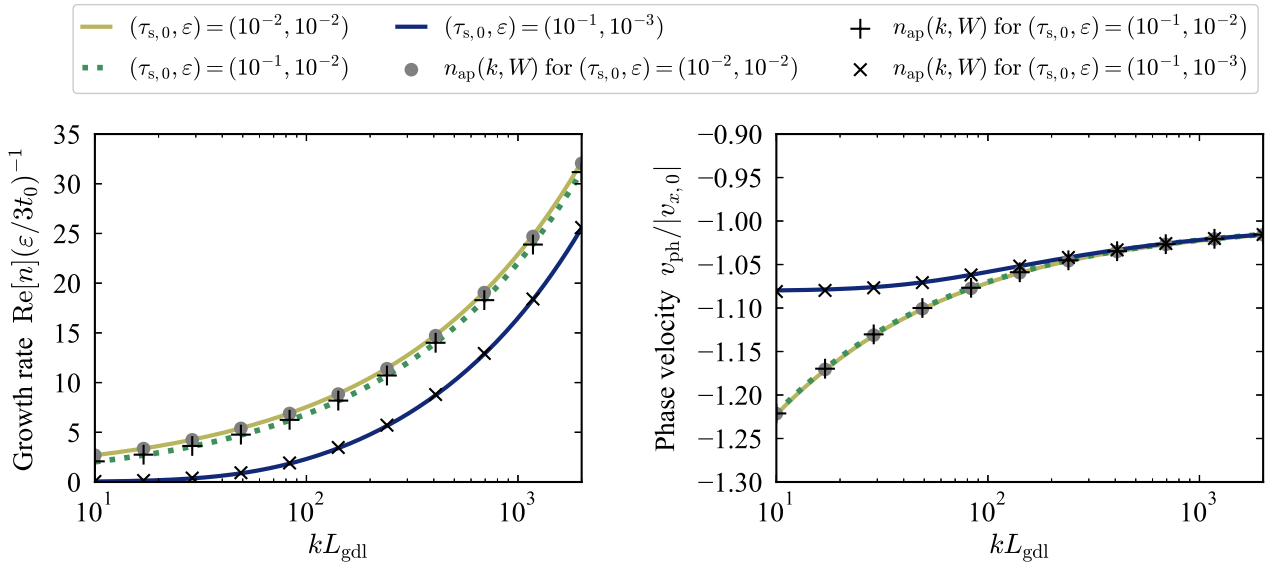}
		%}
	\end{center}
	%\vspace{-30pt}
\caption{Growth rate and phase velocity for $R=20\;\mathrm{au}$. Yellow solid, green dashed, and blue solid lines show the results of the two-fluid analyses for $(\tausup,\varepsilon)=(10^{-2},10^{-2}),\;(10^{-1},10^{-2}),$ and $(10^{-1},10^{-3})$, respectively. Filled circles, plus marks and cross marks represent growth rates and phase velocities obtained from the modified one-fluid analyses (Equation (\ref{eq:nap_k_T})). We find that the results of the two-fluid analyses are well reproduced by the modified one-fluid analyses.}
\label{fig:grw_r20_woExF_compare}
\end{figure*}

Parameters in this analysis are $\varepsilon,\;\tausup$, $\eta_{\mathrm{2D}}$, and the power law index of $\cs^2$ (see Equation (\ref{eq:sigmagup_eta_relation})). As noted above, we substitute the value of $\eta$ into $\eta_{\mathrm{2D}}$ to compare the one-fluid and two-fluid analyses. We refer the minimum mass solar nebula (MMSN) model \citep[][]{Hayashi1981} with a solar-mass star to assume the value of $\eta$ and $\partial\ln\cs^2/\partial R$. In the MMSN model, the gas surface density and the temperature $T$ are given by $\Sigma_{\gas,\mathrm{MMSN}}=1700(R/1\;\mathrm{au})^{-3/2}\;\mathrm{g\;cm}^{-2}$ and $T=280(R/1\;\mathrm{au})^{-1/2}\;\mathrm{K}$. Based on the disk model with the midplane gas density $\rho_{\gas,\mathrm{MMSN}}(0)=\Sigma_{\gas,\mathrm{MMSN}}/\sqrt{2\pi}H$ and the mean molecular weight of 2.34, one obtains $\eta\simeq1.81\times 10^{-3}(R/1\;\mathrm{au})^{1/2}$ and $\partial\ln\cs^2/\partial\ln R=-0.5$. 

Substituting the above value of $\eta$ at a given $R$ in $\eta_{\mathrm{2D}}$ in Equation (\ref{eq:sigmagup_eta_relation}), we calculate the growth rate and the phase velocity. As noted in Section \ref{subsec:2f_unp}, $\sigmagup'/\sigmagup$ used in this section is smaller than $\Sigma_{\gas,\mathrm{MMSN}}'/\Sigma_{\gas,\mathrm{MMSN}}$ because of the replacement of $\eta_{\mathrm{2D}}$ by $\eta$. At $R=20\;\mathrm{au}$, we obtain $\sigmagup'/\sigmagup\simeq-0.19H^{-1}$ and $\Sigma_{\gas,\mathrm{MMSN}}'/\Sigma_{\gas,\mathrm{MMSN}}\simeq-0.11H^{-1}$.

Figure \ref{fig:grw_r20_st01_dgr1e-3_woExF} shows the growth rate $\mathrm{Re}[n]$ normalized by $\varepsilon/3t_0$ and the phase velocity $v_{\mathrm{ph}}\equiv-\mathrm{Im}[n]/k$ normalized by $|v_{x,0}|$ in the case of $\tausup=0.1,\;\varepsilon=10^{-3},$ and $R=20\;\mathrm{au}$. The value of $\eta$ in this case is $\eta\simeq0.00809$. We also plot the growth rate and the phase velocity derived from the one-fluid analyses (Equation (\ref{eq:1fdispersion})). The $k$-dependences of the growth rates from both analyses are similar although the two-fluid analyses give lower growth rates. The phase velocities at high wavenumbers show a good agreement between the one-fluid and two-fluid analyses while the difference is visible at low wavenumbers.

The quantitative differences come from the third term on the right hand side of Equation (\ref{eq:dstdt_full}). The term represents the gas surface density variation during the dust drift. To understand this effect, we perform modified one-fluid analyses with Equation (\ref{eq:dsiddt}) and the following equation:
\begin{equation}
\ddt{\taus}+v_x\ddx{\taus}=\frac{\sigmad}{\sigmag}\frac{\taus}{3t_0}-\frac{\taus}{\sigmag}v_x\ddx{\sigmag},\label{eq:dstdt_modi}
\end{equation}
In the same way in Section \ref{sec:simpleana}, one can derive a dispersion relation of the growing mode:
\begin{equation}
n_{\app}(k,W)\equiv-ikv_{x,0}+\frac{\varepsilon}{6t_0}\left(W+\sqrt{W^2-\frac{12t_0}{\varepsilon}\frac{1-\tausup^2}{1+\tausup^2}ikv_{x,0}}\right),\label{eq:nap_k_T}
\end{equation}
\begin{equation}
W\equiv 1+\frac{2}{1+\tausup^2}\frac{L_{\gdl}}{\sigmagup/\sigmagup'}.\label{eq:def_T}
\end{equation}
Considering $\tausup\ll1$, one obtains
\begin{equation}
n_{\app}(k,W)\simeq-ikv_{x,0}+\frac{\varepsilon}{6t_0}\left(W+\sqrt{W^2-\frac{12t_0}{\varepsilon}ikv_{x,0}}\right)\label{eq:nap_k_T_app}
\end{equation}
\begin{equation}
W\simeq1+2\frac{L_{\gdl}}{\sigmagup/\sigmagup'}.\label{eq:def_T_app}
\end{equation}
Figure \ref{fig:grw_r20_woExF_compare} compares the modified one-fluid dispersion relation $n_{\app}(k,W)$ (Equation (\ref{eq:nap_k_T})) and two-fluid dispersion relation $n$. The modified one-fluid analyses well reproduce the results of the two-fluid analyses. Since the gas surface density gradient is negative, the newly introduced factor $W$ is less than unity, which reduces the growth rate. When the growth-drift length is much larger than $|\sigmagup/\sigmagup'|$, the factor $W$ becomes negative and its absolute value can become larger than unity. In such a case, the growth rate approaches toward 0 for $k\to0$ as shown in Figure \ref{fig:grw_r20_st01_dgr1e-3_woExF}. The phase velocity is independent of $k$ at small wavenumbers for $W<0$ and becomes $v_{\mathrm{ph}}\simeq v_{x,0}(1+1/|W|)$ for $kL_{\gdl}\ll W^2/4$ and $1\pm\tausup^2\simeq1$. This saturation of the phase velocity can be seen for cross marks of Figure \ref{fig:grw_r20_woExF_compare} where the growth-drift length is about $33.8H$ and $W\simeq-12.0$ from Equation (\ref{eq:def_T}), which gives $v_{\mathrm{ph}}/v_{x,0}\simeq1.08$.

%

%---- for double column ---
\begin{figure*}
%\begin{center}
%---- for single column ---
%\begin{figure}
%	\hspace{30pt}
%--------------------------------
	\begin{tabular}{c}
		\begin{minipage}{0.5\hsize}
			\begin{center}
				%\hspace{-20pt}\raisebox{0pt}{
				%\hspace{-30pt}\raisebox{-30pt}{
				\includegraphics[width=\columnwidth]{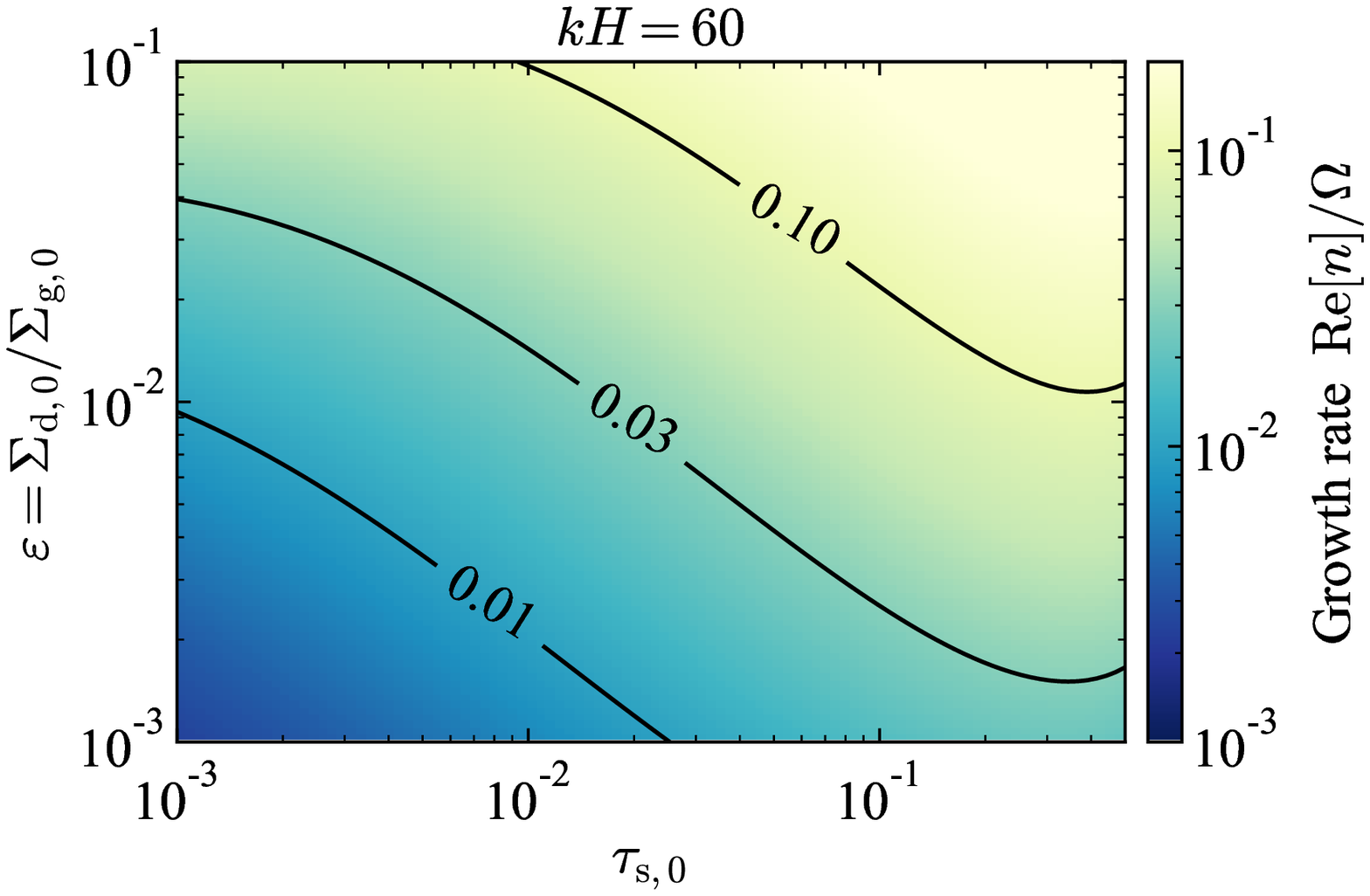}
				%}
			\end{center}
		\end{minipage}
		\begin{minipage}{0.5\hsize}
			\begin{center}
				%\hspace{-20pt}\raisebox{0pt}{
				%\hspace{-30pt}\raisebox{-30pt}{
				\includegraphics[width=\columnwidth]{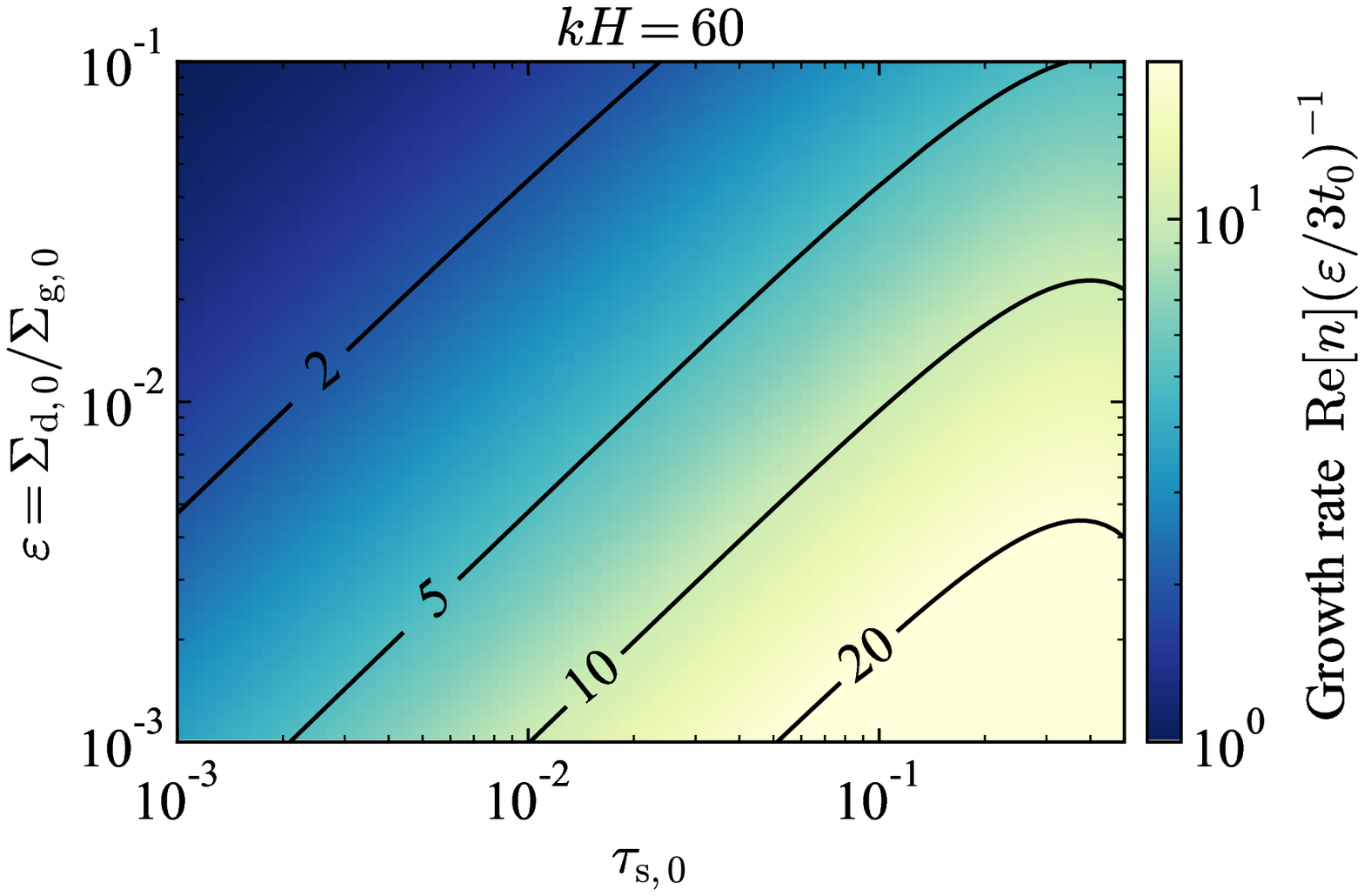}
				%}
			\end{center}
		\end{minipage}
	\end{tabular}
%	}
%\end{center}
%\vspace{15pt}
\caption{Growth rates obtained from the two-fluid analyses at $kH=60$ as a function of the normalized stopping time $\tausup$ and the unperturbed dust-to-gas surface density ratio $\varepsilon=\sigmadup/\sigmagup$. Here we assume $R=20\;\mathrm{au}$. The left figure shows the growth rate in the unit of $\Omega$. The growth rates become larger for larger dust sizes, i.e., larger $\taus$ because the faster drift speed results in the stronger traffic jam. The growth rates also increase as the dust-to-gas ratio increases as expected from the one-fluid analyses (see Equation (\ref{eq:1fdispersion})). The right figure shows the growth rate normalized by the coagulation timescale $3t_0/\varepsilon$. The growth rates increase as $\tausup/\varepsilon$ increases because $kL_{\gdl}\propto k\tausup/\varepsilon$ increases, which is also seen in the one-fluid analyses (see Equation (\ref{eq:normal_n_ap_pre})).}
\label{fig:grw_r20_cont_map}
\end{figure*}
%===

Regardless of the quantitative difference as shown in Figure \ref{fig:grw_r20_st01_dgr1e-3_woExF}, the simple one-fluid analyses roughly reproduce $k$-dependence of the growth rate obtained in two-fluid analyses. The modified one-fluid analyses with steady gas surface density gradient better reproduce the two-fluid dispersion relation. Therefore, the coagulation instability is essentially a one-fluid instability (see also Appendix \ref{app:sublayer}) and different from the streaming instability and the draft instability \citep[e.g.,][]{Youdin2005,Lambrechts2016}.

Figure \ref{fig:grw_r20_cont_map} shows the growth rate as a function of $\tausup$ and $\sigmadup/\sigmagup$. Since the growth rate monotonically increases as $kH$ increases, we fix the wavenumber and take $kH=60$ as the reference value. The left panel of Figure \ref{fig:grw_r20_cont_map} shows that the growth rate in the unit of $\Omega$ increases as the normalized stopping time $\tausup$ or the dust-to-gas ratio $\varepsilon=\sigmadup/\sigmagup$ increases. These trends are consistent with the $\varepsilon$- and $\tausup$-dependences of the dispersion relation from the modified one-fluid analyses. As $\tausup$ increases, one obtains larger dust drift speed $|v_{x,0}|$. Because the velocity perturbation is proportional to $v_{x,0}$ (Equation (\ref{eq:simpl_vx})), the faster drift speed results in the stronger concentration, and thus larger growth rates. As $\varepsilon$ increases, dust coagulation becomes effective and the coagulation instability grows faster. 

The right panel of Figure \ref{fig:grw_r20_cont_map} shows the growth rate normalized by the coagulation timescale $3t_0/\varepsilon$. As shown in Section \ref{sec:simpleana}, the dispersion relation obtained from the one-fluid analyses becomes self-similar for $\tausup\ll1$ when one normalizes the growth rate by $3t_0/\varepsilon$ and wavenumbers by the growth-drift length $L_{\gdl}$ (Equation (\ref{eq:normal_n_ap})). The growth rate $\mathrm{Re}[n]3t_0/\varepsilon$ should be constant for constant $\tausup/\varepsilon$ because $L_{\gdl}$ is proportional to $|v_{x,0}|/\varepsilon$ and the drift speed is proportional to $\tausup$. We find that the right panel of Figure \ref{fig:grw_r20_cont_map} also shows this trend. The modified one-fluid analyses also explain the self-similarity in the two-fluid dispersion relation. From Equation (\ref{eq:nap_k_T_app}) that is valid for $\tausup\ll1$, one obtains
\begin{equation}
n_{\app}(k,W)\left(\frac{3t_0}{\varepsilon}\right)\simeq i\tilde{k}+\frac{1}{2}\left(W+\sqrt{W^2+4i\tilde{k}}\right).
\end{equation}
According to Equation (\ref{eq:def_T_app}), the factor $W$ depends on $\tausup$ and $\varepsilon$ only through their ratio $\tausup/\varepsilon$ in $L_{\gdl}$ for $\tausup\ll1$. Therefore, the growth rate is self-similar for a given gas surface density gradient $\sigmagup'/\sigmagup$.

As already mentioned, we use Equation (\ref{eq:sigmagup_eta_relation}) to calculate $\sigmagup'/\sigmagup$ with the replacement of $\eta_{\mathrm{2D}}$ by $\eta$, leading to $\sigmagup'/\sigmagup<\Sigma_{\gas,\mathrm{MMSN}}'/\Sigma_{\gas,\mathrm{MMSN}} (<0)$. According to the definition of $W$, negative $\sigmagup'$ and its larger magnitude result in smaller $W$, which stabilizes the coagulation instability. Thus, the growth rates are underestimated in the current two-fluid analyses to some extent. Because the analyses of this section show that coagulation instability is essentially one-fluid mode (see also Appendix \ref{app:sublayer}), we conclude that it will be better to use dispersion relations from modified one-fluid analyses with a ``true" gas surface density gradient $(\Sigma_{\gas,\mathrm{MMSN}}'/\Sigma_{\gas,\mathrm{MMSN}})$ instead of using two-fluid dispersion relation. We thus adopt one-fluid approach with a gas surface density gradient in the evolutionary equation for $\taus$ in the following section.

\section{Discussion}\label{sec:discussion}

\subsection{Stabilization due to dust diffusion}\label{subsec:diffusion}
While turbulent motion of gas drives dust collisions, dust grains are also subject to turbulent diffusion that smooths out dust density perturbations. The diffusion is expected to suppress the coagulation instability especially at short wavelengths and limits the growth rate. In this subsection, we discuss to what extent the diffusion suppresses the instability performing one-fluid analyses.

The dust diffusion is often modeled by a diffusion term introduced in the dust continuity equation. The form of the diffusive mass flux depends on a closure relation between turbulent fluctuations and mean-field components. The usually assumed diffusive mass flux is proportional to gradient of either dust density or dust-to-gas ratio \citep[e.g.,][]{Cuzzi1993,Dubrulle1995}. However, simply adding the diffusion term to the continuity equation introduces unphysical momentum flux, which violates the total angular momentum conservation \citep[][]{Goodman2000,Tominaga2019}. \citet{Tominaga2019} shows that replacing the dust velocity by a sum of the mean-flow velocity and the diffusive flux terms recovers the angular momentum conservation. The equations including such terms are mathematically and physically derived from the mean-field approximation of usual momentum equations \citep[see Appendix A in][]{Tominaga2019}. Following \citet[][]{Tominaga2019} in order to discuss the effects of the diffusion, we simply replace $v_x$ in Equations (\ref{eq:dsiddt}) and (\ref{eq:dstdt_modi}) by
\begin{equation}
v_x=\left<v_x\right>-\frac{D}{\sigmad}\ddx{\sigmad},\label{eq:vx_and_diff1}
\end{equation}
or
\begin{equation}
v_x=\left<v_x\right>-\frac{D\sigmag}{\sigmad}\frac{\partial}{\partial x}\left(\frac{\sigmad}{\sigmag}\right).\label{eq:vx_and_diff2}
\end{equation}
where $D$ is a diffusion coefficient and $\left<v_x\right>$ is the mean-flow component representing the so-called drift and defined by
\begin{equation}
\left<v_x\right>\equiv-\frac{2\taus}{1+\taus^2}\eta R\Omega.
\end{equation}
The resultant equations with Equation (\ref{eq:vx_and_diff1})  are
\begin{equation}
\ddt{\sigmad}+\ddx{\sigmad\left<v_x\right>}=\frac{\partial}{\partial x}\left(D\ddx{\sigmad}\right)\label{eq:dsiddt_w_diff1}
\end{equation}
\begin{align}
\ddt{\taus}+&\left(\left<v_x\right>-\frac{D}{\sigmad}\ddx{\sigmad}\right)\ddx{\taus}\notag\\
&=\frac{\sigmad}{\sigmag}\frac{\taus}{3t_0}-\frac{\taus}{\sigmag}\left(\left<v_x\right>-\frac{D}{\sigmad}\ddx{\sigmad}\right)\ddx{\sigmag}.\label{eq:dstdt_modi_w_diff1}
\end{align}
In the case we use Equation (\ref{eq:vx_and_diff2}), we obtain the following equations:
\begin{equation}
\ddt{\sigmad}+\ddx{\sigmad\left<v_x\right>}=\frac{\partial}{\partial x}\left(D\sigmag\frac{\partial}{\partial x}\left(\frac{\sigmad}{\sigmag}\right)\right)\label{eq:dsiddt_w_diff2}
\end{equation}
\begin{align}
\ddt{\taus}+&\left(\left<v_x\right>-\frac{D\sigmag}{\sigmad}\frac{\partial}{\partial x}\left(\frac{\sigmad}{\sigmag}\right)\right)\ddx{\taus}\notag\\
&=\frac{\sigmad}{\sigmag}\frac{\taus}{3t_0}-\frac{\taus}{\sigmag}\left(\left<v_x\right>-\frac{D\sigmag}{\sigmad}\frac{\partial}{\partial x}\left(\frac{\sigmad}{\sigmag}\right)\right)\ddx{\sigmag}.\label{eq:dstdt_modi_w_diff2}
\end{align}
We perform linear analyses based on Equations (\ref{eq:dsiddt_w_diff1}) and (\ref{eq:dstdt_modi_w_diff1}) and those based on Equations (\ref{eq:dsiddt_w_diff2}) and (\ref{eq:dstdt_modi_w_diff2}). In both cases, we set unperturbed states with uniform dust surface density and normalized stopping time, and take into account the dust growth equation (Equation (\ref{eq:dstdt_modi_w_diff1}) or (\ref{eq:dstdt_modi_w_diff2})) only for perturbed quantities as in Section \ref{sec:simpleana} and \ref{sec:linana_woExF}.

%---- for double column ---
\begin{figure*}
%\begin{center}
%---- for single column ---
%\begin{figure}
%	\hspace{30pt}
%--------------------------------
	\begin{tabular}{c}
		\begin{minipage}{0.5\hsize}
			\begin{center}
				%\hspace{-20pt}\raisebox{0pt}{
				%\hspace{-30pt}\raisebox{-30pt}{
				\includegraphics[width=1.0\columnwidth]{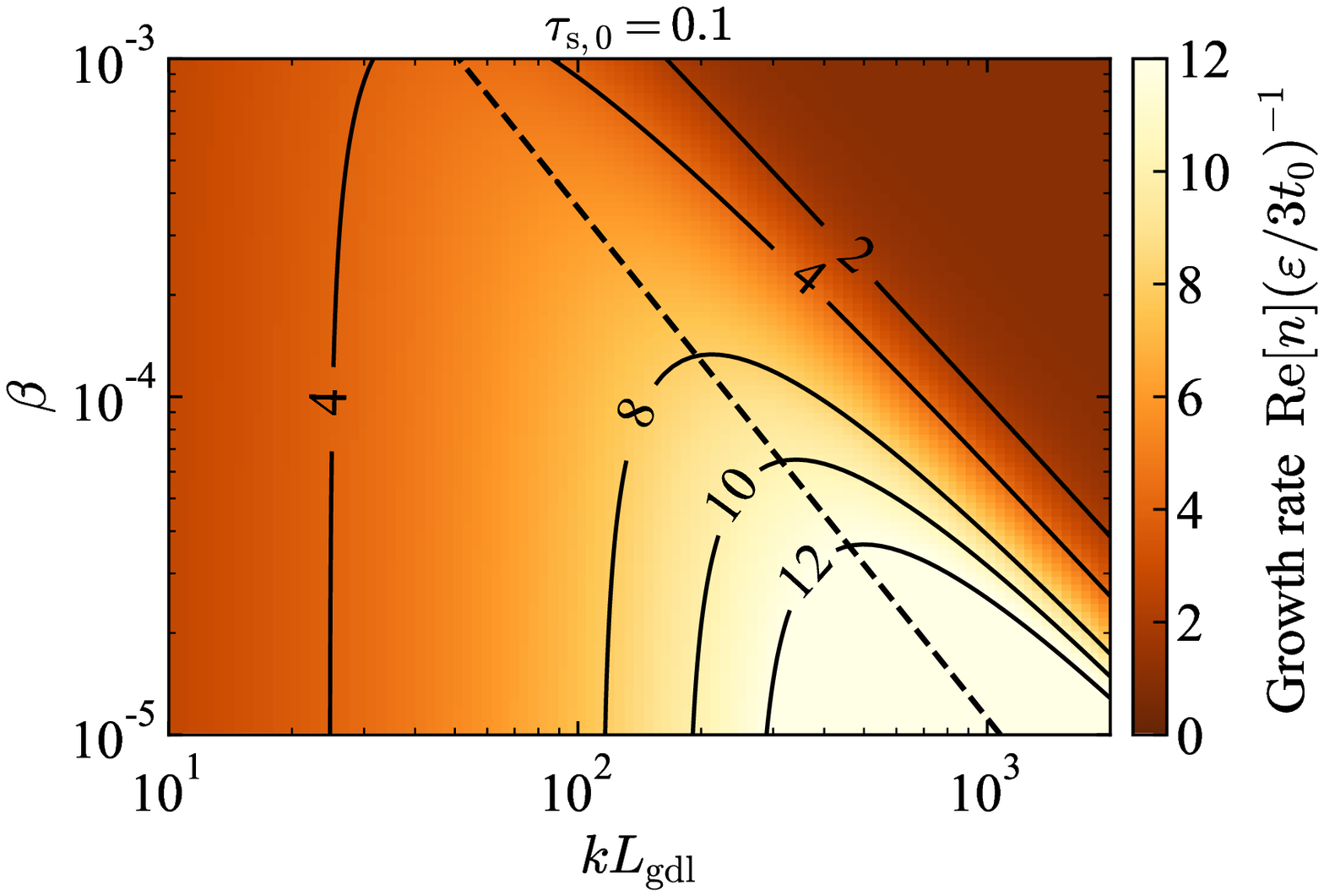}
				%}
			\end{center}
		\end{minipage}
		\begin{minipage}{0.5\hsize}
			\begin{center}
				%\hspace{-20pt}\raisebox{0pt}{
				%\hspace{-30pt}\raisebox{-30pt}{
				\includegraphics[width=1.0\columnwidth]{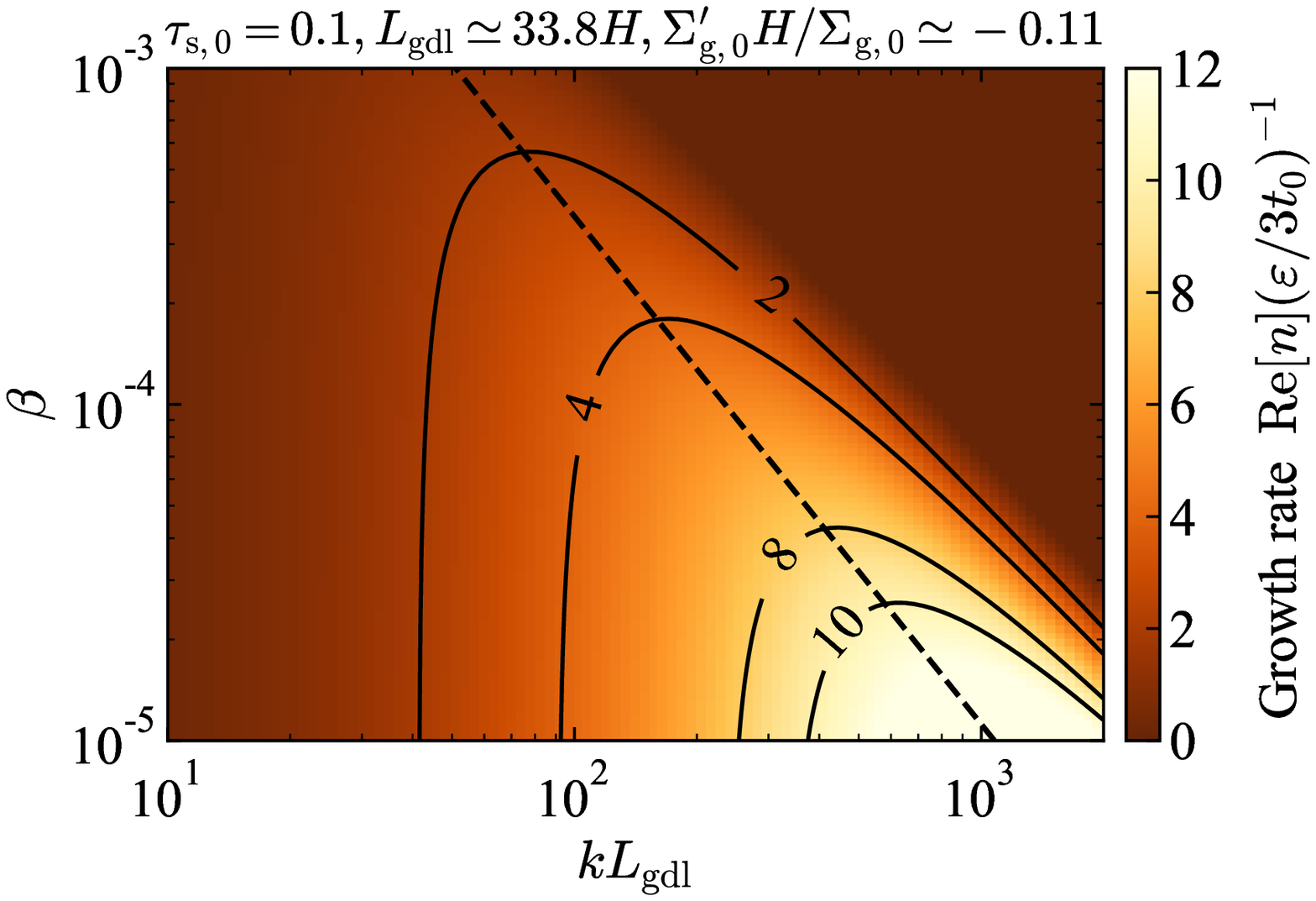}
				%}
			\end{center}
		\end{minipage}
	\end{tabular}
%	}
%\end{center}
%\vspace{15pt}
\caption{Growth rates in the cases with dust diffusion. In both panels, the horizontal axis is wavenumber normalized by the growth-drift length $L_{\gdl}$, and the vertical axis is the dimensionless diffusivity $\beta=DL_{\gdl}^{-2}(3t_0/\varepsilon)$. The left panel shows the growth rate calculated without the terms proportional to $\sigmagup'$ in Equation (\ref{eq:disp_diff1}). The right panel shows the growth rate with all terms in Equation (\ref{eq:disp_diff1}). In the presence of dust diffusion due to weak turbulence, growth rates of coagulation instability are larger than the dust growth rate $\varepsilon/3t_0$.  Dashed lines in both panels show the wavenumber given by Equation (\ref{eq:app_kmax}). The most unstable wavenumber is well described by Equation (\ref{eq:app_kmax}).}
\label{fig:grw_w_diffusion_ts01}
\end{figure*}
%===

First, we present results of the linear analyses with Equations (\ref{eq:dsiddt_w_diff1}) and (\ref{eq:dstdt_modi_w_diff1}). In this case, the unperturbed velocity $v_{x,0}$ is $v_{x,0}=\left<v_{x,0}\right>=-2\tausup\eta R\Omega/(1+\tausup^2)$. Based on the above unperturbed state with uniform $\sigmadup$ and $\tausup$, we obtain the following linearized equations:
\begin{equation}
(n+ik\left<v_{x,0}\right>+Dk^2)\frac{\delta\sigmad}{\sigmadup}+\frac{1-\tausup^2}{1+\tausup^2}ik\left<v_{x,0}\right>\frac{\delta\taus}{\tausup}=0,
\end{equation}
\begin{align}
\biggl(n+ik\left<v_{x,0}\right>&-\frac{\varepsilon}{3t_0}+\frac{2\left<v_{x,0}\right>}{1+\tausup^2}\frac{\sigmagup'}{\sigmagup}\biggr)\frac{\delta\taus}{\tausup}\notag\\
&+\left(-\frac{\varepsilon}{3t_0}-ikD\frac{\sigmagup'}{\sigmagup}\right)\frac{\delta\sigmad}{\sigmadup}=0.
\end{align}
The full dispersion relation is
\begin{equation}
(n+ik\left<v_{x,0}\right>)^2+A_1(n+ik\left<v_{x,0}\right>)+A_0=0,\label{eq:disp_diff1_pre}
\end{equation}
\begin{equation}
A_1=-\frac{\varepsilon}{3t_0}+\frac{2\left<v_{x,0}\right>}{1+\tausup^2}\frac{\sigmagup'}{\sigmagup}+Dk^2,\label{eq:a1}
\end{equation}
\begin{align}
A_0=&ik\left<v_{x,0}\right>\left(\frac{1-\tausup^2}{1+\tausup^2}\right)\left(\frac{\varepsilon}{3t_0}+ikD\frac{\sigmagup'}{\sigmagup}\right)\notag\\
&+Dk^2\left(-\frac{\varepsilon}{3t_0}+\frac{2\left<v_{x,0}\right>}{1+\tausup^2}\frac{\sigmagup'}{\sigmagup}\right).\label{eq:a0}
\end{align}
The terms proportional to $\sigmagup'$ originate from the last term on the right-hand side of Equation (\ref{eq:dstdt_modi_w_diff1}). Using $L_{\gdl}=3t_0\left<v_{x,0}\right>/\varepsilon$ and $W$ defined by Equation (\ref{eq:def_T}), we obtain the dispersion relation of a growing mode in the dimensionless form
\begin{align}
n\left(\frac{3t_0}{\varepsilon}\right)&=i\tilde{k}+\frac{1}{2}\left(W-\beta\tilde{k}^2\right)\notag\\
&+\frac{1}{2}\sqrt{\left(W+\beta\tilde{k}^2\right)^2+4i\tilde{k}\frac{1-\tausup^2}{1+\tausup^2}\left(1+i\tilde{k}\beta L_{\gdl}\frac{\sigmagup'}{\sigmagup}\right)},\label{eq:disp_diff1}
\end{align}
where $\beta=DL_{\gdl}^{-2}(3t_0/\varepsilon)$ is the diffusion coefficient normalized by the growth-drift length $L_{\gdl}$ and the dust coagulation timescale $3t_0/\varepsilon$. Assuming small dust particles ($\tausup\ll1$) and $D\simeq\alpha\cs^2\Omega^{-1}$ \citep[see][]{YL2007}, we can relate $\beta$ to $\alpha$ as follows (see also Equation (\ref{eq:Lgdl_H})):
\begin{align}
\beta\simeq&\alpha\left(\frac{H}{L_{\gdl}}\right)^2\frac{3t_0\Omega}{\varepsilon}\notag\\
\simeq&1.7\times10^{-4}\left(\frac{\alpha}{1.0\times10^{-4}}\right)\left(\frac{t_0\Omega}{0.50}\right)^{-1}\notag\\
&\times\left(\frac{\varepsilon}{1.0\times10^{-3}}\right)\left(\frac{|v_{x,0}|/\cs}{2.0
\times10^{-2}}\right)^{-2}.\label{eq:beta}
\end{align}
We note that $\beta$ depends on dust sizes since $|v_{x,0}|$ depends on $\tausup$. To compare the simple and modified one-fluid analyses (see also Section \ref{subsec:result2f}), we calculate both (1) growth rates without the terms proportional to $\sigmagup'$ and (2) growth rates with all terms in Equation (\ref{eq:disp_diff1}).

Figure \ref{fig:grw_w_diffusion_ts01} shows the growth rates for $\taus=0.1$. The color shows the growth rates normalized by the dust coagulation timescale $3t_0/\varepsilon$. The left panel shows the growth rate calculated without the last term proportional to $\sigmagup'$ on the right-hand side of Equation (\ref{eq:dstdt_modi_w_diff1}) while the right panel shows the growth rate with all terms. We also assume $W=1$ on the left panel. In both cases, the coagulation instability is stabilized by dust diffusion at short wavelengths. As a result, the coagulation instability has the most unstable wavenumber $k_{\mathrm{max}}$ in contrast to the diffusion-free case (Sections \ref{sec:simpleana} and \ref{sec:linana_woExF}). We find that the most unstable wavenumber can be well estimated by the following relation when $|W|$ is not so large ($\sim1$):
\begin{equation}
k_{\mathrm{max}}L_{\gdl} \simeq \frac{1}{3}\left(\frac{4}{\beta^2}\frac{1-\tausup^2}{1+\tausup^2}\right)^{1/3}.\label{eq:app_kmax}
\end{equation}
The dashed line in Figure \ref{fig:grw_w_diffusion_ts01} represents the wavenumber calculated with Equation (\ref{eq:app_kmax}). In both panels, Equation (\ref{eq:app_kmax}) well represents the most unstable wavenumber. Considering small dust particles and adopting $1\pm\tausup^2\simeq 1$, we obtain the most unstable wavelength $\lambda_{\mathrm{max}}\equiv 2\pi/k_{\mathrm{max}}$ as
\begin{align}
\lambda_{\mathrm{max}}\simeq 1.1H&\left(\frac{\alpha}{1.0\times 10^{-4}}\right)^{2/3}\left(\frac{t_0\Omega}{0.50}\right)^{1/3}\notag\\
&\times\left(\frac{\varepsilon}{1.0\times 10^{-3}}\right)^{-1/3}\left(\frac{|v_{x,0}|/\cs}{2.0\times10^{-2}}\right)^{-1/3}
\end{align}
As in the case without the dust diffusion (Section \ref{subsec:result2f}), including the effect of $\sigmagup'$ in the dust growth equation reduces the growth rate by a factor of a few for $L_{\gdl}\simeq33.8H$ (see the right panel of Figure \ref{fig:grw_w_diffusion_ts01}). When a gas disk is less turbulent and $\beta$ becomes less than $1\times10^{-4}$, the coagulation instability grows 2-10 times faster than dust particles coagulate. Such a situation is realized in a region where $\alpha\times\varepsilon\lesssim6\times10^{-8}$ for $|v_{x,0}|/\cs=0.02$ (see Equation (\ref{eq:beta})).

%---- for double column ---
\begin{figure*}
%\begin{center}
%---- for single column ---
%\begin{figure}
%	\hspace{30pt}
%--------------------------------
	\begin{tabular}{c}
		\begin{minipage}{0.5\hsize}
			\begin{center}
				%\hspace{-20pt}\raisebox{0pt}{
				%\hspace{-30pt}\raisebox{-30pt}{
				\includegraphics[width=1.0\columnwidth]{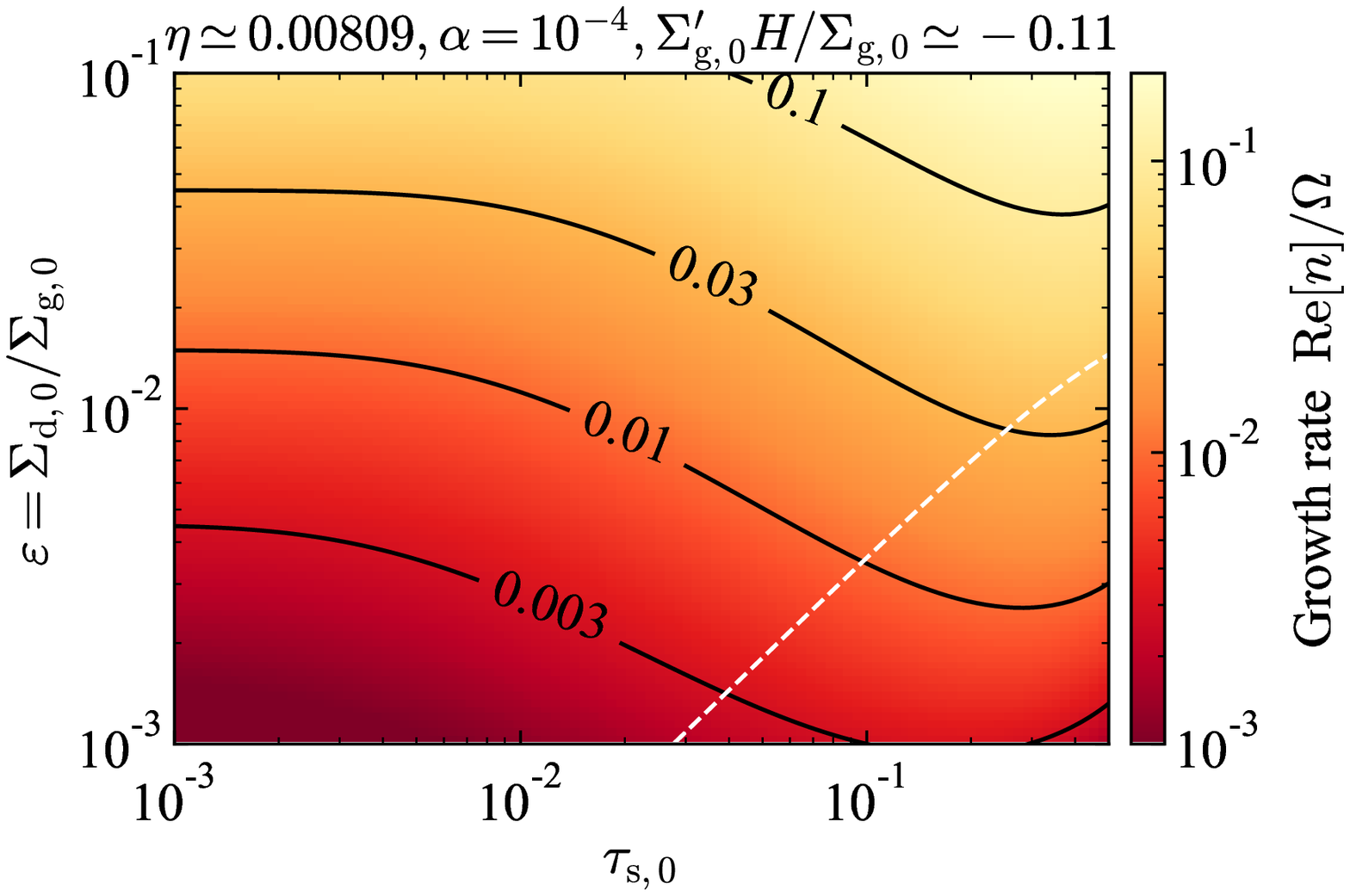}
				%}
			\end{center}
		\end{minipage}
		\begin{minipage}{0.5\hsize}
			\begin{center}
				%\hspace{-20pt}\raisebox{0pt}{
				%\hspace{-30pt}\raisebox{-30pt}{
				\includegraphics[width=1.0\columnwidth]{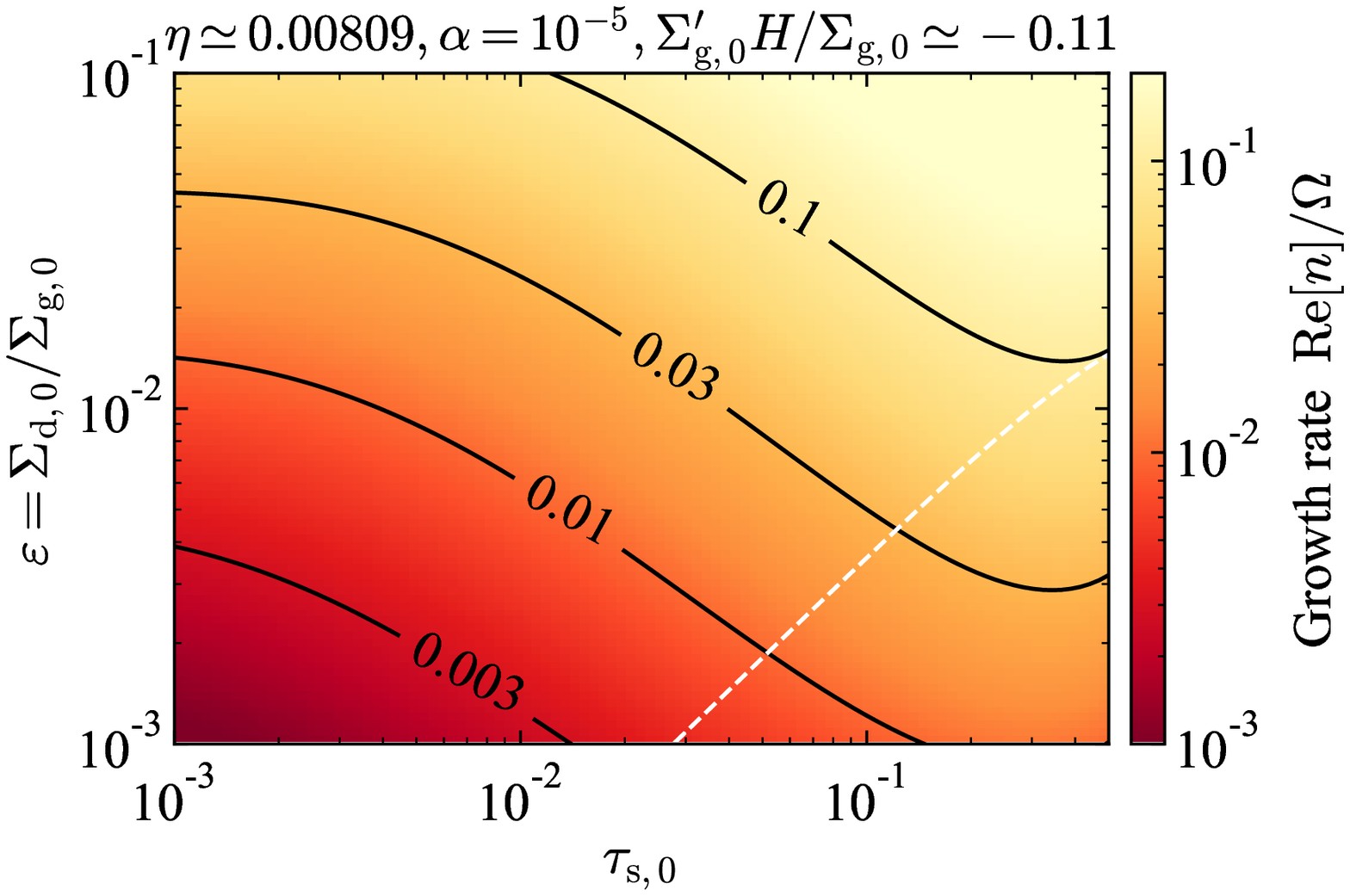}
				%}
			\end{center}
		\end{minipage}
	\end{tabular}
%	}
%\end{center}
%\vspace{15pt}
\caption{Growth rate with the dust diffusion (Equation (\ref{eq:disp_diff1})) in the unit of $\Omega$ at $kL_{\gdl}=(4\beta^{-2}(1-\tausup^2)/27(1+\tausup^2))^{1/3}$ (Equation (\ref{eq:app_kmax})) as a function of the normalized stopping time $\tausup$ and the unperturbed dust-to-gas surface density ratio $\varepsilon=\sigmadup/\sigmagup$. The left figure shows the growth rate for $\alpha=10^{-4}$, and the right figure shows the growth rate for $\alpha=10^{-5}$. The white dashed lines on both panels show sets of $(\tausup,\;\varepsilon)$ that satisfy $\sigmadup/3t_0=\left<v_{x,0}\right>\sigmagup'$, meaning that the equation for $\taus$ is exactly satisfied in the steady unperturbed state with uniform $\tausup$ and $\sigmadup$, and thus our analysis is rigorous. For given $\tausup$ and $\varepsilon$, the growth rate of coagulation instability is larger for smaller $\alpha$.  Even for $\varepsilon\sim10^{-3}$, the growth timescale $1/\mathrm{Re}[n]$ can be a few tens of the orbital period ($2\pi/\Omega$) if dust particles are large and $\tausup\simeq10^{-1}$.}
\label{fig:grw_w_diff_comp}
\end{figure*}
%===

Figure \ref{fig:grw_w_diff_comp} shows growth rates in the unit of $\Omega$ calculated with Equations (\ref{eq:disp_diff1}) and (\ref{eq:app_kmax}). Following \citet{YL2007} and \citet{Youdin2011}, we calculate the diffusion coefficient as
\begin{equation}
D=\frac{1+\tausup+4\tausup^2}{\left(1+\tausup^2\right)^2}\alpha\cs H.
\end{equation}
The white dashed lines on both panels show sets of $(\tausup,\;\varepsilon)$ satisfying $\sigmadup/3t_0=\left<v_{x.0}\right>\sigmagup'$. This means that the equation for $\taus$ (Equation (\ref{eq:dstdt_modi_w_diff1})) is satisfied with uniform $\tausup$ and $\sigmadup$ in the steady unperturbed state. In other words, the adopted uniform state is not an approximated solution as explained in the beginning of Section \ref{subsec:1flinana} but an exact steady solution in the present one-fluid formalism. Thus, the present one-fluid analysis is rigorous on the white dashed lines. The strength of turbulence $\alpha$ is assumed to be $10^{-4}$ on the left panel of Figure \ref{fig:grw_w_diff_comp} and $10^{-5}$ on the right panel. In these cases, the coagulation instability can develop within a few tens of the orbital period even when the dust-to-gas ratio is less than 0.01. Because we use Equation (\ref{eq:app_kmax}) to obtain the approximated maximum growth rates in Figure \ref{fig:grw_w_diff_comp}, one finds not only $\beta$ larger but also $kL_{\gdl}$ smaller for smaller $\tausup$. The long-wavelength limit ($kL_{\gdl}\to0$) gives the growth rate of $\varepsilon/3t_0$ as also discussed in Section \ref{sec:simpleana}. Note that small $\tausup$ makes $L_{\gdl}\propto|v_{x,0}|$ smaller, leading to $n\times\left(3t_0/\varepsilon\right)=W\to1$ at $k=0$ as $\tausup\to0$ (see also Equation (\ref{eq:def_T})). This is the reason why the growth rates remains relatively constant at small $\tausup$ and depends on only $\varepsilon$ in Figure \ref{fig:grw_w_diff_comp}.

Linear analyses with Equations (\ref{eq:dsiddt_w_diff2}) and (\ref{eq:dstdt_modi_w_diff2}) show little difference from the above analyses with Equations (\ref{eq:dsiddt_w_diff1}) and (\ref{eq:dstdt_modi_w_diff1}). We thus simply show a dispersion relation derived. The unperturbed state with uniform $\sigmagup$ has the dust drift velocity $v_{x,0}=\left<v_{x,0}\right>+D\sigmagup'/\sigmagup$. Linearizing Equations (\ref{eq:dsiddt_w_diff2}) and (\ref{eq:dstdt_modi_w_diff2}), we obtain the following dispersion relation: 
\begin{equation}
(n+ikv_{x,0})^2+B_1(n+ikv_{x,0})+B_0=0,\label{eq:disp_diff2}
\end{equation}
\begin{equation}
B_1=-\frac{\varepsilon}{3t_0}+\frac{2v_{x,0}}{1+\tausup^2}\frac{\sigmagup'}{\sigmagup}+Dk^2 -D\left(\frac{\sigmagup'}{\sigmagup}\right)^2\frac{1-\tausup^2}{1+\tausup^2},\label{eq:b1}
\end{equation}
\begin{align}
B_0=&ikv_{x,0}\left(\frac{1-\tausup^2}{1+\tausup^2}\right)\left(\frac{\varepsilon}{3t_0}+ikD\frac{\sigmagup'}{\sigmagup}\right)\notag\\
&+Dk^2\left(-\frac{\varepsilon}{3t_0}+\frac{2v_{x,0}}{1+\tausup^2}\frac{\sigmagup'}{\sigmagup}\right)-\frac{ik\varepsilon D}{3t_0}\frac{\sigmagup'}{\sigmagup}\frac{1-\tausup^2}{1+\tausup^2}.\label{eq:b0}
\end{align}
The differences from Equations (\ref{eq:disp_diff1_pre})-(\ref{eq:a0}) are twofold: (1) $\left<v_{x,0}\right>$ is replaced with $v_{x,0}$, (2) the last terms on the right-hand side of Equations (\ref{eq:b1}) and (\ref{eq:b0}). The former difference is small when $\tausup$ is larger than $\alpha$ and $|\left<v_{x,0}\right>|>D|\sigmagup'/\sigmagup|$. The latter difference is also small. Because the coagulation instability grows at $kH\gtrsim1$ and $kR\gg1$ in weakly turbulent disks, the last term of Equation (\ref{eq:b1}) is much smaller than the third term $Dk^2$. The last term on the right-hand side of Equation (\ref{eq:b0}) is also small compared to the first term when $\tausup>\alpha$ and $|v_{x,0}|>D|\sigmagup'/\sigmagup|$ are satisfied. Because of $\hd/H\simeq\sqrt{\alpha/\tausup}$, the condition $\tausup>\alpha$ is equivalent to $\hd<H$, which we usually expect.

\subsection{Backreaction at the midplane}\label{subsec:BR_midplane}
In Section \ref{sec:simpleana}, we neglect the frictional  backreaction to the gas and use the drift velocity of the test particle limit. In Section \ref{sec:linana_woExF}, we take into account the backreaction in the two-fluid height-integrated system but the backreaction terms are proportional to the dust-to-gas surface density ratio. In reality, dust grains settle around the midplane and the backreaction is determined by the midplane dust-to-gas ratio $\epmid$:
\begin{equation}
\epmid\equiv\frac{H\sigmad}{\hd\sigmag}=\sqrt{\frac{\taus}{\alpha}}\varepsilon,\label{eq:midplane_dgr}
\end{equation}
where we approximate the dust scale height with $\hd=H\sqrt{\alpha/\taus}$ (see also Appendix \ref{app:sublayer} for sublayer modeling). The midplane dust-to-gas ratio $\epmid$ is larger than $\varepsilon=\sigmad/\sigmag$ for small $\hd/H$, which brakes dust grains and can limit the efficiency of traffic jam, and thus coagulation instability. \footnote{ One could modify the backreaction terms in gas equations by multiplying a numerical factor of $\epmid/\varepsilon$ to mimic the midplane backreaction in full-disk averaging model. However, such a modeling will violate the momentum conservation. Violation of the conservation law leads to unphysical mode properties, which is discussed in another context (diffusion modeling) in \citet{Tominaga2019}. Thus, we only consider the dust motion and regard the gas as a momentum reservoir in this subsection. We describe another model based on a sublayer-averaging in Appendix \ref{app:sublayer}. }  Moreover, the degree of dust settling depends on stopping time and thus time-dependent and spatially dependent during the growth of coagulation instability. Therefore, in this subsection, utilizing one-fluid equations, we discuss coagulation instability under the radial drift mediated by the $\taus$-dependent midplane dust-to-gas mass ratio $\epmid$.

The dust velocity is given by
\begin{equation}
\left<v_x\right>=-\frac{2\taus}{(1+\epmid)^2+\taus^2}\eta R\Omega.\label{eq:vx_w_midplaneBR}
\end{equation}
Taking into account the diffusion, we use Equations (\ref{eq:dsiddt_w_diff1}) and (\ref{eq:dstdt_modi_w_diff1}) in this subsection. We note that the coagulation rate is still determined by the dust-to-gas surface density ratio $\sigmad/\sigmag$ regardless of the modification of the velocity with the midplane dust-to-gas ratio. Linearizing Equations  (\ref{eq:midplane_dgr}) and (\ref{eq:vx_w_midplaneBR}) with $\sigmag=\sigmagup$ gives
\begin{equation}
\frac{\delta\epmid}{\epmidup}=\frac{\delta\sigmad}{\sigmadup}+\frac{1}{2}\frac{\delta\taus}{\tausup},
\end{equation}
\begin{equation}
\frac{\delta\left<v_x\right>}{\left<v_{x,0}\right>}=\frac{1+\epmidup-\tausup^2}{(1+\epmidup)^2+\tausup^2}\frac{\delta\taus}{\tausup}-\frac{2\epmidup(1+\epmidup)}{(1+\epmidup)^2+\tausup^2}\frac{\delta\sigmad}{\sigmadup},\label{eq:linearized_vx_w_midplaneBR}
\end{equation}
where $\epmidup\equiv(\sigmadup/\sigmagup)\sqrt{\tausup/\alpha}=\varepsilon\sqrt{\tausup/\alpha}$ and
\begin{equation}
\left<v_{x,0}\right>=-\frac{2\tausup}{(1+\epmidup)^2+\tausup^2}\eta R\Omega.
\end{equation}
One can derive a dispersion relation using Equation (\ref{eq:linearized_vx_w_midplaneBR}) and
\begin{equation}
(n+ik\left<v_{x,0}\right>+Dk^2)\frac{\delta\sigmad}{\sigmadup}+ik\delta\left<v_x\right>=0,
\end{equation}
\begin{align}
\bigg(n+ik\left<v_{x,0}\right>&-\frac{\varepsilon}{3t_0}+\frac{\left<v_{x,0}\right>\sigmagup'}{\sigmagup}\bigg)\frac{\delta\taus}{\tausup}\notag\\
&=\left(\frac{\varepsilon}{3t_0}+\frac{Dik\sigmagup'}{\sigmagup}\right)\frac{\delta\sigmad}{\sigmadup}-\frac{\sigmagup'}{\sigmagup}\delta\left<v_x\right>,
\end{align}
which are the linearized equations of Equations (\ref{eq:dsiddt_w_diff1}) and (\ref{eq:dstdt_modi_w_diff1}), respectivley. We obtain the following dispersion relation of coagulation instability
\begin{align}
n\left(\frac{3t_0}{\varepsilon}\right)&=i\tilde{k}+\frac{1}{2}\left(\wmod-\beta\tilde{k}^2-\frac{2i\tilde{k}\epmidup(1+\epmidup)}{(1+\epmidup)^2+\tausup^2}\right)\notag\\
&+\frac{1}{2}\Bigg[\left(\wmod+\beta\tilde{k}^2+\frac{2i\tilde{k}\epmidup(1+\epmidup)}{(1+\epmidup)^2+\tausup^2}\right)^2\notag\\
&+4i\tilde{k}\frac{1+\epmidup-\tausup^2}{(1+\epmidup)^2+\tausup^2}\notag\\
&\times\left\{1+ L_{\gdl}\frac{\sigmagup'}{\sigmagup}\left(i\tilde{k}\beta-\frac{2\epmidup(1+\epmidup)}{(1+\epmidup)^2+\tausup^2}\right)\right\}\Bigg]^{1/2},\label{eq:disp_diff1_wmidplaneBR}
\end{align}
where $\wmod$ is given by
\begin{equation}
\wmod\equiv1+\frac{(1+\epmidup)(2+\epmidup)}{(1+\epmidup)^2+\tausup^2}\frac{L_{\gdl}\sigmagup'}{\sigmagup}.\label{eq:wmod}
\end{equation}
As already discussed in Sections \ref{subsec:result2f} and \ref{subsec:diffusion}, $\wmod$ and the terms proportional to $L_{\gdl}\sigmagup'/\sigmagup$ represent the stabilizing effects due to the second term on the right had side of Equation (\ref{eq:dstdt_modi_w_diff1}). To focus on stabilizing effects due to the backreaction with $\epmid$, we first set $\wmod=1$ and neglect the other terms proportional to $L_{\gdl}\sigmagup'/\sigmagup$. Thus, the dispersion relation discussed below is
\begin{align}
n\left(\frac{3t_0}{\varepsilon}\right)&=i\tilde{k}+\frac{1}{2}\left(1-\beta\tilde{k}^2-\frac{2i\tilde{k}\epmidup(1+\epmidup)}{(1+\epmidup)^2+\tausup^2}\right)\notag\\
&+\frac{1}{2}\Bigg[\left(1+\beta\tilde{k}^2+\frac{2i\tilde{k}\epmidup(1+\epmidup)}{(1+\epmidup)^2+\tausup^2}\right)^2\notag\\
&\;\;\;\;\;\;\;\;\;+4i\tilde{k}\frac{1+\epmidup-\tausup^2}{(1+\epmidup)^2+\tausup^2}\Bigg]^{1/2}.\label{eq:disp_diff1_wmidplaneBR_woW}
\end{align}
The parameters determining the growth rates are $\tausup$, $\varepsilon$, and $\alpha$. The midplane dust-to-gas ratio $\epmidup$ is determined by those parameters through Equation (\ref{eq:midplane_dgr}).

%\begin{figure*}
%\gridline{
%\fig{Compare_r20_woExF_modi.eps}{0.8\textwidth}{ }
%}
%\vspace{-20pt}
\begin{figure*}[htp]%[htp] or [H]
	\begin{center}
		%\hspace{-20pt}\raisebox{0pt}{
		%\hspace{100pt}\raisebox{20pt}{
		\includegraphics[width=2.0\columnwidth]{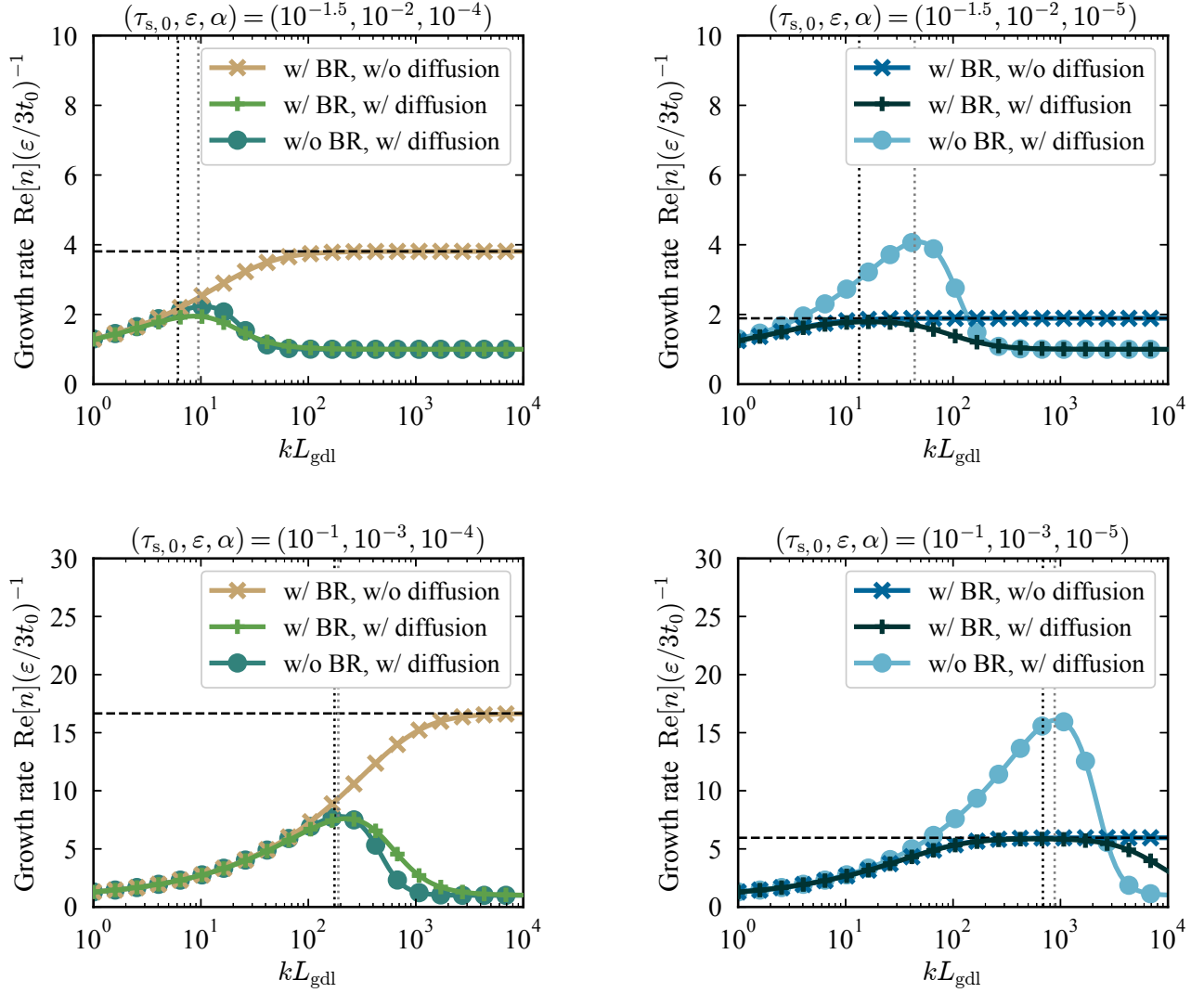}
		%}
	\end{center}
	%\vspace{-30pt}
\caption{Growth rate obtained from Equation (\ref{eq:disp_diff1_wmidplaneBR_woW}). The ``BR" in the legends stands for backreaction calculated with the midplane dust-to-gas  ratio $\epmid$. Cross marks, plus marks, and filled circles show different growth rates and represents whether the backreaction or diffusion is considered or not. Each panel shows growth rates for different parameters: $(\tausup,\varepsilon,\alpha)$. The top panels and the bottom panels are results of $(\tausup,\varepsilon)=(10^{-1.5}, 10^{-2})$ and $(10^{-1}, 10^{-3})$, respectively. We use $\alpha=10^{-4}$ for the left column and $\alpha=10^{-5}$ for the right column. The black dashed line shows the upper limit of the growth rate given by Equation (\ref{eq:ngrw_wmidplaneBR_woW_woDiff_highk}). In the presence of the diffusion, the maximum growth rates are similar to those obtained in the absence of the backreaction (see the left column), or limited by the black dashed line (see the right column). The black and gray vertical dotted lines represent wavelengths estimated from Equation (\ref{eq:app_kmax}) for the cases with and without the backreaction, respectively. We note that $\beta$ changes by the inclusion of the backreaction because the drift velocity is reduced. The estimated wavenumbers (the black dotted lines) match the most unstable wavenumbers within errors of a few tens percent in the four panels. Therefore, Equation (\ref{eq:app_kmax}) can be used for rough estimation of the most unstable wavenumber in the present analyses with the backreaction.
}
\label{fig:grw_wmidplaneBR_compare}
\end{figure*}
%
%

%---- for double column ---
\begin{figure*}
%\begin{center}
%---- for single column ---
%\begin{figure}
%	\hspace{30pt}
%--------------------------------
	\begin{tabular}{c}
		\begin{minipage}{0.5\hsize}
			\begin{center}
				%\hspace{-20pt}\raisebox{0pt}{
				%\hspace{-30pt}\raisebox{-30pt}{
				\includegraphics[width=1.0\columnwidth]{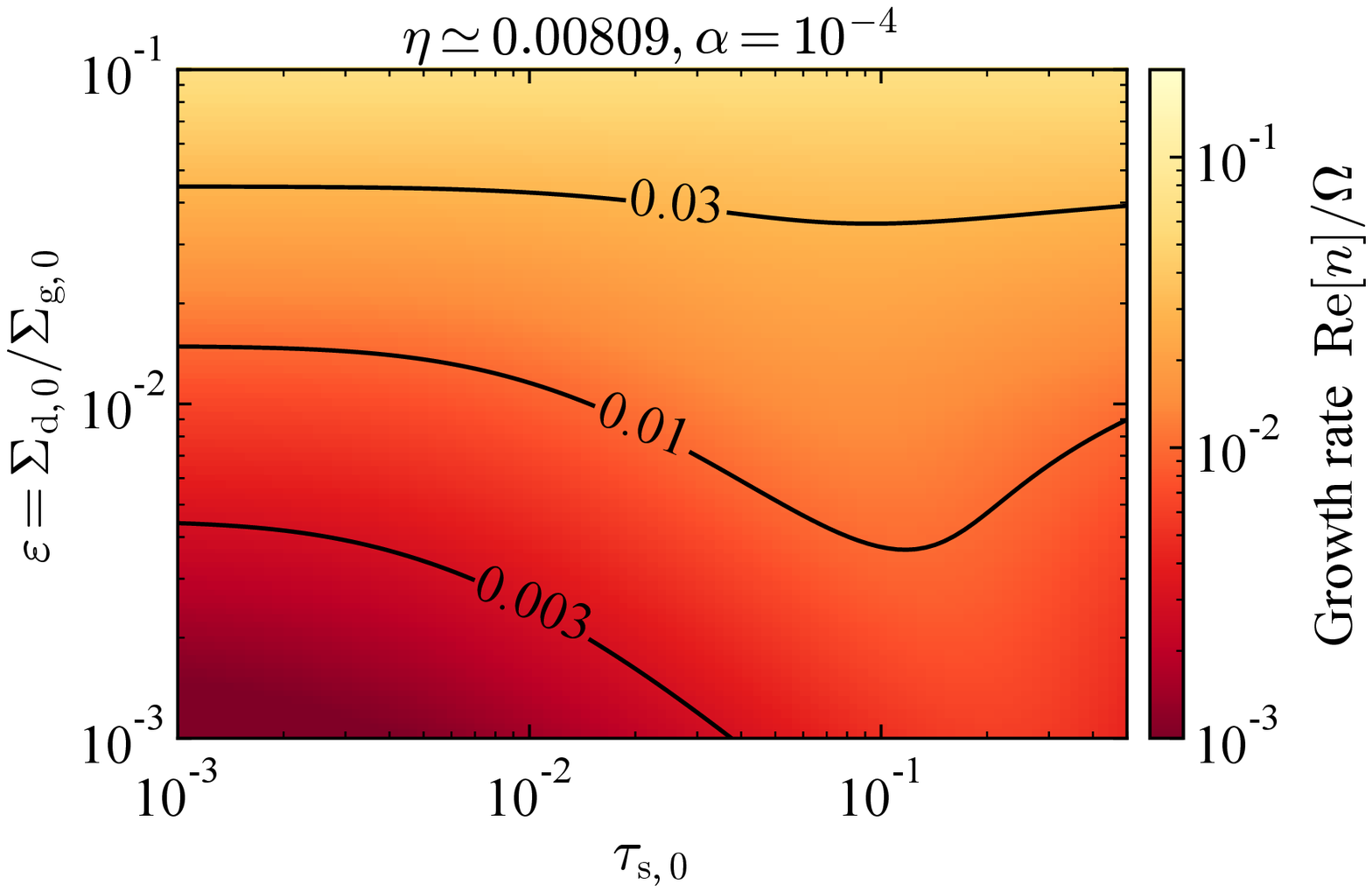}
				%}
			\end{center}
		\end{minipage}
		\begin{minipage}{0.5\hsize}
			\begin{center}
				%\hspace{-20pt}\raisebox{0pt}{
				%\hspace{-30pt}\raisebox{-30pt}{
				\includegraphics[width=1.0\columnwidth]{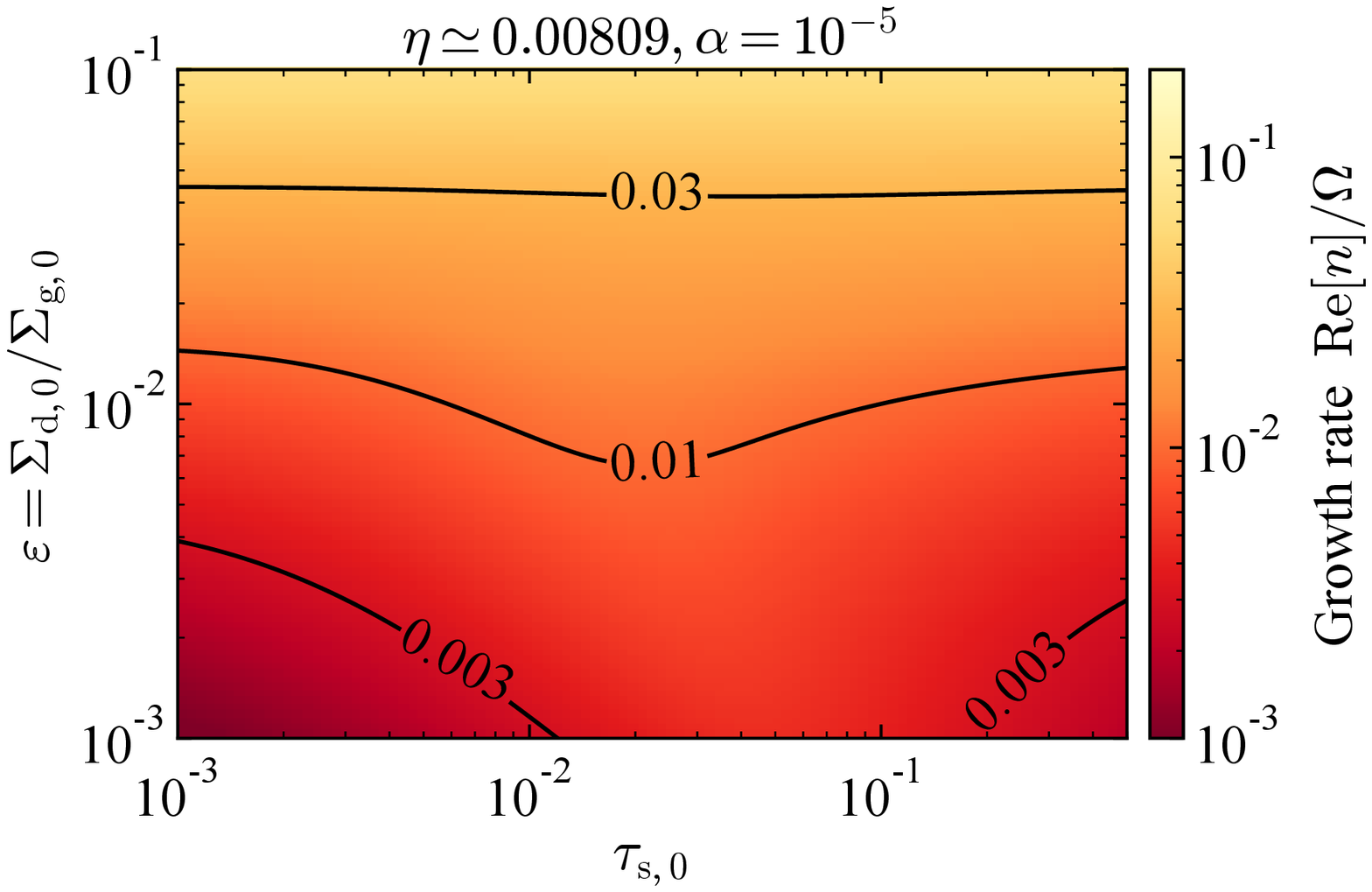}
				%}
			\end{center}
		\end{minipage}
	\end{tabular}
%	}
%\end{center}
%\vspace{15pt}
\caption{Maximum growth rates with the effect of backreaction (Equation (\ref{eq:disp_diff1_wmidplaneBR_woW})) as a function of the normalized stopping time $\tausup$ and the unperturbed dust-to-gas surface density ratio $\varepsilon=\sigmadup/\sigmagup$. The left figure shows the growth rate for $\alpha=10^{-4}$, and the right figure shows the growth rate for $\alpha=10^{-5}$. The reduction of the growth rates are less significant for $\alpha=10^{-4}$ than for $\alpha=10^{-5}$. The reduction is also less significant for smaller $\tausup$ (e.g., $\tausup\lesssim0.1$ on the left panel). } 
\label{fig:grw_w_diff_comp_wmidplaneBR_woW}
\end{figure*}
%===

Figure \ref{fig:grw_wmidplaneBR_compare} shows growth rates for four different parameter sets: $(\tausup,\varepsilon,\alpha)$ $=(10^{-1.5},\;10^{-2},\;10^{-4})$, $(10^{-1.5},\;10^{-2},\;10^{-4})$, $(10^{-1},\;10^{-3},\;10^{-4})$, and $(10^{-1},\;10^{-3},\;10^{-5})$. Each panel shows three dispersion relations whose difference is whether the backreaction and diffusion are included or not. The left column shows the results for $\alpha=10^{-4}$. In the absence of the diffusion, the growth rate is independent from wavenumbers at $kL_{\gdl}>10^2$ for the left top panel and $kL_{\gdl}>10^3$ for the left bottom panel, respectively. In other words, a combined effect of $\taus$-dependent midplane dust-gas ratio and the deceleration due to the backreaction limits the growth of coagulation instability. Equation (\ref{eq:disp_diff1_wmidplaneBR_woW}) with $\beta=0$ is reduced at $\epmidup^2\tilde{k}\gg1$ as follows
\begin{align}
n\left(\frac{3t_0}{\varepsilon}\right)\simeq& i\tilde{k} + \frac{1}{2}\left(1+\frac{(1+\epmidup)^2-\tausup^2}{\epmidup(1+\epmidup)}\right)\notag\\
&+\frac{i}{\tilde{k}}\frac{\left((1+\epmidup)^4-\tausup^4\right)((1+\epmidup)^2-\tausup^2)}{8\epmidup^3(1+\epmidup)^3}.\label{eq:ngrw_wmidplaneBR_woW_woDiff_highk}
\end{align}
The upper limit of the growth rate is given by the second term on the right hand side, which is shown by dashed lines in Figure \ref{fig:grw_wmidplaneBR_compare}.

The growth rates in the presence of the diffusion are shown by plus marks and filled circles in Figure \ref{fig:grw_wmidplaneBR_compare}.  In the left two panels, the diffusion significantly stabilizes coagulation instability, and the growth rate is smaller than the maximum growth rate shown by the black dashed line (Equation (\ref{eq:ngrw_wmidplaneBR_woW_woDiff_highk})) even in the absence of the backreaction. In those cases, including the backreaction insignificantly affects the maximum growth rates. On the other hand, the right two panels show larger growth rates in the absence of the backreaction (see filled circles) because the diffusion is ineffective. In this case, the backreaction limits the growth rates at larger $kL_{\gdl}$, and the maximum growth rates are well reproduced by the black dashed line. 

At larger wavenumbers than the most unstable wavenumber, the growth rate is slightly increased by the backreaction in the presence of the dust diffusion (see the plus marks and filled circles). This trend is more prominent in the lower two panels in Figure \ref{fig:grw_wmidplaneBR_compare}. We find that this increase originates from a cross term of $\beta\tilde{k}^2$ and $2i\tilde{k}\epmidup(1.0+\epmidup)/((1.0+\epmidup)^2+\tausup^2)$ in the square root of the dispersion relation (see Equations (\ref{eq:disp_diff1_wmidplaneBR}) and (\ref{eq:disp_diff1_wmidplaneBR_woW})). The physical interpretation may be as follows. Including the dust diffusion can be regarded as modification in the velocity perturbation $\delta v_x$ by $-ikD\delta\sigmad/\sigmadup\propto\exp(-i\pi/2)\delta\sigmad$ as represented in Equation (\ref{eq:vx_and_diff1}). Including the backreaction also modifies $\delta v_x$ by the term proportional to $ -2\epmidup\left<v_{x,0}\right>\delta\sigmad/\sigmadup$ (see Equation (\ref{eq:linearized_vx_w_midplaneBR})). Because coagulation instability in the absence of diffusion and backreaction has the phase difference of $3\pi/4$ between $\delta\sigmad$ and $\delta v_x$ (see Equations (\ref{eq:simpl_vx}) and (\ref{eq:4pi_ts_sigmad})), the above two modifications act as a phase shift of $\delta v_x$ in the opposite direction and cancel each other.

The black and gray vertical dotted lines in the four panels represent wavenumbers estimated by Equation (\ref{eq:app_kmax}). Note that $\beta$ on the right hand side of Equation (\ref{eq:app_kmax}) changes by the inclusion of the backreaction because the drift velocity is reduced (see Equation (\ref{eq:beta})). The cases with the backreaction show larger $\beta$ and smaller wavenumber given by Equation (\ref{eq:app_kmax}). The black dotted lines match the most unstable wavenumbers within errors of a few tens percent. Therefore, we can use Equation (\ref{eq:app_kmax}) for the rough estimation of $k_{\mathrm{max}}L_{\gdl}$.

As in Figure \ref{fig:grw_w_diff_comp}, we also plot the maximum growth rate as a function of $\tausup$ and $\varepsilon$ in the presence of backreaction. For simplicity and clarity, we first show the growth rates derived without the $\sigmagup'$ and with $W_{\mathrm{mod}}=1$ (Equation (\ref{eq:disp_diff1_wmidplaneBR_woW})). Figure \ref{fig:grw_w_diff_comp_wmidplaneBR_woW} shows the results for $\alpha=10^{-4}$ (on the left panel) and $\alpha=10^{-5}$ (on the right panel). In contrast to the previous sections, we numerically derive the maximum growth rate and the most unstable wavenumber instead of using Equation (\ref{eq:app_kmax}) for comparison with the case of nonzero $\sigmagup'$ (see below) In most of the parameter space, the backreaction makes the growth rates only a few times smaller. The growth rates with $\alpha=10^{-5}$ are smaller than those with $\alpha=10^{-4}$ because weaker turbulence results in denser midplane dust layer and makes the backreaction more effective. In a region of lower dust-to-gas ratio (e.g., $\sigmadup/\sigmagup\sim10^{-3}$), the reduction of the growth rates is less significant than in a region of higher dust-to-gas ratio. The reduction is also less significant for smaller dust sizes, especially for the case of $\alpha=10^{-4}$ (e.g., $\tausup\lesssim0.1$).

%---- for double column ---
\begin{figure*}
%\begin{center}
%---- for single column ---
%\begin{figure}
%	\hspace{30pt}
%--------------------------------
	\begin{tabular}{c}
		\begin{minipage}{0.5\hsize}
			\begin{center}
				%\hspace{-20pt}\raisebox{0pt}{
				%\hspace{-30pt}\raisebox{-30pt}{
				\includegraphics[width=1.0\columnwidth]{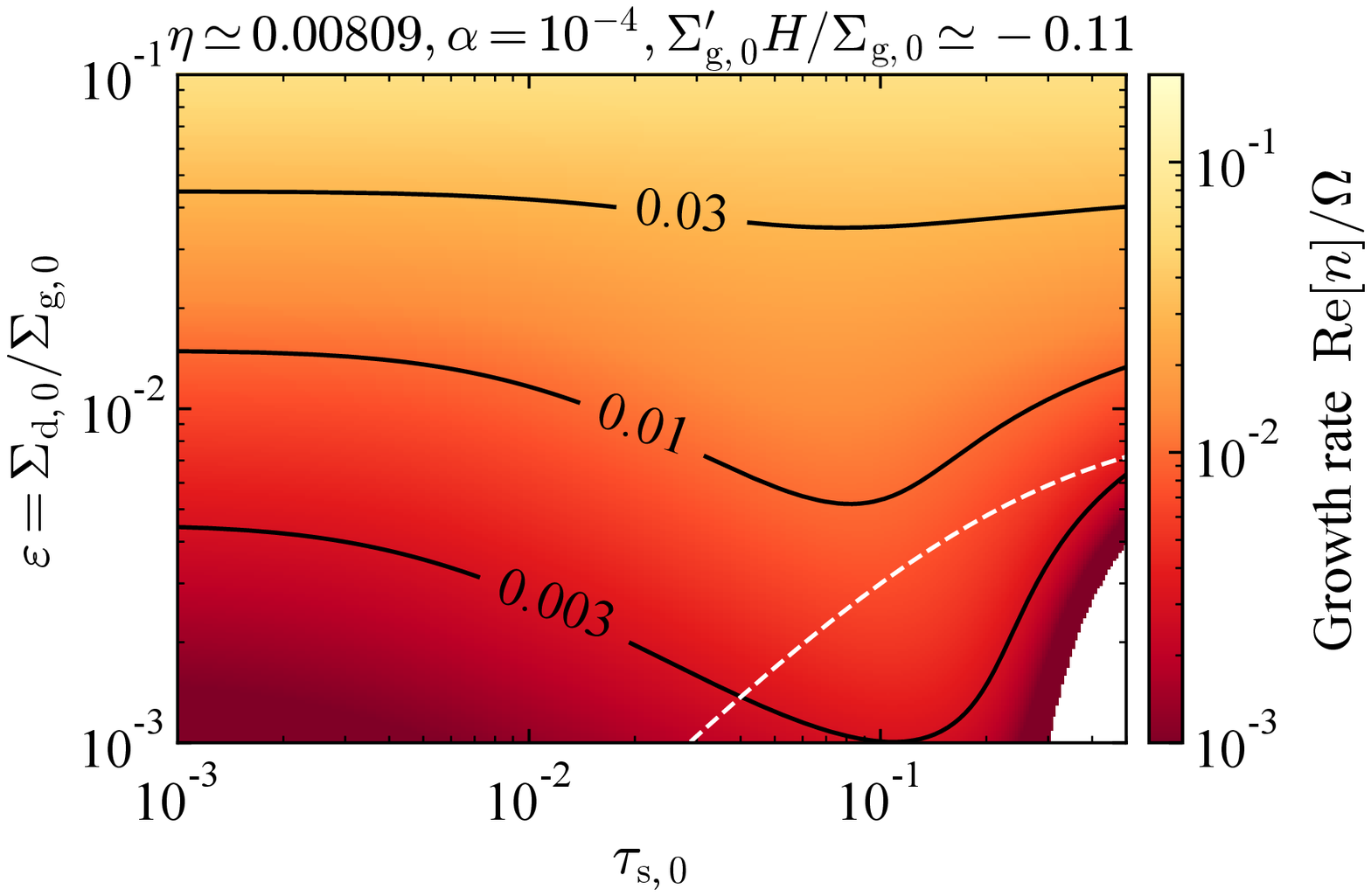}
				%}
			\end{center}
		\end{minipage}
		\begin{minipage}{0.5\hsize}
			\begin{center}
				%\hspace{-20pt}\raisebox{0pt}{
				%\hspace{-30pt}\raisebox{-30pt}{
				\includegraphics[width=1.0\columnwidth]{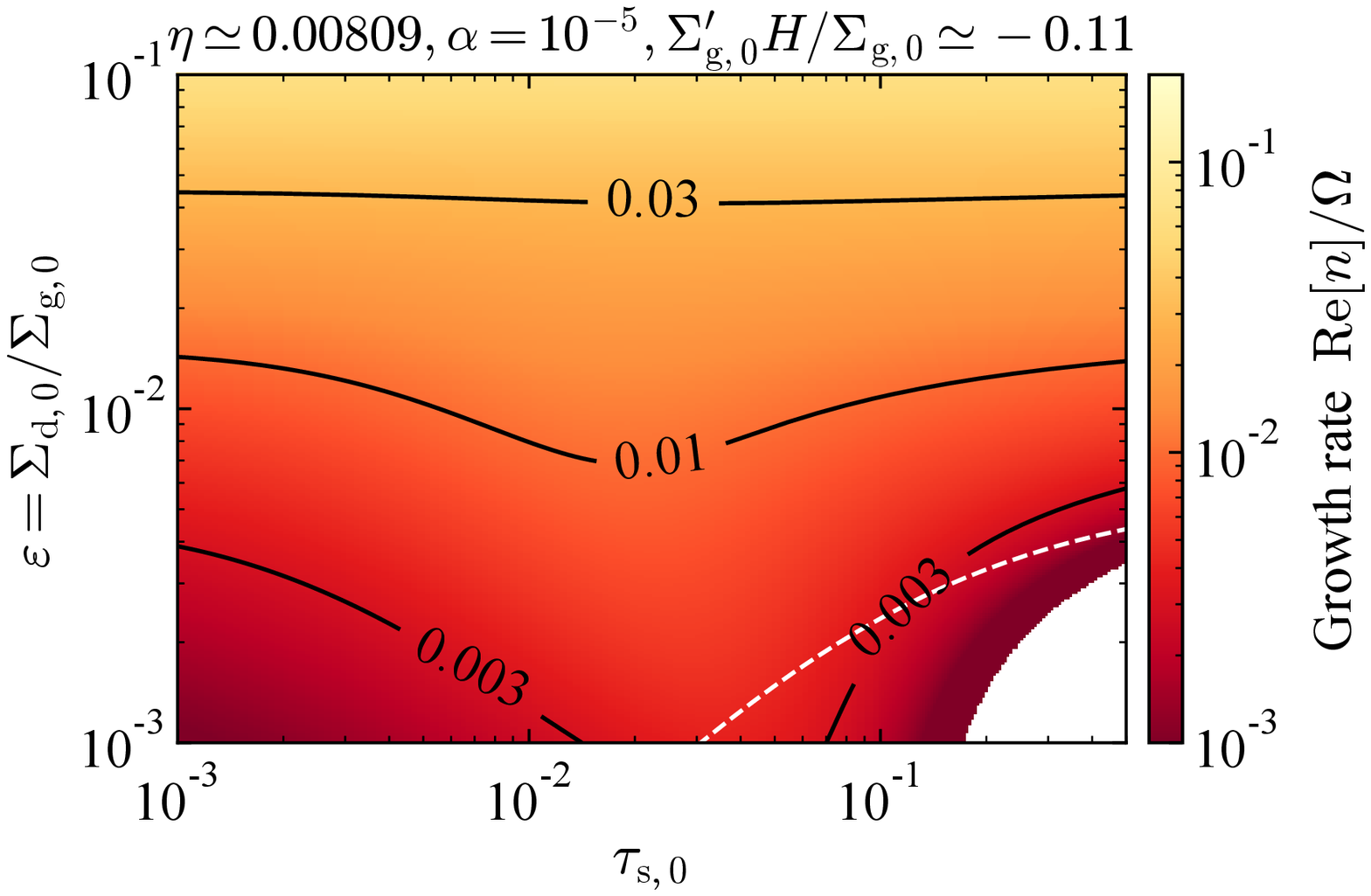}
				%}
			\end{center}
		\end{minipage}
	\end{tabular}
%	}
%\end{center}
%\vspace{15pt}
\caption{Maximum growth rates with all the stabilizing effects (Equation (\ref{eq:disp_diff1_wmidplaneBR})). The left figure shows the growth rate for $\alpha=10^{-4}$, and the right figure shows the growth rate for $\alpha=10^{-5}$. The white dashed lines on both panels show sets of $(\tausup,\;\varepsilon)$ that satisfy $\sigmadup/3t_0=\left<v_{x,0}\right>\sigmagup'$, meaning that the equation for $\taus$ is exactly satisfied in the steady unperturbed state with uniform $\tausup$ and $\sigmadup$, and thus our analysis is rigorous. Comparing with Figures \ref{fig:grw_w_diff_comp} and \ref{fig:grw_w_diff_comp_wmidplaneBR_woW}, we find the growth rate with $\alpha=10^{-4}$ to be similar for $\tausup\lesssim10^{-1}$ and $\varepsilon\lesssim10^{-2}$. This trend can be also found on the right panel (i.e., $\alpha=10^{-5}$). In the white regions in both panels, the coagulation instability is stable and the maximum value of $\mathrm{Re}[n]$ is 0. The area of the white regions are relatively small in the plotted parameter space. Therefore, we expect that the coagulation instability can grow in most cases.}
\label{fig:grw_w_diff_comp_wmidplaneBR_wW}
\end{figure*}
%===

\begin{figure}[htp]%[htp] or [H]
	\begin{center}
		%\hspace{-20pt}\raisebox{0pt}{
		%\hspace{100pt}\raisebox{-20pt}{
		\includegraphics[width=1.0\columnwidth]{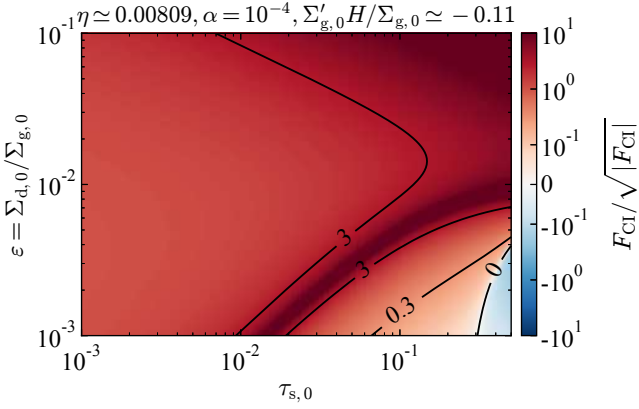}
		%}
	\end{center}
\caption{This figure shows $F_{\mathrm{CI}}$ normalized by the square root of its magnitude as a function of $\tausup$ and $\varepsilon$ for $\alpha=10^{-4}$. The blue region shows a region of $F_{\mathrm{CI}}<0$, which matches with the stable region shown on the left panel of Figure \ref{fig:grw_w_diff_comp_wmidplaneBR_wW}. }
\label{fig:FCI}
\end{figure}

Next, we show the maximum growth rates with all the stabilizing effects including the gas surface density gradient (Equation (\ref{eq:disp_diff1_wmidplaneBR})). We found that the most unstable wavenumber can deviate from the wavelength given by Equation (\ref{eq:app_kmax}) when $W_{\mathrm{mod}}$ is negative and $|W_{\mathrm{mod}}|$ is relatively large. Thus, we numerically derive the most unstable wavelength instead of using the approximated formula. Figure \ref{fig:grw_w_diff_comp_wmidplaneBR_wW} shows the results for $\alpha=10^{-4}$ (on the left panel) and $\alpha=10^{-5}$ (on the right panel). Comparing the derived growth rates with those in Figure \ref{fig:grw_w_diff_comp}, we find that the stabilizing effect due to the backreaction is insignificant for $\tausup\lesssim10^{-1}$ and $\varepsilon\lesssim10^{-2}$, especially in the case of $\alpha=10^{-4}$. One can see the stabilizing effect due to the gas surface density gradient by comparing Figures \ref{fig:grw_w_diff_comp_wmidplaneBR_woW} and \ref{fig:grw_w_diff_comp_wmidplaneBR_wW}. The maximum growth rates become 0 in the white region in Figure \ref{fig:grw_w_diff_comp_wmidplaneBR_wW}. Including the gas gradient stabilizes the coagulation instability the most significantly for larger $\tausup$ and smaller $\varepsilon$. According to the definition of $W_{\mathrm{mod}}$ (Equation (\ref{eq:wmod})), a combination of large $\tausup$ and small $\varepsilon$ leads to negative $W_{\mathrm{mod}}$ with its large magnitude, which stabilizes the instability. We find that the boundary between the colored and white regions in Figure \ref{fig:grw_w_diff_comp_wmidplaneBR_wW} is approximately given by the following function $F_{\mathrm{CI}}$:
\begin{align}
F_{\mathrm{CI}}\equiv &-\epmidup^2(1+\epmidup)^2\notag\\
&+\wmod^{-2}\Bigg((1+\epmidup)^2-\tausup^2+L_{\gdl}\frac{\sigmagup'}{\sigmagup}\notag\\
&\;\;\;\;\;\;\times\frac{\epmidup(1+\epmidup)\left(\epmidup^2+\epmidup+2\tausup^2\right)}{(1+\epmidup)^2+\tausup^2}\Bigg)\label{eq:FCI}
\end{align}
Figure \ref{fig:FCI} shows $F_{\mathrm{CI}}$ normalized by $\sqrt{|F_{\mathrm{CI}}|}$ as a function of $\tausup$ and $\varepsilon$ for $\alpha=10^{-4}$. The value of $F_{\mathrm{CI}}$ is maximized for $\wmod=0$ that can be seen as a color-saturated curve in the bottom half region. One can see that a region of $F_{\mathrm{CI}}<0$ matches with the stable region shown on the left panel of Figure \ref{fig:grw_w_diff_comp_wmidplaneBR_wW}\footnote{One can analytically derive the stability condition ($F_{\mathrm{CI}}<0$) from Equation (\ref{eq:disp_diff1_wmidplaneBR}) by seeking the condition that the growth rate is decreasing function of $\tilde{k}$ at $\tilde{k}\ll1$ with assuming $\beta \tilde{k}^2\ll1$.}. 
From Equation (\ref{eq:FCI}), we can also confirm that large $|\wmod |$ stabilizes the instability because the second term in Equation (\ref{eq:FCI}) decreases and $F_{\mathrm{CI}}$ becomes negative.

As above, we found the stabilization of the coagulation instability by the backreaction and the combination of the backreaction and the gas surface density gradient. We also found that a combination of diffusion and the gas surface density gradient stabilizes the instability for much smaller dust-to-gas ratio ($\varepsilon\lesssim 10^{-4}$). Nevertheless, the stabilized region is limited in the plotted parameter region of Figure \ref{fig:grw_w_diff_comp_wmidplaneBR_wW} ($\varepsilon\geq 10^{-3}$). Besides, the stabilization is insignificant in a dust-depleted region ($\varepsilon\sim10^{-3}$) unless the dust size is not so large (e.g., $\tausup\lesssim0.1$). The previous studies on dust coagulation in protoplanetary disks show that dust-to-gas ratio decreases as dust grows in size. The dust depletion prevents $\tausup$ from becoming large. This indicates that the growth condition of the coagulation instability will be naturally satisfied. Therefore, we expect that coagulation instability can operate even when one considers the deceleration due to high dust-to-gas ratio at the midplane.

\subsection{Effects of enhanced collision velocities}\label{subsec:disp_drift}
Although we consider turbulence-induced collisions in the previous sections, the collision velocity in reality is determined by multiple components:
\begin{equation} 
\Delta \vpp=\sqrt{\left(\Delta v_{\mathrm{t}}\right)^2+\left(\Delta v_{\mathrm{B}}\right)^2+\left(\Delta v_{r}\right)^2+\left(\Delta v_{\phi}\right)^2+\left(\Delta v_{z}\right)^2},
\end{equation}
where $\Delta v_{\mathrm{t}}$ is the turbulence-induced velocity, $\Delta v_{\mathrm{B}}$ is the collision velocity due to Brownian motion, $\Delta v_{r}$, $\Delta v_{\phi}$, and $\Delta v_z$ are the relative velocities due to drift motion in radial, azimuthal, and vertical directions, respectively. The coagulation instability is expected to have larger growth rates than those obtained in the Sections \ref{sec:simpleana} and \ref{sec:linana_woExF} because the coagulation timescale becomes shorter for the collision velocity larger than $\Delta v_{\mathrm{t}}$.

We estimate how much the coagulation instability will be accelerated. For a collision of particles with masses of $m_1$ and $m_2$, $\Delta v_{\mathrm{B}}$ is given by $\sqrt{8k_{\mathrm{B}}T(m_1+m_2)/\pi m_1m_2}$, where $k_{\mathrm{B}}$ is Boltzmann constant. Because collisional growth due to Brownian motion is effective for much small particles that are subject to insignificant drift \citep[][]{Brauer2008}, we can neglect $\Delta v_{\mathrm{B}}$ for the coagulation instability, which requires significant drift. In the absence of backreaction from dust to gas, the radial, azimuthal, vertical drift velocities of one dust particle are as follows \citep[][]{Nakagawa1986}:
\begin{equation}
v_r(\taus)=-\frac{2\taus}{1+\taus^2}\eta R\Omega
\end{equation}
\begin{equation}
v_{\phi}(\taus)=R\Omega-\frac{1}{1+\taus^2}\eta R\Omega,
\end{equation}
\begin{equation}
v_z(\taus)=-\frac{\taus}{1+\taus}z\Omega.
\end{equation}
Note that $v_r(\taus)$ refers to velocity of a dust particle while $v_x$ given by Equation (\ref{eq:vxNSH}) refers to ensemble-averaged velocity derived from the vertically integrated equations. Because the moment equation tested by \citet{Sato2016} (Equation (\ref{eq:dmpdt})) assumes collisions of particles whose size ratio is 0.5, the relative velocities due to drift motion are
\begin{equation}
|\Delta v_r|=\left|\frac{2-\taus^2}{4+\taus^2}v_r(\taus)\right|,
\end{equation}
\begin{equation}
|\Delta v_{\phi}|=\frac{3\taus^2}{(4+\taus^2)(1+\taus^2)}\eta R\Omega%v_{\phi}(\taus),
\end{equation}
\begin{equation}
|\Delta v_z|=\frac{1}{2+\taus}|v_z(\taus)|.
\end{equation}

Assuming $\Delta v_{\mathrm{t}}=\sqrt{C\alpha\taus}\cs$, one finds $\Delta v_r/\Delta v_{\mathrm{t}}\propto\Delta v_{z}/\Delta v_{\mathrm{t}}\propto\sqrt{\taus/\alpha}$ and $\Delta v_{\phi}/\Delta v_{\mathrm{t}}\propto\taus\sqrt{\taus/\alpha}$ for a leading-order term. Because of such a $\taus$-dependence, $\Delta v_r$ and $\Delta v_z$ dominate over $\Delta v_{\phi}$ for $\taus< 1$, and we may approximate the collision velocity $\Delta \vpp$ with the following equation:
\begin{equation}
\Delta\vpp\simeq\Delta v_{\mathrm{t}}\sqrt{1+f\frac{\taus}{\alpha}},\label{eq:rough_vpp}
\end{equation}
where $f$ is defined as
\begin{align}
f&\equiv\frac{\alpha}{\taus}\left(\left(\frac{\Delta v_r}{\Delta v_{\mathrm{t}}}\right)^2+\left(\frac{\Delta v_z}{\Delta v_{\mathrm{t}}}\right)^2\right),\notag\\
&=\frac{4}{C}\left(\frac{2-\taus^2}{(4+\taus^2)(1+\taus^2)}\right)^2\left(\frac{\eta R\Omega}{\cs}\right)^2\notag\\
&\;\;\;\;\;\;\;\;\;\;\;\;\;\;\;\;\;\;+\frac{1}{C(2+\taus)^2(1+\taus)^2}\left(\frac{z}{H}\right)^2,\\
&\simeq\frac{1}{C}\left[\left(\frac{\eta R\Omega}{\cs}\right)^2+\left(\frac{z}{2H}\right)^2\right].
\end{align}
In the last equality, we make use of $\taus\ll1$. 

Substituting Equation (\ref{eq:rough_vpp}) into Equation (\ref{eq:dstdt1f_formal}), we obtain
\begin{equation}
\ddt{\taus}+v_x\ddx{\taus}=\frac{\sqrt{\pi}}{4}\left(\frac{\Delta v_{\mathrm{t}}}{\taus\hd}\right)\sqrt{1+f\frac{\taus}{\alpha}}\frac{\sigmad}{\sigmagup}\taus.\label{eq:dstdt1f_allvpp_pre}
\end{equation}
When turbulent motion determines the dust scale height and we can approximate $\hd\simeq\sqrt{\alpha/\taus}$, Equation (\ref{eq:dstdt1f_allvpp_pre}) is reduced to
\begin{equation}
\ddt{\taus}+v_x\ddx{\taus}=\sqrt{1+f\frac{\taus}{\alpha}}\frac{\sigmad}{\sigmagup}\frac{\taus}{3t_0}.\label{eq:dstdt1f_allvpp}
\end{equation}
In this case, we can roughly expect that growth rates of the coagulation instability increase by $\sqrt{1+f\tausup/\alpha}$. We should note that the moment equation is derived from the vertically integrated coagulation equation \citep[see][]{Sato2016}. Thus, the enhancement factor $\sqrt{1+f\tausup/\alpha}$ should be vertically averaged because $\Delta v_z$ depends on $z$. Assuming the Gaussian profile for the vertical dust distribution, one finds the vertical collision rate can be expressed in terms of $\Delta v_{z}(z\sim\hd)$.\footnote{Using the Gaussian function for the vertical dust density, one can calculate the vertically averaged collision velocity due to the dust settling $\overline{\Delta v_{z}}$. Assuming that $\taus$ and $\Omega$ is almost constant within $|z|<H_{\dst}$, we obtain the approximated velocity $\overline{\Delta v_{z}}=\Delta v_{z}(z=\sqrt{2/\pi}H_{\dst,12})$ for a collision of dust grains with $\taus=\tau_{\mathrm{s},1}$ and $\tau_{\mathrm{s},2}$, where $H_{\dst,12}\equiv\left(\hd(\tau_{\mathrm{s},1})^{-2}+\hd(\tau_{\mathrm{s},2})^{-2}\right)^{-1/2}$ \citep[see][]{Taki2021}.}  Thus, we just take $z=\hd\simeq\sqrt{\alpha/\tausup}H$ to evaluate $\Delta v_z$ rather than exactly integrating and taking the vertical average of the enhancement factor. We then obtain $\sqrt{1+f\tausup/\alpha}\simeq2.5$ for $\tausup=0.1$, $\alpha=10^{-4}$, $C=2.3$, and $\eta R\Omega=0.11\cs$ that corresponds to a value at $R=20\;\mathrm{au}$ in the minimum-mass solar nebula. 

When gas turbulence is sufficiently weak, the dust scale height is determined by the settling motion. If the vertical relative velocity dominates over the other components, i.e., $\Delta\vpp\simeq\Delta v_z$, we expect $\Delta\vpp/\taus\hd\sim\Omega$, where we substitute $z=\hd$ to calculate $\Delta v_z$ \citep[see][]{Nakagawa1981}. In this case, Equation (\ref{eq:dstdt1f_allvpp_pre}) is reduced to
\begin{align}
\ddt{\taus}+v_x\ddx{\taus}&\sim\frac{\sqrt{\pi}}{4}\frac{\sigmad}{\sigmagup}\Omega\taus=\frac{\sigmad}{\sigmagup}\frac{\taus}{3\sqrt{C}t_0}.
\end{align}
The growth rate of coagulation instability in this case is comparable to that expected in Section \ref{sec:simpleana} because of $\sqrt{C}\simeq1.5$. Including the radial relative velocity in $\Delta \vpp$ gives
\begin{align}
\ddt{\taus}+v_x\ddx{\taus}&\sim\sqrt{1+\left(\frac{\Delta v_r}{\Delta v_z}\right)^2}\frac{\sigmad}{\sigmagup}\frac{\taus}{3\sqrt{C}t_0}\\
&\sim\sqrt{1+\left(\frac{2\eta R}{\hd}\right)^2}\frac{\sigmad}{\sigmagup}\frac{\taus}{3\sqrt{C}t_0}.
\end{align}
Thus, we may expect larger growth rates by the factor of $\sim\sqrt{1+4\eta^2R^2/\hd^2}$.

The present analyses describe the disk evolution with the height-integrated equations. We thus take into account $\Delta v_z$ substituting the representative height ($z\sim\hd$). To consider the vertical structure and dust motion might be important for more precise discussion on coagulation instability because of the $z$-dependence of the collision velocity. Our future work will address this issue.

\subsection{Growth versus drift}\label{subsec:vs_drift}

Perturbations growing via the coagulation instability travel inward with phase velocity $-\mathrm{Im}[n]/k$ (see the left panels of Figure \ref{fig:grw_r20_st01_dgr1e-3_woExF} and \ref{fig:grw_r20_woExF_compare}). The coagulation instability plays a significant role in disk evolution when the growth timescale of the instability is shorter than the timescale on which perturbations reach the central star. In this subsection, we examine whether the coagulation instability can grow significantly in a disk or not.

\begin{figure}[htp]%[htp] or [H]
	\begin{center}
		%\hspace{-20pt}\raisebox{0pt}{
		%\hspace{100pt}\raisebox{-20pt}{
		\includegraphics[width=1.0\columnwidth]{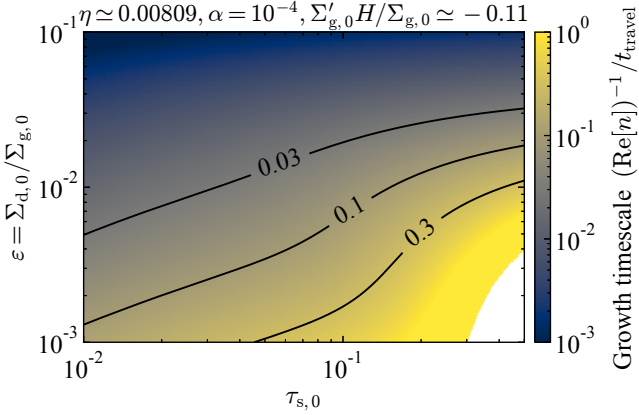}
		%}
	\end{center}
\caption{Growth timescale of the coagulation instability (Equation (\ref{eq:disp_diff1_wmidplaneBR})) normalized by the traveling timescale of perturbations $t_{\mathrm{travel}}$. We numerically find the shortest growth timescale as in Figure \ref{fig:grw_w_diff_comp_wmidplaneBR_wW}. The horizontal and vertical axes are the normalized stopping time $\tausup$ and the unperturbed dust-to-gas surface density ratio $\varepsilon=\sigmadup/\sigmagup$, respectively. The strength of turbulence is assumed to be $\alpha=10^{-4}$. The growth timescale is basically shorter than the traveling time except for the white region where the growth rate becomes 0. Thus, we expect that the coagulation instability develops before perturbations reach the central star.}
\label{fig:grw_r20_alpha1e-4}
\end{figure}

Although the phase velocity slightly deviates from dust drift velocity $v_{x,0}$ depending on wavelengths, we use $v_{x,0}$ to evaluate the traveling time of perturbations:
\begin{equation}
t_{\mathrm{travel}}\equiv\frac{R}{|v_{x,0}|}=\frac{(1+\epmid)^2+\taus^2}{2\tausup\eta\Omega},
\end{equation}
where we use the drift velocity mediated by the midplane dust-to-gas ratio. From this equation, we obtain $t_{\mathrm{travel}}(2\pi/\Omega)^{-1}\simeq100$ for $\tausup=0.1$, $\epmid\ll 1$, and $\eta=0.008$. Thus, perturbations reach a central star within one hundred orbital periods. Figure \ref{fig:grw_r20_alpha1e-4} shows growth timescale of the instability normalized by $t_{\mathrm{travel}}$ as a function of dust-gas ratio $\varepsilon=\sigmadup/\sigmagup$ and normalized stopping time $\tausup$. We use Equation (\ref{eq:disp_diff1_wmidplaneBR}) to calculate the growth rate in Figure \ref{fig:grw_r20_alpha1e-4}. The growth timescale is shorter than the traveling timescale in most of the plotted region. Although the growth timescale relative to the traveling timescale decreases as dust grains grow ($\tausup\sim0.1$), the growth timescale can be still 3 times shorter than the traveling timescale.

Because the growth rate just represents $e$-folding time of perturbation amplitudes, it generally takes multiple times of the growth timescale for perturbations to grow significantly and develop into nonlinear stage. The required time for perturbations to grow into nonlinear stage $t_{\mathrm{nonlin}}$ is represented in terms of initial amplitudes $\delta\Sigma_{\dst,\mathrm{ini}}$ and amplitudes at nonlinear stage $\delta\sigmad(t_{\mathrm{nonlin}})$ as follows
\begin{equation}
t_{\mathrm{nonlin}}=\frac{\ln(\delta\sigmad(t_{\mathrm{nonlin}})/\delta\Sigma_{\dst,\mathrm{ini}})}{\mathrm{Re}[n]}.
\end{equation}
For significant growth of the coagulation instability, the timescale $t_{\mathrm{nonlin}}$ should be shorter than the traveling timescale $t_{\mathrm{travel}}$. We can expect nonlinearity for $\delta\sigmad(t_{\mathrm{nonlin}})\sim\sigmadup$. Therefore, $t_{\mathrm{nonlin}}$ is roughly given by
\begin{align}
t_{\mathrm{nonlin}}&\sim\frac{\ln(\sigmadup/\delta\Sigma_{\dst,\mathrm{ini}})}{\mathrm{Re}[n]},\notag\\
&\simeq 2.3\mathrm{Re}[n]^{-1}\left(\frac{\ln(\sigmadup/\delta\Sigma_{\dst,\mathrm{ini}})}{\ln(10)}\right).
\end{align}
When the initial amplitudes are 10 \% of $\sigmadup$, $t_{\mathrm{nonlin}}$ becomes 2.3 times larger than the growth timescale, which satisfies the condition $t_{\mathrm{nonlin}}<t_{\mathrm{travel}}$ in most of the parameter space shown in Figure \ref{fig:grw_r20_alpha1e-4}. Initial amplitudes of 1 \% gives $t_{\mathrm{nonlin}}\sim4.6\mathrm{Re}[n]^{-1}$ and still satisfies the condition in most parameters ($\tausup,\varepsilon$) shown in Figure \ref{fig:grw_r20_alpha1e-4}.

\subsection{Effects of imperfect sticking and fragmentation}\label{subsec:fragmentation}
In the previous sections, we assume perfect sticking through particle collisions. Dust particles however experience imperfect sticking or fragment when collision speed becomes close to or larger than a critical speed, respectively \citep[e.g.,][]{Wada2009,Wada2013}. Erosive collisions also result in a decrease in the peak-mass \citep[][]{Krijt2015}.

We test if coagulation instability operates in the presence of imperfect sticking and collisional fragmentation. We modify Equation (\ref{eq:dstdt1f}) introducing a sticking efficiency $p_{\mathrm{eff}}$ as in previous studies \citep[e.g.,][]{OH2012,Okuzumi2016,Ueda2019}:
\begin{equation}
\ddt{\taus}+v_x\ddx{\taus}=p_{\mathrm{eff}}\frac{\sigmad}{\sigmagup}\frac{\taus}{3t_0},\label{eq:dstdt1f_modi}
\end{equation}
They prescribe the sticking efficiency $p_{\mathrm{eff}}$\footnote{The previous studies use $\epsilon_{\mathrm{eff}}$ to denote the sticking efficiency \citep{Okuzumi2016,Ueda2019}. Because we already use similar variables, $\varepsilon$ and $\epmid$, to denote the dust-gas ratios, we use $p_{\mathrm{eff}}$ to avoid confusion.} by fitting data of collision simulations of \citet{Wada2009} (e.g., see Figure 11 therein):
\begin{equation}
p_{\mathrm{eff}}=\mathrm{min}\left(1,-\frac{\ln(\Delta\vpp/v_{\mathrm{frag}})}{\ln5}\right),
\end{equation}
where $v_{\mathrm{frag}}$ denotes a critical fragmentation velocity \citep[e.g., see Equation (12) in][]{Ueda2019}. We refer to imperfect sticking as a process of $0\leq p_{\mathrm{eff}}<1$ and fragmentation as a process of $p_{\mathrm{eff}}<0$. In this study, we just formally show how the growth rates change because of $p_{\mathrm{eff}}$. 

We obtain dispersion relation by replacing $t_0$ with $t_0/p_{\mathrm{eff}}$ in Equation (\ref{eq:1fdispersion})
\begin{align}
n_{\app,\pm}\equiv&-ikv_{x,0}+\notag\\
&p_{\mathrm{eff}}\frac{\varepsilon}{6t_0}\left(1\pm\sqrt{1-p_{\mathrm{eff}}^{-1}\frac{12t_0}{\varepsilon}\frac{1-\tausup^2}{1+\tausup^2}ikv_{x,0}}\right).\label{eq:1fdispersion_modi}
\end{align}
When dust particles are subject to imperfect sticking ($0<p_{\mathrm{eff}}<1$),  growth rate of coagulation instability decreases and is proportional to $\sqrt{p_{\mathrm{eff}}}$ at short wavelengths.

In the case of fragmentation ($p_{\mathrm{eff}}<0$), we find that a mode with complex growth rate $n_{\app,-}$ becomes unstable while the coagulation instability is stable (i.e., $\mathrm{Re}[n_{\app,+}]<0$). The growth rate $\mathrm{Re}[n_{\app,-}]$ at short wavelengths is
\begin{equation}
\mathrm{Re}[n_{\app,-}]\simeq|p_{\mathrm{eff}}|\frac{\varepsilon}{6t_0}\sqrt{|p_{\mathrm{eff}}|^{-1}\frac{6t_0}{\varepsilon}\frac{1-\tausup^2}{1+\tausup^2}k|v_{x,0}|}.
\end{equation}
At short wavelengths, relative amplitude of $\delta\taus$ and $\delta\sigmad$ is approximately given as
\begin{equation}
\frac{\delta\taus/\tausup}{\delta\sigmad/\sigmadup}\simeq-\exp\left(i\frac{\pi}{4}\right)\sqrt{\frac{|p_{\mathrm{eff}}|\varepsilon}{3t_0 k|v_{x,0}|}\frac{1+\tausup^2}{1-\tausup^2}}. \label{eq:dstdsig_frag}
\end{equation}
Although Equation (\ref{eq:dstdsig_frag}) shows negative correlation except for $\exp(i\pi/4)$ in contrast to the coagulation instability, its physical mechanism is similar to coagulation instability presented in Section \ref{sec:simpleana}. Dust particles are subject to more significant fragmentation in regions of $\delta\sigmad>0$, leading to radial variation of $\delta \taus$ and $\delta v_x$. Radial gradient of $\delta v_x$ leads to radial concentration and promotes further fragmentation in the concentrated regions. This positive feedback results in the instability. 

The single size approximation tends to be invalid in the presence of fragmentation and erosion. For comprehensive studies, we should explicitly discuss time evolution of size distribution during the development of the instabilities with fragmentation and erosion, which will be the scope of our future studies.

\subsection{Coevolution with other dust-gas instabilities}\label{subsec:SI_SGI}
Because the coagulation instability is triggered by dust coagulation, the instability is entirely different from any other dust-gas instabilities including streaming instability \citep[e.g.,][]{Youdin2005,Youdin2007,Jacquet2011}, resonant drag instability \citep[][]{Squire2018a,Squire2018b}, secular GI \citep[e.g.,][]{Youdin2005a,Youdin2011,Takahashi2014,Tominaga2019}, and TVGI \citep[][]{Tominaga2019}. This is confirmed by taking limit of $t_0\to\infty$ with Equation (\ref{eq:1fdispersion}), which leads to $\mathrm{Re}[n_{\app,+}]\to0$. 

Previous studies on dust collisional growth showed dust depletion when dust grains are spherical and grow into larger sizes ($\taus\sim0.1$) \citep[e.g.,][]{Brauer2008,Okuzumi2012}. The coagulation instability can grow even when dust-to-gas surface density ratio is $10^{-3}$ while streaming instability, secular GI, and TVGI insufficiently grow in such a dust-poor region. Development of the coagulation instability leads to dust concentration at small scale $\sim k^{-1}$. Nonlinear development will results in significant increase in dust surface density by an order of magnitude. If such a nonlinear development is achieved, coagulation timescale becomes short and collisional growth toward planetesimals might be expected.

If nonlinear development of the coagulation instability increases dust-to-gas ratio from $10^{-3}$ to 0.02 or even higher, streaming instability subsequently operates in the resultant dust-rich region. Recent studies show that streaming instability can be substantially stabilized if dust particles have a power-law size distribution \citep[][]{Krapp2019,Paardekooper2020,Paardekooper2021,Zhu2021,McNally2021}. \cite{McNally2021} shows that adding a bump at the high mass end of the power-law size distribution makes the stabilization less efficient. Therefore, it is important to study how coagulation instability affects dust size distributions, which will be explored in future work.

If the above dust enrichment and size growth occur in radially wide region ($\sim H$), we can also expect growth of secular GI and TVGI. Those secular instabilities lead to further dust enrichment at a scale $\lesssim H$, and eventually result in gravitational fragmentation of enriched regions and planetesimal formation \citep[][]{Tominaga2018,Tominaga2019,Tominaga2020,Pierens2021}. Resultant planetesimals can be the origin of debris disks and Kuiper Belt objects. We will also investigate planetesimal formation via a combination of coagulation instability, secular GI, and TVGI in future work.

\section{Conclusions}\label{sec:conclusion}
In this work, we show that collisional growth of drifting dust leads to a new instability operating in a protoplanetary disk. We refer to the new instability as coagulation instability. Coagulation instability grows at short wavelengths ($< H$) in the absence of dust diffusion, and its growth timescale is much shorter than coagulation timescale determined by dust-to-gas mass ratio (Figure \ref{fig:grw_r20_cont_map}). Even in the presence of dust diffusion, coagulation instability can grow at a wavelength of $\sim H$ within a few tens of the orbital periods (Figure \ref{fig:grw_w_diff_comp}). Although the previous studies showed that dust surface density decreases as dust grows, coagulation instability can grow even in a dust-depleted region ($\varepsilon\sim10^{-3}$) and locally increases dust abundance. Therefore, coagulation instability provide preferable sites for other dust-gas instabilities that require high dust abundance of large dust particles (e.g., $\taus\sim0.1$). In other words, coagulation instability plays an important role for connecting collisional dust growth and hydrodynamical clumping at the early evolutionary stage of protoplanetary disks.

In the present study, we consider time evolution of only the single moment $\mpar$ although we roughly consider the size dispersion by following the formalism proposed by \cite{Sato2016}. Considering higher moments and/or evolution of dust size distribution is significantly important to discuss whether coagulation instability is viable in protoplanetary disk or not. Those effects will be important at nonlinear growth phase before the onset of other dust-gas instabilities (e.g., secular GI). Besides, the present analyses are based on the vertically averaged equations and thus neglect vertical shear and stratification. To investigate coagulation instability with vertical direction will be important for discussing its viability in realistic disks (cf. Appendix \ref{app:sublayer}). Our future work addresses these issues.

\acknowledgments
We thank the anonymous referee for the detailed review, which helped to improve the manuscript. We also thank Hidekazu Tanaka, Takeru K. Suzuki and Sanemichi Z. Takahashi for fruitful discussions. This work was supported by JSPS KAKENHI Grant Nos. JP18J20360 (R.T.T.), 16H02160, 18H05436, 18H05437 (S.I.), 17H01103, 17K05632, 17H01105, 18H05438, 18H05436 and 20H04612 (H.K.). R.T.T. is also supported by RIKEN Special Postdoctoral Researchers Program.


\software{Mathematica \citep{Wolfram}, Matplotlib \citep{Hunter2007}, SciPy \citep{Virtanen2020}, NumPy \citep{Harris2020}}


%\newpage

\clearpage
\appendix

\section{Derivation of the moment equations}\label{app:derivation_momenteq}
The moment equation for dust coagulation we adopted was derived in \citet{Sato2016} from the vertically integrated Smoluchowski equation \citep[see also][]{Taki2021}. In this appendix, we review the derivation and mention its validity for the analysis on coagulation instability. We refer readers to \cite{Sato2016} for numerical validation based on comparison with full-size simulations.

In the case of perfect sticking, time evolution of a column number density $N(r,m)$ per unit dust particle mass $m$  is governed by the Smoluchowski equation (\citealp{Smoluchowski1916}; \citealp{Schumann1940}; \citealp{Safronov1972}):
\begin{align}
\ddt{mN}&=\frac{m}{2}\int_{0}^{m}dm'K(r,m',m-m')N(r,m')N(r,m-m')\notag\\
&-mN(r,m)\int_0^{\infty}dm'K(r,m,m')N(r,m')-\frac{1}{r}\frac{\partial}{\partial r}\left(rv_r(r,m)mN(r,m)\right),\label{eq:smoluchowski}
\end{align}
where $K(r,m_1,m_2)$ represents collision rates of dust particles with masses of $m_1=4\pi\rhoint a_1^3/3$ and $m_2=4\pi\rhoint a_2^3/3$. The expression of the collision kernel is given as follows \citep[e.g.,][]{Brauer2008,Okuzumi2012}
\begin{equation}
K(r,m_1,m_2)\equiv \frac{\sigma_{\mathrm{coll}}}{2\pi H_{\dst}(m_1)H_{\dst}(m_2)}\int_{-\infty}^{\infty}\Delta \vpp\exp\left[-\frac{z^2}{2}\left(\frac{1}{H_{\dst}(m_1)^2}+\frac{1}{H_{\dst}(m_2)^2}\right)\right]dz,
\end{equation}
where $\Delta\vpp$ is collision velocity and $\sigma_{\mathrm{coll}}$ is a collisional cross section:
\begin{equation}
\sigma_{\mathrm{coll}}\equiv \pi(a_1+a_2)^2.
\end{equation}
We note that the radial velocity $v_r$ and the dust scale height $H_{\dst}$ depend on dust particle mass. 
Describing coagulation in terms of a column density is valid when vertical mixing of dust particles is faster than coagulation. The timescale of the vertical mixing is of the order of $\hd^2/D\simeq\Omega^{-1}/\taus$ \citep[see also][]{YL2007}. As mentioned in Section \ref{sec:simpleana}, the coagulation timescale is $3t_0/\varepsilon\simeq1.5\Omega^{-1}/\varepsilon$. The mixing is thus faster than coagulation when dust sizes are large enough ($\taus\sim0.1$) and/or dust-to-gas ratio is decreased ($\varepsilon\sim10^{-3}$). Such a situation is one preferable for coagulation instability. According to Section \ref{sec:discussion}, the growth rate of the instability in the presence of diffusion is $\sim1\times10^{-3}\Omega-3\times10^{-3}\Omega$ for $\tausup\simeq0.1$ and $\varepsilon\simeq10^{-3}$ (e.g., see Figure \ref{fig:grw_w_diff_comp}). Thus, the growth timescale is $\sim10^2\Omega^{-1}-10^3\Omega^{-1}$, which is longer than the mixing timescale. This validates our methodology.

Following \cite{Sato2016}, we define the $i$th moment $M_i(r)$ as follows:
\begin{equation}
M_i(r)\equiv\int_0^{\infty}m^{i+1}N(r,m)dm. \label{eq:ith-moment}
\end{equation}
Using Equations (\ref{eq:smoluchowski}) and (\ref{eq:ith-moment}), one obtains the $i$th moment equation \citep[see][]{Estrada2008,Ormel2008}:
\begin{align}
\ddt{M_i}&=\frac{1}{2}\int_0^{\infty}dm\int_0^{\infty}dm'K(r,m,m')N(m)N(m')\left[(m+m')^{i+1}-(m^{i+1}+m'^{i+1})\right]-\frac{1}{r}\frac{\partial}{\partial r}\left(r\left<m^iv_r\right>_m\sigmad\right),
\end{align}
where $\left<m^iv_r\right>_m$ is defined as follows
\begin{equation}
\left<m^iv_r\right>_m\equiv\frac{1}{\sigmad}\int_0^{\infty}m^{i+1}v_r(m)N(r,m)dm.
\end{equation}
One obtains the continuity equation from the 0th moment equation
\begin{equation}
\ddt{\sigmad}+\frac{1}{r}\frac{\partial}{\partial r}\left(r\left<v_r\right>_m\sigmad\right)=0.\label{eq:0thmomenteq}
\end{equation}
The 1st moment equation yields 
\begin{equation}
\ddt{\mpar\sigmad}=\int_0^{\infty}dm\int_0^{\infty}dm'mm'K(r,m,m')N(m)N(m')-\frac{1}{r}\frac{\partial}{\partial r}\left(r\left<mv_r\right>_m\sigmad\right),\label{eq:1stmomenteq}
\end{equation}
where $\mpar\equiv M_1/\sigmad$ is the weighted average mass, called the peak mass \citep{Ormel2008}.

One needs closure relations to solve 0th and 1st moment equations. Following \cite{Sato2016} and \citet{Taki2021}, we assume the following closure relation to approximate $\left<mv_r\right>_m$ in terms of $v_r(\mpar)$:
\begin{equation}
\left<mv_r\right>_m\simeq\mpar\left<v_r\right>_m\simeq\mpar v_r(\mpar).\label{eq:closure}
\end{equation}
We note that Equation (\ref{eq:closure}) is more general than the single-size approximation where one explicitly assumes the delta function for dust size distribution. In addition, we further approximate the integral on the right hand side of Equation (\ref{eq:1stmomenteq}). \cite{Sato2016} found that the moment equations reproduce full-size simulations when one approximates the integral by the equal-sized kernel 
\begin{align}
\int_0^{\infty}dm\int_0^{\infty}dm'mm' & K(r,m,m')N(m)N(m')\simeq K(\mpar,\mpar)=\frac{2a^2}{H_{\dst}(\mpar)}\int_{-\infty}^{\infty}\Delta \vpp\exp\left(-\frac{z^2}{H_{\dst}(\mpar)^2}\right)dz,
\end{align}
where we use $\mpar=4\pi\rhoint a^3/3$. \citet{Sato2016} further approximated the vertical integration taking $\Delta \vpp$ outside the integral and obtained
\begin{equation}
K(\mpar,\mpar)=\frac{2\sqrt{\pi}a^2\Delta\vpp}{H_{\dst}(\mpar)}.\label{eq:closure:Kmpmp}
\end{equation}
They showed that using a collision speed $\Delta\vpp$ of dust particles of size ratio 0.5 show better agreement between full-size simulations and simulations that consider evolution of only $\sigmad$ and $\mpar$ (see Appendix A therein). It should be noted that the choice of the size ratio for $\Delta \vpp$ is not so important for $\tstop\Omega\sim0.1-1$, with which the present analysis is concerned. For example, the turbulence-driven collision velocity that we consider in the main analyses is $\simeq\sqrt{2.3\alpha\tstop\Omega}\cs$ for the size ratio of 0.5 and $\simeq\sqrt{1.9\alpha\tstop\Omega}\cs$ for the size ratio of 1 \citep[][]{Ormel2007}. The mass distribution is characterized by $\mpar$ until the onset of runaway growth \citep[][]{Kobayashi2016}.
As the first analyses of coagulation instability, we follow \cite{Sato2016} and only consider the evolution of $\sigmad$ and $\mpar$ with Equation (\ref{eq:closure:Kmpmp}). Equations (\ref{eq:0thmomenteq}), (\ref{eq:1stmomenteq}), (\ref{eq:closure}), and (\ref{eq:closure:Kmpmp}) give the following equations:
\begin{equation}
\ddt{\sigmad}+\frac{1}{r}\frac{\partial}{\partial r}\left(rv_r\sigmad\right)=0,
\end{equation}
\begin{equation}
\Lddt{\mpar}=\ddt{\mpar}+v_r\frac{\partial\mpar}{\partial r}=\frac{2\sqrt{\pi}a^2\Delta\vpp}{\hd}\sigmad.
\end{equation}
One can derive Equations (\ref{eq:dsiddt}) and (\ref{eq:dmpdt}) adopting the local cartesian coordinates for these equations.

\section{Unperturbed state for two-fluid analyses without an external force}\label{app:unpert_state}

In this appendix, we describe the derivation of the unperturbed state adopted in Section \ref{sec:linana_woExF}. The gas surface density and the pressure are assumed to have non-zero but small gradients: $\sigmagup'$, $(\cs^2\sigmagup)'$. We assume $|x|/R<\Delta R/R\ll 1$ so that we can approximate the unperturbed quantities as constants or linear functions of $x$:
\begin{equation}
u_x=u_{x,0}+u_{x,0}'x,\label{eq:ux0_linx}
\end{equation}
\begin{equation}
u_y=-\frac{3}{2}\Omega x+u_{y,0}+u_{y,0}'x,\label{eq:uy0_linx}
\end{equation}
\begin{equation}
\sigmad=\sigmadup+\sigmadup'x,\label{eq:sid0_linx}
\end{equation}
\begin{equation}
v_x=v_{x,0}+v_{x,0}'x,\label{eq:vx0_linx}
\end{equation}
\begin{equation}
v_y=-\frac{3}{2}\Omega x+v_{y,0}+v_{y,0}'x,\label{eq:vy0_linx}
\end{equation}
\begin{equation}
\taus=\tausup+\tausup'x.\label{eq:ts0_linx}
\end{equation}
We also assume that the pressure $\cs^2\sigmag$ is a smooth function in $x$ and can be approximated as a linear function, i.e., $\cs^2\sigmag=\cs^2\sigmagup+(\cs^2\sigmagup)'x$. The temperature structure is assumed in the present analyses. Thus, we have 12 variables to determine the unperturbed state.

To analytically solve the set of 12 equations (Equations (\ref{eq:eocgas_steady})-(\ref{eq:steadyeomy_dust})) is difficult. To approximately derive unperturbed state values, we first make use of the assumption that spatial gradients of the gas pressure and density ($(\cs^2\sigmagup)'$, $\sigmagup'$) are so small that second- and higher-order terms of those gradients can be neglected. This assumption is valid for the following cases: (1) those gradients originate from a global disk profile without a small-scale characteristic structure (e.g., a pressure bump), i.e., $\sigmagup'x/\sigmagup\sim(\cs^2\sigmagup)'x/(\cs^2\sigmagup)\sim x/R$ and (2) the local domain with a width of $\Delta R$ satisfies $x\ll \Delta R\ll R$.
Considering given gas profiles, $\sigmag,\;\cs^2\sigmag$, with such small gradients, we seek a drift solution that requires the velocities, $u_{x,0},\;u_{y,0},\;v_{x,0},\;v_{y,0}$, to be linearly proportional to the gas pressure gradient $(\cs^2\sigmag)'=(\cs^2\sigmagup)'$. In other words, those velocities are already of the first order in $(\cs^2\sigmagup)'$ and $\sigmagup'$ because of the nature of the drift, which one should keep in mind below. We also assume that the spatial gradients of the other variables (e.g., $\sigmadup'$) are so small and satisfies $A'x/A\sim x/R\ll 1$, where $A=\{\sigmadup,\;u_{x,0},\;v_{x,0},\;u_{y,0},\;v_{y,0},\;\tausup$\}. Based on those assumptions, we derive an approximated steady drift solution that is valid at the first order in the gradients, i.e. $\mathcal{O}(x/R)$, by neglecting the higher order terms.

First, the continuity equation for gas (Equation (\ref{eq:eocgas_steady})) gives
\begin{equation}
\sigmagup'u_{x,0}+\sigmagup u_{x,0}'+2\sigmagup'u_{x,0}'x=0.\label{eq:eocgas_steady_b}
\end{equation}
Equation (\ref{eq:eocgas_steady_b}) is satisfied for arbitrary $x$ when the coefficients of the polynominal of $x$ are zero at the first order in $x/R$: 
\begin{equation}
\sigmagup'u_{x,0}+\sigmagup u_{x,0}'=\mathcal{O}\left((x/R)^2\right), \label{eq:eoc_order1}
\end{equation}
\begin{equation}
\sigmagup'u_{x,0}'=\mathcal{O}\left((x/R)^2\right).\label{eq:eoc_order2}
\end{equation}
As noted above, $u_{x,0}$ is the first order in $(\cs^2\sigmagup)'$ for the drift solution. Therefore, the first term in Equation (\ref{eq:eoc_order1}) is the second order. For Equation (\ref{eq:eoc_order1}) to be satisfied at the first order in the gradients, the spatial derivative of $u_x$ should be zero or, more precisely, the higher order in the gradients. This is reasonable for the drift solution because the spatial derivative of $u_{x,0}$ will introduce combinations of $(\cs^2\sigmagup)'$ and the first-order gradient of other variables, meaning that $u_{x,0}'$ is the higher-order term. Therefore, as long as we consider the drift solution, the left hand side of Equation (\ref{eq:eoc_order1}) includes only the higher-order terms of the gradients, and the equality is automatically satisfied. One then finds that Equation (\ref{eq:eoc_order2}) is also satisfied because $\sigmagup'u_{x,0}'$ is the third-order for the drift solution. For simplicity, we set $u_{x,0}'=0$ in the following, which are justified as long as we conduct analyses that are valid at the first order in $x/R$. In the same way, one finds that the continuity equation for dust (Equation (\ref{eq:eocdust_steady})) is also satisfied at the first order for the drift solution we seek. For example, we have $\sigmadup'v_{x,0}\sim\mathcal{O}((x/R)^2)$. We set $v_{x,0}'=0$ for simplicity as in the case of $u_{x,0}'$. This choice is valid in the present analyses of the first order of $x/R$.

Equations (\ref{eq:steadyeomx_gas}), (\ref{eq:steadyeomy_gas}), (\ref{eq:steadyeomx_dust}) and (\ref{eq:steadyeomy_dust}) generally introduce 8 equations to determine the unperturbed state from the coefficients of the polynominal of $x$, i.e., $x^0$ and $x^1$. One can freely fix four variables because the number of variables is larger than the number of equations. In the present analyses, we regard $\sigmadup$ and $\tausup$ as input parameters and set $\sigmadup'=\tausup'=0$. The dust surface density gradient can be safely neglected in a steady state that is valid at the first order in $x/R$. This is because the dust continuity equation is automatically satisfied as long as we consider the drift solution with the small gradients, which includes $\sigmadup'$ (see the above discussion). The assumption of $\tausup'=0$ mimics the drift-limited coagulation, where the dimensionless stopping time tends to be uniform \citep[e.g., see][]{Okuzumi2012}. For the drift solution we seek, the spatial derivative of $\tausup$ only appear in the equations when we keep terms of the higher-order in the gradients, and thus we safely neglect $\tausup'$ in the first-order analyses. In this case, we find the following velocity fields from the equations of motion:
\begin{equation}
u_{x,0}=\frac{2\varepsilon\tausup}{(1+\varepsilon)^2+\tausup^2}\left(-\frac{(\cs^2\sigmagup)'}{2\Omega\sigmagup}\right),
\end{equation}
\begin{equation}
u_{y,0}=-\left[1+\frac{\varepsilon\tausup^2}{(1+\varepsilon)^2+\tausup^2}\right]\left(-\frac{(\cs^2\sigmagup)'}{2\Omega\sigmagup(1+\varepsilon)}\right),
\end{equation}
\begin{equation}
v_{x,0}=-\frac{2\tausup}{(1+\varepsilon)^2+\tausup^2}\left(-\frac{(\cs^2\sigmagup)'}{2\Omega\sigmagup}\right),
\end{equation}
\begin{equation}
v_{y,0}=-\left[1-\frac{\tausup^2}{(1+\varepsilon)^2+\tausup^2}\right]\left(-\frac{(\cs^2\sigmagup)'}{2\Omega\sigmagup(1+\varepsilon)}\right),
\end{equation}
and $v_{y,0}'=u_{y,0}'=0$ at the first order of $x/R$. These are the unperturbed velocity fields presented in Section \ref{sec:linana_woExF}. In this way, instead of invoking external force we find a simple solution valid at the first order of $x/R$ for the unperturbed state with finite drift velocities where $x$-dependence appears only in the gas surface density and pressure.

\section{Sublayer model with uniform surface densities}\label{app:sublayer}

In Section \ref{sec:linana_woExF}, we perform two-fluid linear analyses with vertically-full averaging despite the large difference in vertical scale heights of dust and gas disks. On the other hand, when vertical momentum mixing/transfer in the gas disk is weak, the dust-gas momentum exchange should be averaged within a dust-gas sublayer. Such a sublayer modeling is invoked in some two-fluid linear analyses on dust-gas instabilities \cite[e.g.,][]{Latter2017}. In this appendix, we consider such a limiting case of weak momentum mixing in gas and discuss its effect on coagulation instability. We find that sublayer analyses show insignificant difference, and results can be well reproduced by the one-fluid dispersion relation as long as the total dust-to-gas ratio $\varepsilon$ is smaller than $10^{-2}$.

\begin{figure*}[htp]%[htp] or [H]
	\begin{center}
		%\hspace{-20pt}\raisebox{0pt}{
		%\hspace{20pt}\raisebox{-20pt}{
		\includegraphics[width=0.7\columnwidth]{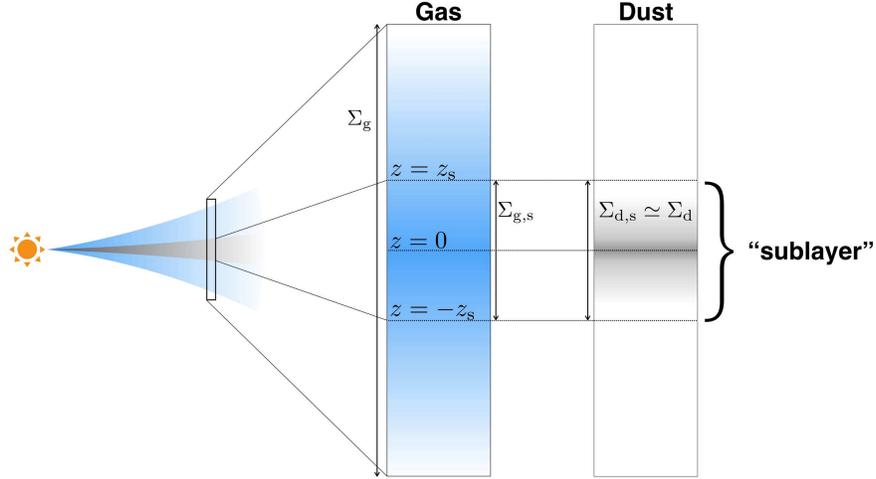}
		%}
	\end{center}
\caption{ Schematic picture of setups in the sublayer model. We assume the upper gas at $|z|>z_{\mathrm{s}}$ to be steady and that dust and gas within the sublayer ($|z|\leq z_{\mathrm{s}}$) evolve during the development of coagulation instability. }
\label{fig:sublayer}
\end{figure*}

In the sublayer model, we divide a gas disk into a sublayer gas and an upper gas that is steady and not perturbed. We denote a surface density of the sublayer gas by $\sigsub$. A surface density of the upper steady gas is given by $\Delta\sigmag\equiv\sigmag-\sigsub$. Vertical boundary positions of the sublayer are denoted by $z=\pm z_{\mathrm{s}}$. Thus, for a stratified gas disk, the sublayer-gas surface density is given as follows
\begin{equation}
\sigsub=\sigmag\mathrm{erf}\left[\frac{z_{\mathrm{s}}}{\sqrt{2}H}\right].
\end{equation}
We adopt $z_{\mathrm{s}}\geq 3\hd$ so that most of dust grains are included in the sublayer. For simplicity, we equate a sublayer-dust surface density $\Sigma_{\dst,\mathrm{s}}$ to the dust surface density $\sigmad$. Larger $z_{\mathrm{s}}$ means that Reynolds stress (or Maxwell stress in a magnetized disk) transfer momentum to higher altitude, and momentum exchange is effectively balanced between dust and larger amount of gas. We neglect momentum transport between sublayer gas and upper gas in order to make analyses as simple as possible. The setup is summarized in Figure \ref{fig:sublayer}.

We assume that $z_{\mathrm{s}}$ is constant in time and space and determined by unperturbed value of dimensionless stopping time $\tausup$. This assumption excludes the effect of $\taus$-dependent midplane dust-gas ratio adopted in Section \ref{subsec:BR_midplane}, which enables us to focus on an effect of sublayer-gas motion on the present instability.

To avoid further complexity, we follow widely-adopted formalism that we add an external force on gas to set up dust-gas drift motion in the unperturbed state \citep[e.g.,][]{Youdin2005,Youdin2007}. Sublayer-gas equations are the following:
\begin{equation}
\ddt{\sigsub}+\ddx{\sigsub u_x}=0,\label{eq:eocsubgas}
\end{equation}
\begin{equation}
\ddt{u_x}+u_x\ddx{u_x}=3\Omega^2x+2\Omega u_y+2\eta R\Omega^2-\frac{\cs^2}{\sigsub}\ddx{\sigsub}+\frac{\sigmad}{\sigsub}\frac{v_x-u_x}{\taus}\Omega,\label{eq:eomxsubgas}
\end{equation}
\begin{equation}
\ddt{u_y}+u_x\ddx{u_y}=-2\Omega u_x+\frac{\sigmad}{\sigsub}\frac{v_y-u_y}{\taus}\Omega,\label{eq:eomysubgas}
\end{equation}
\begin{equation}
\ddt{\Delta\sigmag}=\ddx{\Delta\sigmag}=0,\label{eq:inert_gas}
\end{equation}
where the third term on the right hand side of Equation (\ref{eq:eomxsubgas}) is the external force introduced in the previous studies. For dust equations, we use Equations (\ref{eq:eocdust}), (\ref{eq:eomxdust}), and (\ref{eq:eomydust}). The moment equation for dust coagulation is
\begin{equation}
\ddt{\taus}+v_x\ddx{\taus}=\frac{\sigmad}{\sigmag}\frac{\taus}{3t_0}+\frac{\taus}{\sigmag}\ddx{\sigsub u_x}-\frac{\taus}{\sigmag}v_x\ddx{\sigmag}.\label{eq:dstdt_sub}
\end{equation}
This equation can be derived from Equations (\ref{eq:dmpdt}), (\ref{eq:stEp}), (\ref{eq:eocsubgas}), and (\ref{eq:inert_gas}). We note that coagulation rate is determined by $\sigmad/\sigmag$ even in the sublayer model.

With the aid of the external force in Equation (\ref{eq:eomxsubgas}), we adopt an unperturbed state with uniform surface densities $\sigmagup$, $\sigmadup$, $\sigsubup$. In the current sublayer model, we exclude the stabilizing effect from the unperturbed surface density profile, i.e. the last term of Equation (\ref{eq:dstdt_sub})), because the term is the same as the term in the full-disk-averaging equation and its effect is already discussed in Section \ref{sec:linana_woExF}. The unperturbed velocities are given by the usual drift solution \citep[e.g.,][]{Nakagawa1986}:
\begin{equation}
u_x=u_{x,0}=\frac{2(\sigmadup/\sigsubup)\tausup}{(1+\sigmadup/\sigsubup)^2+\tausup^2}\eta R\Omega,
\end{equation}
\begin{equation}
u_y=-\frac{3}{2}\Omega x+u_{y,0},
\end{equation}
\begin{equation}
u_{y,0}=-\left[1+\frac{(\sigmadup/\sigsubup)\tausup^2}{(1+\sigmadup/\sigsubup)^2+\tausup^2}\right]\frac{\eta R\Omega}{1+\sigmadup/\sigsubup},
\end{equation}
\begin{equation}
v_{x,0}=-\frac{2\tausup}{(1+\sigmadup/\sigsubup)^2+\tausup^2}\eta R \Omega,\label{eq:submodel_vx}
\end{equation}
\begin{equation}
v_y=-\frac{3}{2}\Omega x+v_{y,0},
\end{equation}
\begin{equation}
v_{y,0}=-\left[1-\frac{\tausup^2}{(1+\sigmadup/\sigsubup)^2+\tausup^2}\right]\frac{\eta R\Omega}{1+\sigmadup/\sigsubup},
\end{equation}
As in Section \ref{sec:linana_woExF}, we adopt a value of $\eta$ obtained for the MMSN disk model and $R=20\;\mathrm{au}$. We note that the drift velocities are related to dust-gas ratio averaged in the sublayer: $\sigmadup/\sigsubup$, which is in contrast to analyses in Section \ref{sec:linana_woExF}. Based on the above unperturbed state value, we linearize Equations (\ref{eq:eocdust}), (\ref{eq:eomxdust}), (\ref{eq:eomydust}), (\ref{eq:eocsubgas}), (\ref{eq:eomxsubgas}), (\ref{eq:eomysubgas}), and (\ref{eq:dstdt_sub}) and derive growth rates $n=n_{\mathrm{2fsub}}$ of coagulation instability. We also assume $\delta\sigsub=\delta\sigmag$ based on the assumption of the steady upper gas (Equation (\ref{eq:inert_gas})).

\begin{figure}[htp]%[htp] or [H]
	\begin{center}
		%\hspace{-20pt}\raisebox{0pt}{
		%\hspace{0pt}\raisebox{-20pt}{
		\includegraphics[width=0.5\columnwidth]{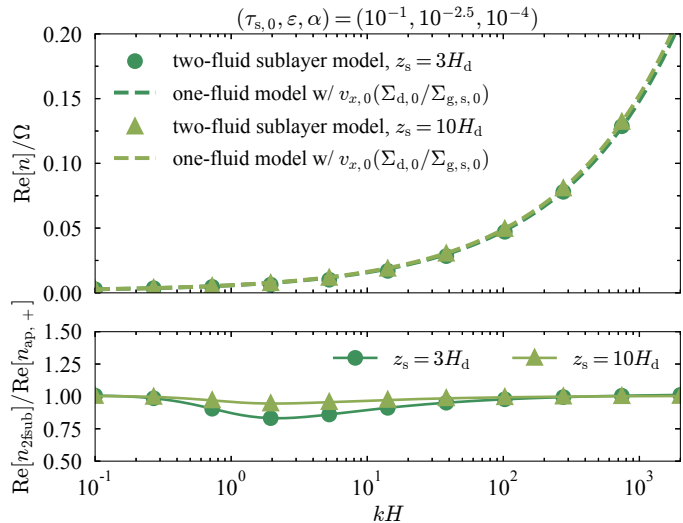}
		%}
	\end{center}
\caption{ The upper panel shows growth rates for $z_{\mathrm{s}}=3\hd,\;10\hd$ and compares the obtained growth rates with the one-fluid growth rates $n_{\app,+}$. We find that the growth rate in the sublayer model can be well reproduced by $n_{\app,+}$. The lower panel shows the ratio of the two growth rates. The ratio has minimum at $kH\simeq2$. We find that the difference from the one-fluid growth rate is only a few tens percent for $z_{\mathrm{s}}=3\hd$ and less for $z_{\mathrm{s}}=10\hd$. }
\label{fig:a1e-4_sublayer}
\end{figure}

An upper panel of Figure \ref{fig:a1e-4_sublayer} shows the growth rates $n_{\mathrm{2fsub}}$ for $z_{\mathrm{s}}=3\hd,\;10\hd$ with the other parameter being fixed: $(\tausup,\varepsilon,\alpha)=(10^{-1},10^{-2.5},10^{-4})$. We also plot one-fluid growth rate $n_{\app,+}$ (Equation (\ref{eq:1fdispersion})) but we substitute the modified unperturbed drift velocity (Equation (\ref{eq:submodel_vx})) in the one-fluid dispersion relation. The growth rates in the two-fluid sublayer model are well reproduced by the one-fluid growth rates. Besides, based on the discussion in Section \ref{sec:linana_woExF}, this also indicates that the sublayer analyses and the full-disk analyses give essentially the same results, and differences are mainly from (1) different drift velocity and (2) the effect of the unperturbed gas density profile (the last term of Equation (\ref{eq:dstdt_sub})). A lower panel of Figure \ref{fig:a1e-4_sublayer} shows a ratio $\mathrm{Re}[n_{\mathrm{2fsub}}]/\mathrm{Re}[n_{\mathrm{ap},+}]$ as a function of $kH$. The difference of two growth rates is maximum at $kH\simeq2$. Nevertheless, the difference from the one-fluid growth rate is only a few tens percent for $z_{\mathrm{s}}=3\hd$ and less for $z_{\mathrm{s}}=10\hd$. The difference monotonically decreases with increasing $z_{\mathrm{s}}$. 

We find that the difference between $\mathrm{Re}[n_{\mathrm{2fsub}}]$ and $\mathrm{Re}[n_{\mathrm{ap},+}]$ for other parameters is also maximum at $kH\simeq1-10$ (e.g., $\alpha=10^{-5}$). The difference from one-fluid growth rates is still about a few tens percent to about 50 percent even for $\alpha=10^{-5}$ when total dust-to-gas ratio is $10^{-3}$ and $\tausup=10^{-1}$, which are realistic values based on the previous dust growth simulations with the MMSN model \citep[e.g., see Section 3.1 of][]{Okuzumi2012}. Therefore, coagulation instability and resulting dust reaccumulation in dust-depleted regions are viable even when the momentum mixing in the gas disk is weak.



%\newpage

\bibliographystyle{aasjournal}
\bibliography{rttominaga2020}

\begin{thebibliography}{}
\expandafter\ifx\csname natexlab\endcsname\relax\def\natexlab#1{#1}\fi

\bibitem[{{Abod} {et~al.}(2019){Abod}, {Simon}, {Li}, {Armitage}, {Youdin}, \&
  {Kretke}}]{Abod2019}
{Abod}, C.~P., {Simon}, J.~B., {Li}, R., {et~al.} 2019, \apj, 883, 192

\bibitem[{{Adachi} {et~al.}(1976){Adachi}, {Hayashi}, \&
  {Nakazawa}}]{Adachi1976}
{Adachi}, I., {Hayashi}, C., \& {Nakazawa}, K. 1976, Progress of Theoretical
  Physics, 56, 1756

\bibitem[{{Auffinger} \& {Laibe}(2018)}]{Auffinger2018}
{Auffinger}, J., \& {Laibe}, G. 2018, \mnras, 473, 796

\bibitem[{{Bai} \& {Stone}(2010)}]{Bai2010c}
{Bai}, X.-N., \& {Stone}, J.~M. 2010, \apjl, 722, L220

\bibitem[{{Bai} \& {Stone}(2014)}]{Bai2014}
---. 2014, \apj, 796, 31

\bibitem[{{Barge} \& {Sommeria}(1995)}]{Barge1995}
{Barge}, P., \& {Sommeria}, J. 1995, \aap, 295, L1

\bibitem[{{Birnstiel} {et~al.}(2009){Birnstiel}, {Dullemond}, \&
  {Brauer}}]{Birnstiel2009}
{Birnstiel}, T., {Dullemond}, C.~P., \& {Brauer}, F. 2009, \aap, 503, L5

\bibitem[{{Birnstiel} {et~al.}(2012){Birnstiel}, {Klahr}, \&
  {Ercolano}}]{Birnstiel2012}
{Birnstiel}, T., {Klahr}, H., \& {Ercolano}, B. 2012, \aap, 539, A148

\bibitem[{{Brauer} {et~al.}(2008){Brauer}, {Dullemond}, \&
  {Henning}}]{Brauer2008}
{Brauer}, F., {Dullemond}, C.~P., \& {Henning}, T. 2008, \aap, 480, 859

\bibitem[{{Carballido} {et~al.}(2006){Carballido}, {Fromang}, \&
  {Papaloizou}}]{Carballido2006}
{Carballido}, A., {Fromang}, S., \& {Papaloizou}, J. 2006, \mnras, 373, 1633

\bibitem[{{Carrera} {et~al.}(2015){Carrera}, {Johansen}, \&
  {Davies}}]{Carrera2015}
{Carrera}, D., {Johansen}, A., \& {Davies}, M.~B. 2015, \aap, 579, A43

\bibitem[{{Carrera} {et~al.}(2020){Carrera}, {Simon}, {Li}, {Kretke}, \&
  {Klahr}}]{Carrera2020}
{Carrera}, D., {Simon}, J.~B., {Li}, R., {Kretke}, K.~A., \& {Klahr}, H. 2020,
  arXiv e-prints, arXiv:2008.01727

\bibitem[{{Casassus} {et~al.}(2015){Casassus}, {Wright}, {Marino}, {Maddison},
  {Wootten}, {Roman}, {P{\'e}rez}, {Pinilla}, {Wyatt}, {Moral}, {M{\'e}nard},
  {Christiaens}, {Cieza}, \& {van der Plas}}]{Casassus2015}
{Casassus}, S., {Wright}, C.~M., {Marino}, S., {et~al.} 2015, \apj, 812, 126

\bibitem[{{Chavanis}(2000)}]{Chavanis2000}
{Chavanis}, P.~H. 2000, \aap, 356, 1089

\bibitem[{{Chen} \& {Lin}(2020)}]{Chen2020}
{Chen}, K., \& {Lin}, M.-K. 2020, \apj, 891, 132

\bibitem[{{Cuzzi} {et~al.}(1993){Cuzzi}, {Dobrovolskis}, \&
  {Champney}}]{Cuzzi1993}
{Cuzzi}, J.~N., {Dobrovolskis}, A.~R., \& {Champney}, J.~M. 1993, \icarus, 106,
  102

\bibitem[{{Dr{\c{a}}{\.z}kowska} \& {Alibert}(2017)}]{Drazkowska2017}
{Dr{\c{a}}{\.z}kowska}, J., \& {Alibert}, Y. 2017, \aap, 608, A92

\bibitem[{{Dr{\c{a}}{\.z}kowska} \& {Dullemond}(2014)}]{Drazkowska2014}
{Dr{\c{a}}{\.z}kowska}, J., \& {Dullemond}, C.~P. 2014, \aap, 572, A78

\bibitem[{{Dubrulle} {et~al.}(1995){Dubrulle}, {Morfill}, \&
  {Sterzik}}]{Dubrulle1995}
{Dubrulle}, B., {Morfill}, G., \& {Sterzik}, M. 1995, \icarus, 114, 237

\bibitem[{{Dzyurkevich} {et~al.}(2010){Dzyurkevich}, {Flock}, {Turner},
  {Klahr}, \& {Henning}}]{Dzyurkevich2010}
{Dzyurkevich}, N., {Flock}, M., {Turner}, N.~J., {Klahr}, H., \& {Henning}, T.
  2010, \aap, 515, A70

\bibitem[{{Estrada} \& {Cuzzi}(2008)}]{Estrada2008}
{Estrada}, P.~R., \& {Cuzzi}, J.~N. 2008, \apj, 682, 515

\bibitem[{{Flock} {et~al.}(2015){Flock}, {Ruge}, {Dzyurkevich}, {Henning},
  {Klahr}, \& {Wolf}}]{Flock2015}
{Flock}, M., {Ruge}, J.~P., {Dzyurkevich}, N., {et~al.} 2015, \aap, 574, A68

\bibitem[{{Fukagawa} {et~al.}(2013){Fukagawa}, {Tsukagoshi}, {Momose}, {Saigo},
  {Ohashi}, {Kitamura}, {Inutsuka}, {Muto}, {Nomura}, {Takeuchi}, {Kobayashi},
  {Hanawa}, {Akiyama}, {Honda}, {Fujiwara}, {Kataoka}, {Takahashi}, \&
  {Shibai}}]{Fukagawa2013}
{Fukagawa}, M., {Tsukagoshi}, T., {Momose}, M., {et~al.} 2013, \pasj, 65, L14

\bibitem[{{Gerbig} {et~al.}(2020){Gerbig}, {Murray-Clay}, {Klahr}, \&
  {Baehr}}]{Gerbig2020}
{Gerbig}, K., {Murray-Clay}, R.~A., {Klahr}, H., \& {Baehr}, H. 2020, \apj,
  895, 91

\bibitem[{{Goldreich} \& {Lynden-Bell}(1965)}]{Goldreich1965}
{Goldreich}, P., \& {Lynden-Bell}, D. 1965, \mnras, 130, 125

\bibitem[{{Gole} {et~al.}(2020){Gole}, {Simon}, {Li}, {Youdin}, \&
  {Armitage}}]{Gole2020}
{Gole}, D.~A., {Simon}, J.~B., {Li}, R., {Youdin}, A.~N., \& {Armitage}, P.~J.
  2020, \apj, 904, 132

\bibitem[{{Gonzalez} {et~al.}(2017){Gonzalez}, {Laibe}, \&
  {Maddison}}]{Gonzalez2017}
{Gonzalez}, J.~F., {Laibe}, G., \& {Maddison}, S.~T. 2017, \mnras, 467, 1984

\bibitem[{{Goodman} \& {Pindor}(2000)}]{Goodman2000}
{Goodman}, J., \& {Pindor}, B. 2000, \icarus, 148, 537

\bibitem[{Harris {et~al.}(2020)Harris, Millman, van~der Walt, Gommers,
  Virtanen, Cournapeau, Wieser, Taylor, Berg, Smith, Kern, Picus, Hoyer, van
  Kerkwijk, Brett, Haldane, del R{\'{i}}o, Wiebe, Peterson,
  G{\'{e}}rard-Marchant, Sheppard, Reddy, Weckesser, Abbasi, Gohlke, \&
  Oliphant}]{Harris2020}
Harris, C.~R., Millman, K.~J., van~der Walt, S.~J., {et~al.} 2020, Nature, 585,
  357

\bibitem[{{Hayashi}(1981)}]{Hayashi1981}
{Hayashi}, C. 1981, Progress of Theoretical Physics Supplement, 70, 35

\bibitem[{{Hunter}(2007)}]{Hunter2007}
{Hunter}, J.~D. 2007, Computing in Science and Engineering, 9, 90

\bibitem[{{Jacquet} {et~al.}(2011){Jacquet}, {Balbus}, \&
  {Latter}}]{Jacquet2011}
{Jacquet}, E., {Balbus}, S., \& {Latter}, H. 2011, \mnras, 415, 3591

\bibitem[{{Johansen} {et~al.}(2007){Johansen}, {Oishi}, {Mac Low}, {Klahr},
  {Henning}, \& {Youdin}}]{Johansen2007nature}
{Johansen}, A., {Oishi}, J.~S., {Mac Low}, M.-M., {et~al.} 2007, \nat, 448,
  1022

\bibitem[{{Johansen} {et~al.}(2009{\natexlab{a}}){Johansen}, {Youdin}, \&
  {Klahr}}]{Johansen2009a}
{Johansen}, A., {Youdin}, A., \& {Klahr}, H. 2009{\natexlab{a}}, \apj, 697,
  1269

\bibitem[{{Johansen} {et~al.}(2009{\natexlab{b}}){Johansen}, {Youdin}, \& {Mac
  Low}}]{Johansen2009b}
{Johansen}, A., {Youdin}, A., \& {Mac Low}, M.-M. 2009{\natexlab{b}}, \apjl,
  704, L75

\bibitem[{{Kataoka} {et~al.}(2013{\natexlab{a}}){Kataoka}, {Tanaka}, {Okuzumi},
  \& {Wada}}]{Kataoka2013a}
{Kataoka}, A., {Tanaka}, H., {Okuzumi}, S., \& {Wada}, K. 2013{\natexlab{a}},
  \aap, 557, L4

\bibitem[{{Kataoka} {et~al.}(2013{\natexlab{b}}){Kataoka}, {Tanaka}, {Okuzumi},
  \& {Wada}}]{Kataoka2013b}
---. 2013{\natexlab{b}}, \aap, 554, A4

\bibitem[{{Kimura} {et~al.}(2015){Kimura}, {Wada}, {Senshu}, \&
  {Kobayashi}}]{Kimura2015}
{Kimura}, H., {Wada}, K., {Senshu}, H., \& {Kobayashi}, H. 2015, \apj, 812, 67

\bibitem[{{Kobayashi} {et~al.}(2016){Kobayashi}, {Tanaka}, \&
  {Okuzumi}}]{Kobayashi2016}
{Kobayashi}, H., {Tanaka}, H., \& {Okuzumi}, S. 2016, \apj, 817, 105

\bibitem[{{Krapp} {et~al.}(2019){Krapp}, {Ben{\'\i}tez-Llambay}, {Gressel}, \&
  {Pessah}}]{Krapp2019}
{Krapp}, L., {Ben{\'\i}tez-Llambay}, P., {Gressel}, O., \& {Pessah}, M.~E.
  2019, \apjl, 878, L30

\bibitem[{{Kretke} \& {Lin}(2007)}]{Kretke2007}
{Kretke}, K.~A., \& {Lin}, D.~N.~C. 2007, \apjl, 664, L55

\bibitem[{{Krijt} {et~al.}(2015){Krijt}, {Ormel}, {Dominik}, \&
  {Tielens}}]{Krijt2015}
{Krijt}, S., {Ormel}, C.~W., {Dominik}, C., \& {Tielens}, A.~G.~G.~M. 2015,
  \aap, 574, A83

\bibitem[{{Lambrechts} {et~al.}(2016){Lambrechts}, {Johansen}, {Capelo},
  {Blum}, \& {Bodenschatz}}]{Lambrechts2016}
{Lambrechts}, M., {Johansen}, A., {Capelo}, H.~L., {Blum}, J., \&
  {Bodenschatz}, E. 2016, \aap, 591, A133

\bibitem[{{Latter} \& {Rosca}(2017)}]{Latter2017}
{Latter}, H.~N., \& {Rosca}, R. 2017, \mnras, 464, 1923

\bibitem[{{Lin} \& {Youdin}(2017)}]{Lin2017}
{Lin}, M.-K., \& {Youdin}, A.~N. 2017, \apj, 849, 129

\bibitem[{{Lyra} \& {Lin}(2013)}]{Lyra2013}
{Lyra}, W., \& {Lin}, M.-K. 2013, \apj, 775, 17

\bibitem[{{McNally} {et~al.}(2021){McNally}, {Lovascio}, \&
  {Paardekooper}}]{McNally2021}
{McNally}, C.~P., {Lovascio}, F., \& {Paardekooper}, S.-J. 2021, \mnras, 502,
  1469

\bibitem[{{Nakagawa} {et~al.}(1981){Nakagawa}, {Nakazawa}, \&
  {Hayashi}}]{Nakagawa1981}
{Nakagawa}, Y., {Nakazawa}, K., \& {Hayashi}, C. 1981, \icarus, 45, 517

\bibitem[{{Nakagawa} {et~al.}(1986){Nakagawa}, {Sekiya}, \&
  {Hayashi}}]{Nakagawa1986}
{Nakagawa}, Y., {Sekiya}, M., \& {Hayashi}, C. 1986, \icarus, 67, 375

\bibitem[{{Okuzumi} \& {Hirose}(2012)}]{OH2012}
{Okuzumi}, S., \& {Hirose}, S. 2012, \apjl, 753, L8

\bibitem[{{Okuzumi} {et~al.}(2016){Okuzumi}, {Momose}, {Sirono}, {Kobayashi},
  \& {Tanaka}}]{Okuzumi2016}
{Okuzumi}, S., {Momose}, M., {Sirono}, S.-i., {Kobayashi}, H., \& {Tanaka}, H.
  2016, \apj, 821, 82

\bibitem[{{Okuzumi} {et~al.}(2012){Okuzumi}, {Tanaka}, {Kobayashi}, \&
  {Wada}}]{Okuzumi2012}
{Okuzumi}, S., {Tanaka}, H., {Kobayashi}, H., \& {Wada}, K. 2012, \apj, 752,
  106

\bibitem[{{Ormel} \& {Cuzzi}(2007)}]{Ormel2007}
{Ormel}, C.~W., \& {Cuzzi}, J.~N. 2007, \aap, 466, 413

\bibitem[{{Ormel} \& {Spaans}(2008)}]{Ormel2008}
{Ormel}, C.~W., \& {Spaans}, M. 2008, \apj, 684, 1291

\bibitem[{{Paardekooper} {et~al.}(2020){Paardekooper}, {McNally}, \&
  {Lovascio}}]{Paardekooper2020}
{Paardekooper}, S.-J., {McNally}, C.~P., \& {Lovascio}, F. 2020, \mnras, 499,
  4223

\bibitem[{{Paardekooper} {et~al.}(2021){Paardekooper}, {McNally}, \&
  {Lovascio}}]{Paardekooper2021}
---. 2021, \mnras, 502, 1579

\bibitem[{{Pierens}(2021)}]{Pierens2021}
{Pierens}, A. 2021, \mnras, arXiv:2101.07762

\bibitem[{{Raettig} {et~al.}(2015){Raettig}, {Klahr}, \& {Lyra}}]{Raettig2015}
{Raettig}, N., {Klahr}, H., \& {Lyra}, W. 2015, \apj, 804, 35

\bibitem[{{Safronov}(1972)}]{Safronov1972}
{Safronov}, V.~S. 1972, {Evolution of the protoplanetary cloud and formation of
  the earth and planets.}

\bibitem[{{Saito} \& {Sirono}(2011)}]{Saito2011}
{Saito}, E., \& {Sirono}, S.-i. 2011, \apj, 728, 20

\bibitem[{{Sato} {et~al.}(2016){Sato}, {Okuzumi}, \& {Ida}}]{Sato2016}
{Sato}, T., {Okuzumi}, S., \& {Ida}, S. 2016, \aap, 589, A15

\bibitem[{{Schoonenberg} \& {Ormel}(2017)}]{Schoonenberg2017}
{Schoonenberg}, D., \& {Ormel}, C.~W. 2017, \aap, 602, A21

\bibitem[{{Schoonenberg} {et~al.}(2018){Schoonenberg}, {Ormel}, \&
  {Krijt}}]{Schoonenberg2018}
{Schoonenberg}, D., {Ormel}, C.~W., \& {Krijt}, S. 2018, \aap, 620, A134

\bibitem[{{Schumann}(1940)}]{Schumann1940}
{Schumann}, T.~E.~W. 1940, Quarterly Journal of the Royal Meteorological
  Society, 66, 195

\bibitem[{{Shakura} \& {Sunyaev}(1973)}]{Shakura1973}
{Shakura}, N.~I., \& {Sunyaev}, R.~A. 1973, \aap, 24, 337

\bibitem[{{Shu}(1992)}]{Shu1992}
{Shu}, F.~H. 1992, {The physics of astrophysics. Volume II: Gas dynamics.}
  (University Science Books, Mill Valley, CA (USA))

\bibitem[{{Simon} {et~al.}(2016){Simon}, {Armitage}, {Li}, \&
  {Youdin}}]{Simon2016}
{Simon}, J.~B., {Armitage}, P.~J., {Li}, R., \& {Youdin}, A.~N. 2016, \apj,
  822, 55

\bibitem[{{Smoluchowski}(1916)}]{Smoluchowski1916}
{Smoluchowski}, M.~V. 1916, Zeitschrift fur Physik, 17, 557

\bibitem[{{Squire} \& {Hopkins}(2018{\natexlab{a}})}]{Squire2018a}
{Squire}, J., \& {Hopkins}, P.~F. 2018{\natexlab{a}}, \mnras, 477, 5011

\bibitem[{{Squire} \& {Hopkins}(2018{\natexlab{b}})}]{Squire2018b}
---. 2018{\natexlab{b}}, \apjl, 856, L15

\bibitem[{{Steinpilz} {et~al.}(2019){Steinpilz}, {Teiser}, \&
  {Wurm}}]{Steinpilz2019}
{Steinpilz}, T., {Teiser}, J., \& {Wurm}, G. 2019, \apj, 874, 60

\bibitem[{{Stevenson} \& {Lunine}(1988)}]{Stevenson1988}
{Stevenson}, D.~J., \& {Lunine}, J.~I. 1988, \icarus, 75, 146

\bibitem[{{Takahashi} \& {Inutsuka}(2014)}]{Takahashi2014}
{Takahashi}, S.~Z., \& {Inutsuka}, S.-i. 2014, \apj, 794, 55

\bibitem[{{Taki} {et~al.}(2016){Taki}, {Fujimoto}, \& {Ida}}]{Taki2016}
{Taki}, T., {Fujimoto}, M., \& {Ida}, S. 2016, \aap, 591, A86

\bibitem[{{Taki} {et~al.}(2021){Taki}, {Kuwabara}, {Kobayashi}, \&
  {Suzuki}}]{Taki2021}
{Taki}, T., {Kuwabara}, K., {Kobayashi}, H., \& {Suzuki}, T.~K. 2021, \apj,
  909, 75

\bibitem[{{Tominaga} {et~al.}(2018){Tominaga}, {Inutsuka}, \&
  {Takahashi}}]{Tominaga2018}
{Tominaga}, R.~T., {Inutsuka}, S.-i., \& {Takahashi}, S.~Z. 2018, \pasj, 70, 3

\bibitem[{{Tominaga} {et~al.}(2019){Tominaga}, {Takahashi}, \&
  {Inutsuka}}]{Tominaga2019}
{Tominaga}, R.~T., {Takahashi}, S.~Z., \& {Inutsuka}, S.-i. 2019, \apj, 881, 53

\bibitem[{{Tominaga} {et~al.}(2020){Tominaga}, {Takahashi}, \&
  {Inutsuka}}]{Tominaga2020}
---. 2020, \apj, 900, 182

\bibitem[{{Ueda} {et~al.}(2019){Ueda}, {Flock}, \& {Okuzumi}}]{Ueda2019}
{Ueda}, T., {Flock}, M., \& {Okuzumi}, S. 2019, \apj, 871, 10

\bibitem[{{Umurhan} {et~al.}(2020){Umurhan}, {Estrada}, \&
  {Cuzzi}}]{Umurhan2020}
{Umurhan}, O.~M., {Estrada}, P.~R., \& {Cuzzi}, J.~N. 2020, \apj, 895, 4

\bibitem[{{van der Marel} {et~al.}(2013){van der Marel}, {van Dishoeck},
  {Bruderer}, {Birnstiel}, {Pinilla}, {Dullemond}, {van Kempen}, {Schmalzl},
  {Brown}, {Herczeg}, {Mathews}, \& {Geers}}]{van-der-Marel2013}
{van der Marel}, N., {van Dishoeck}, E.~F., {Bruderer}, S., {et~al.} 2013,
  Science, 340, 1199

\bibitem[{{Virtanen} {et~al.}(2020){Virtanen}, {Gommers}, {Oliphant},
  {Haberland}, {Reddy}, {Cournapeau}, {Burovski}, {Peterson}, {Weckesser},
  {Bright}, {van der Walt}, {Brett}, {Wilson}, {Millman}, {Mayorov}, {Nelson},
  {Jones}, {Kern}, {Larson}, {Carey}, {Polat}, {Feng}, {Moore}, {VanderPlas},
  {Laxalde}, {Perktold}, {Cimrman}, {Henriksen}, {Quintero}, {Harris},
  {Archibald}, {Ribeiro}, {Pedregosa}, {van Mulbregt}, \& {SciPy 1. 0
  Contributors}}]{Virtanen2020}
{Virtanen}, P., {Gommers}, R., {Oliphant}, T.~E., {et~al.} 2020, Nature
  Methods, 17, 261

\bibitem[{{V\"{o}lk} {et~al.}(1980){V\"{o}lk}, {Jones}, {Morfill}, \&
  {Roeser}}]{Volk1980}
{V\"{o}lk}, H.~J., {Jones}, F.~C., {Morfill}, G.~E., \& {Roeser}, S. 1980,
  \aap, 85, 316

\bibitem[{{Wada} {et~al.}(2013){Wada}, {Tanaka}, {Okuzumi}, {Kobayashi},
  {Suyama}, {Kimura}, \& {Yamamoto}}]{Wada2013}
{Wada}, K., {Tanaka}, H., {Okuzumi}, S., {et~al.} 2013, \aap, 559, A62

\bibitem[{{Wada} {et~al.}(2009){Wada}, {Tanaka}, {Suyama}, {Kimura}, \&
  {Yamamoto}}]{Wada2009}
{Wada}, K., {Tanaka}, H., {Suyama}, T., {Kimura}, H., \& {Yamamoto}, T. 2009,
  \apj, 702, 1490

\bibitem[{{Ward}(2000)}]{Ward2000}
{Ward}, W.~R. 2000, {On Planetesimal Formation: The Role of Collective Particle
  Behavior}, ed. R.~M. {Canup}, K.~{Righter}, \& {et al.} (Tucson, AZ: Univ.
  Arizona Press), 75--84

\bibitem[{{Weidenschilling}(1977)}]{Weidenschilling1977}
{Weidenschilling}, S.~J. 1977, \mnras, 180, 57

\bibitem[{{Whipple}(1972)}]{Whipple1972}
{Whipple}, F.~L. 1972, in From Plasma to Planet, ed. A.~{Elvius}, 211

\bibitem[{{{Wolfram Research, Inc.}}(2019)}]{Wolfram}
{{Wolfram Research, Inc.}} 2019, Mathematica, {V}ersion 12.0, , , champaign,
  IL, 2019

\bibitem[{{Yang} {et~al.}(2017){Yang}, {Johansen}, \& {Carrera}}]{Yang2017}
{Yang}, C.~C., {Johansen}, A., \& {Carrera}, D. 2017, \aap, 606, A80

\bibitem[{{Youdin} \& {Johansen}(2007)}]{Youdin2007}
{Youdin}, A., \& {Johansen}, A. 2007, \apj, 662, 613

\bibitem[{{Youdin}(2005)}]{Youdin2005a}
{Youdin}, A.~N. 2005, ArXiv Astrophysics e-prints, astro-ph/0508659

\bibitem[{{Youdin}(2011)}]{Youdin2011}
---. 2011, \apj, 731, 99

\bibitem[{{Youdin} \& {Goodman}(2005)}]{Youdin2005}
{Youdin}, A.~N., \& {Goodman}, J. 2005, \apj, 620, 459

\bibitem[{{Youdin} \& {Lithwick}(2007)}]{YL2007}
{Youdin}, A.~N., \& {Lithwick}, Y. 2007, \icarus, 192, 588

\bibitem[{{Zhu} \& {Yang}(2021)}]{Zhu2021}
{Zhu}, Z., \& {Yang}, C.-C. 2021, \mnras, 501, 467

\end{thebibliography}

%\begin{thebibliography}{}
%
%\bibitem[Astropy Collaboration et al.(2013)]{2013A&A...558A..33A} Astropy Collaboration, Robitaille, T.~P., Tollerud, E.~J., et al.\ 2013, \aap, 558, A33 
%\bibitem[Bertin \& Arnouts(1996)]{1996A&AS..117..393B} Bertin, E., \& Arnouts, S.\ 1996, \aaps, 117, 393 
%\bibitem[Corrales(2015)]{2015ApJ...805...23C} Corrales, L.\ 2015, \apj, 805, 23
%\bibitem[Ferland et al.(2013)]{2013RMxAA..49..137F} Ferland, G.~J., Porter, R.~L., van Hoof, P.~A.~M., et al.\ 2013, \rmxaa, 49, 137
%\bibitem[Hanisch \& Biemesderfer(1989)]{1989BAAS...21..780H} Hanisch, R.~J., \& Biemesderfer, C.~D.\ 1989, \baas, 21, 780 
%\bibitem[Lamport(1994)]{lamport94} Lamport, L. 1994, LaTeX: A Document Preparation System, 2nd Edition (Boston, Addison-Wesley Professional)
%\bibitem[Schwarz et al.(2011)]{2011ApJS..197...31S} Schwarz, G.~J., Ness, J.-U., Osborne, J.~P., et al.\ 2011, \apjs, 197, 31  
%\bibitem[Vogt et al.(2014)]{2014ApJ...793..127V} Vogt, F.~P.~A., Dopita, M.~A., Kewley, L.~J., et al.\ 2014, \apj, 793, 127  
%
%\end{thebibliography}


\end{document}